# Title

## Physical phenomena during nanoindentation deformation of amorphous glassy polymers

First author: Prakash Sarkar

PhD student, Department of Metallurgical Engineering and Materials Science, Indian Institute of Technology Bombay, Mumbai- 400076, Maharashtra, India

ORCID: 0000-0002-4404-916X

Second author: Prita Pant

Professor, Department of Metallurgical Engineering and Materials Science, Indian Institute of Technology Bombay, Mumbai- 400076, Maharashtra, India

ORCID: 0000-0002-0807-4157

Corresponding author: Hemant Nanavati

Professor, Department of Chemical Engineering, Indian Institute of Technology Bombay, Mumbai- 400076, Maharashtra, India

*E-mail: hnanavati@iitb.ac.in

ORCID: 0000-0002-5982-6531

# Graphical Abstract

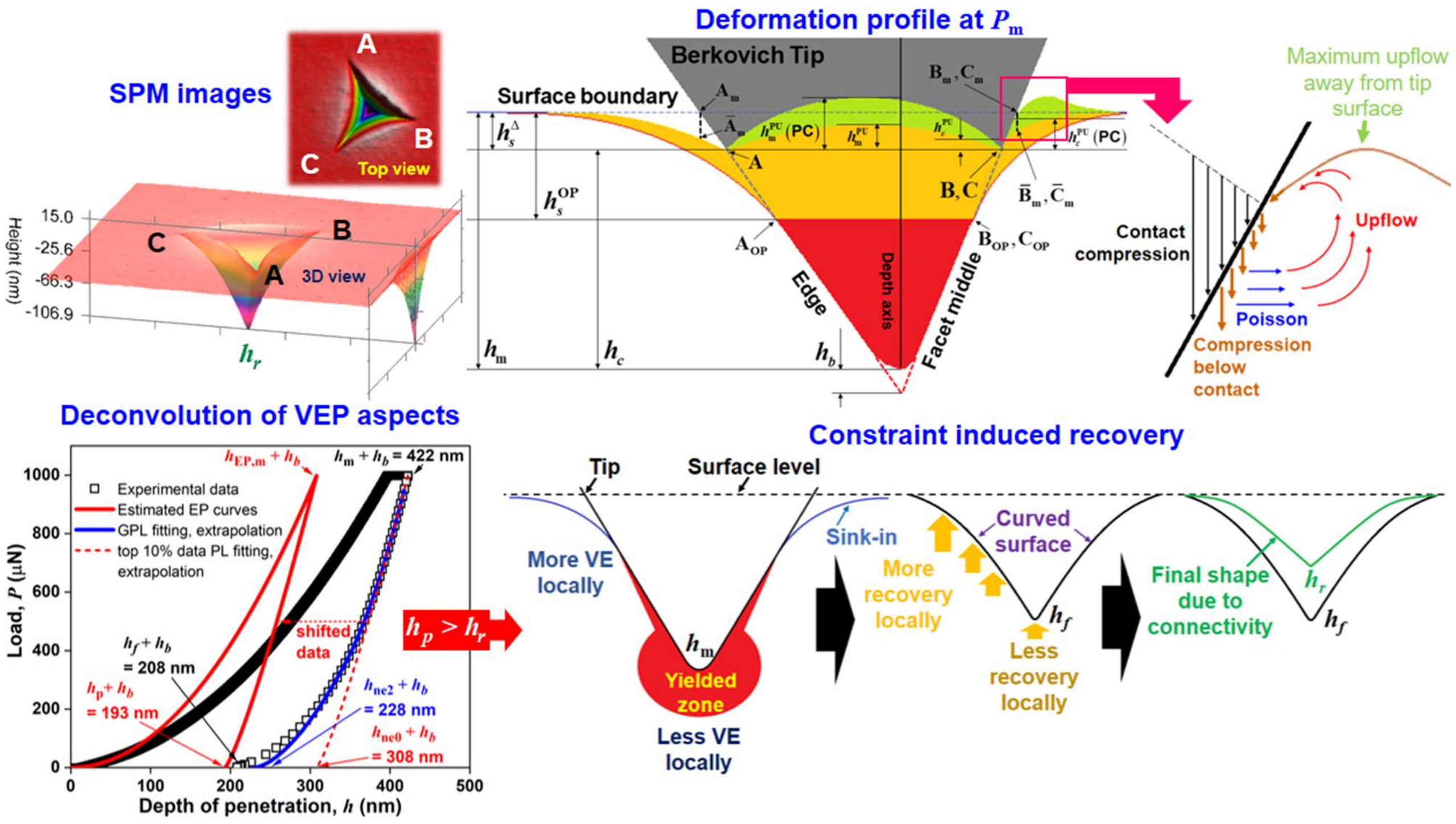

# Abstract

We identify for visco-elasto-plastic (VEP) glassy polymers, physical phenomena during Berkovich nanoindentation, a locally imposed deformation. Live visuals via *in situ* nanoindentation indicate mainly sink-in during loading, with pile-up after unloading. Scanning Probe Microscopy (SPM) indicates significant volume conserving upflow below the tip, for these high $\nu$, compliant materials, with compliance correlated high geometric fractional contact, $h_c^{\Delta} / h_{\mathrm{m}}^{\Delta} \sim 0.86 - 0.95$. We adapt the ideal conical indentation framework to VEP Berkovich nanoindentation, to calculate the contact area and visually depict the upflow and the displacement paths, in the material. The combination of SPM and $P-h$ data, indicates a mixed comparison with uniaxial modulus and yield stress, with conventionally defined hardness, $H < 3\sigma_y$, and nanoindentation modulus, $E_{\mathrm{N}} > E$. By rationally removing viscoelastic (VE) effects from the loading $P-h$ data, we find instant, zero-time hardness $H_{L0} > 3\sigma_y$ . We apply the power law model to only the recovery onset, to estimate pure elastic recovery. We then deconvolute the VEP nanoindentation into the conventional EP and elastic contributions, isolating the VE component. Constraint-induced sink-in, pile-up and VE recovery of the highly yielded tip-apex region, mirror the converse constrained deformation effects, governing the trends in conventionally defined $E_{\mathrm{N}}$ and $H_{L0}$ estimates for glassy polymers.

# 1. Introduction

Nanoindentation enables determination of material behaviour of very small systems and devices [1] over a region ~ 10 μm$^2$ – 100 μm$^2$, and up to a depth of ~100nm – 5 μm. The material response to nanoindentation loads, is significantly richer in variety and distribution, than that to bulk loading methods. Deformation via bulk loads is broadly uniform, i.e., either tensile or compressive for uniaxial loads, or well graded from tensile to compressive for bending experiments, or measurable shear. Only if the deformation is beyond the elastic limit, does plastic flow occur. Viscous effects, which participate, depending on the rate of deformation, do so uniformly over the bulk.

In contrast, nanoindentation is locally imposed deformation. Load controlled (LC) nanoindentation experiments involve the relationship between the applied load, $P$ and the resultant penetration, $h$, followed by that in the unloading process. Localized elastic behaviour and viscoplastic (VP) flow can occur simultaneously, and exhibit a complex distribution, from the onset of measurable deformation. The full load contact depth, $h_c = h_{\mathrm{m}} - h_s$; $h_s$ is the non-contact sink-in depth for the maximum depth, $h_{\mathrm{m}}$. The stressed, contact region deformation is constrained by the surrounding non-contact region. Conversely, the sink-in in the immediate surrounding non-contact region, is a constraint-induced deformation, caused by the contact region deformation.

The framework of Oliver and Pharr [2]–[4] (OP), has been employed almost universally, to analyse the $P-h$ data, since they have demonstrated its validity over a broad range of elasto-plastic (EP) materials. The OP framework follows the elastic nanoindentation analysis of Love [5] and Sneddon [6]–[8] (LS). The LS analysis essentially correlates the elastic loading $P-h$ data in terms of the elastic constants, which include the modulus, $E$, and the Poisson ratio, $\nu$. In their conical indenter analysis, $\left(h_c/h_{\mathrm{m}}\right) = \left(2/\pi\right)$, is a boundary condition. The displacements in the contact and non-contact regions, contain axial and radial components [9], [10]. $A_c(h_c)$ is the geometry based, projected contact area. The strain is implicitly correlated to $h_{\mathrm{m}}$ and the effective true stress or hardness (at maximum load, $P_{\mathrm{m}}$), $H = P_{\mathrm{m}}/A_c$.

The OP method applies the ideas of Bulychev *et al.* [11], employing stiffness, $S = (dP/dh)\big|_{P_{\mathrm{m}},h_{\mathrm{m}}}$, the unloading onset slope, to yield $h_c$. This relates the elastically recoverable strain measure due to $h_{\mathrm{m}}$, to its EP deformation-based contact area measure. Thus, the nanoindentation modulus, $E_{\mathrm{N}} \approx (1-\nu^2)S\sqrt{\pi}\big/\big(2\sqrt{A_c}\big)$ (for relatively compliant materials such as polymers). For a broad set of EP materials, $E_{\mathrm{N}} \sim E$ and $H \sim 3\sigma_y$ ($\sigma_y$ is the yield stress) [12], [13]. In addition to sink-in, nanoindentation also causes a surface level pile-up, adjacent to the indent [14]–[25], which, if small, is ignored by the OP framework. For significant pile-up, the framework by Loubet [26] disregards any sink-in and considers $h_c > h_{\mathrm{m}}$.

In our work, we examine Berkovich nanoindentation of amorphous glassy polymers. These are visco-elasto-plastic (VEP) compliant materials, and they were not among the materials which led to OP framework [2]–[4]. However, they have still undergone nanoindentation analyses via the OP method, although empirical corrections have been reported (e.g., [27], [28]).

In contrast to the OP framework, numerical simulation approaches to examining nanoindentation, implement Finite Element (FE) Methods (FEM) or Finite Element Analyses (FEA) [29]–[57]. They broadly identify material parameters (including via user-defined models) and phenomena which match the $P-h$ data, to infer insights such as stress and strain distributions in the sample. However, in case of a complex nanoindentation deformation of such compliant VEP materials, predictive modelling and FEA need to be consistent with a greater range of experimental observations, beyond $P-h$ data.

Our experimental analysis recognizes the high compliance and high $\nu$ of amorphous glassy polymers, to identify relevant aspects of consequent material behaviour. Direct observation from i*n situ* nanoindentation and topographical imaging, provide explicit evidence of sink-in, pile-up and high contact fraction. We analyse our observations by first providing indentation profiles, to suggest likely deformation flow paths. Combining with the $P-h$ data, we progress to examine the VE effects on estimates of the conventionally defined EP nanoindentation properties, $E_{\mathrm{N}}$ and $H$, to identify the VE, EP and elastic components of the VEP material deformation. This decomposition, then, extends our analysis to interactions between the surrounding region and the locally imposed deformation region. Thus, we systematically traverse multiple levels and provide phenomenological perpectives. An analysis such as ours, would thus provide guidelines to the way forward for formal characterization of polymer nanoindentation, via predictive modelling and numerical simulations.

# 2. Background

The OP method provides the metrics, reduced modulus, $E_r\left(\approx E_{\mathrm{N}}/\left(1-\nu^2\right)\right)$ and $H$, of EP nanoindentation phenomena. It is applicable for $h_f/h_{\mathrm{m}}$ values up to 0.7 [2] or 0.83 [58], where $h_f$ is the final depth at full unload. This method considers that (i) both, the elastic as well as plastic components of the loading are parabolic, i.e., $P \propto h^2$ [15] (ii) the unloading data (from 0.95 $P_{\mathrm{m}}$ to 0.2 $P_{\mathrm{m}}$), correspond to elastic recovery, and satisfy the power law (PL), $P = B\left(h-h_f\right)^m$ (iii) the resulting estimate of the stiffness, *S*, yields $h_s = \left(\varepsilon P_{\mathrm{m}}/S\right)$. For a Berkovich tip, $\varepsilon$ = 0.75 [2].

For an ideal Berkovich tip, $A_c = C_0 h_c^2$, where $C_0$ = $\left(3\sqrt{3}\tan^2\alpha\right)$ ≈ 24.5; $\alpha$ = 65.27°, is the semi angle of the Berkovich tip. The wear of the tip apex with usage, is accounted for by regular calibration with standard materials, whose $H$ and $E_r$ are known (usually quartz). The back-calculated contact areas are fit to $A_c = \sum_{i=1}^{8} C_i h_c^{2^{-(i-1)}}$, where $C_i$ are the bluntness constants [2].

OP validated their relationship between $S$ and $h_c$, by comparing their $A_c$ estimates, with those obtained by direct imaging of the imprint area, on six EP materials: aluminum (modulus ~68GPa), fused silica, soda lime glass, quartz, tungsten, sapphire (modulus ~440GPa); we will call these materials the OP materials. A triangular imprint area is formed by connecting by lines (length=*a*) [2], [13], [59], the outer corners of the permanent Berkovich edge imprints; $A_c = A_c^{\Delta} = \sqrt{3}a^2/4$.

For indentation of compliant VEP materials such as glassy polymers (moduli <~5 GPa), $h_c$ values would be much greater than for quartz, and the $A_c$ calibration needs to extend to corresponding depths. Therefore, $A_c$ calibration in terms of OP-based $h_c$, has been reported via Polycarbonate (PC) [60]–[62], other soft materials such as PS-4™ [63] and PS-6™ [64] photoelastic polymer coatings (Vishay Micro-Measurements, USA), and Polymethyl methacrylate (PMMA) [65]. The tensile modulus, *E*, considered for OP-based estimates of $A_c$, was that of the material being indented when the polymer was a vendor-supplied standard [61]–[64]; in some cases [60], [65], it was a literature value. In all cases, the underlying assumption was that $E_{\mathrm{N}} \sim E$. Therefore, the use of polymers as standards has been critiqued by Tranchida *et al.* [28]. In this context, a depth-independent calibration considers the wear via different approaches of estimating the blunted height, $h_b$, so that $A_c(h_c) \approx 24.5 \times (h_c + h_b)^2$ [30], [66]–[72].

Progress beyond the OP framework, in describing the nanoindentation deformation phenomena, has been in two broad categories. One category is in terms of numerical simulations such as FEA, as described in the previous section. The second category, over the broad gamut of the literature, considers the physical phenomena based on nanoindentation properties, $E_r$ and $H$. These in turn, have been obtained via the OP framework, by estimating $S$ and $A_c$; e.g., modifications such as polynomial fits (as well as normalizations $P/P_{\mathrm{m}}$ and $h/h_{\mathrm{m}}$) to the unloading data [73]–[76], have been reported. References [75], [76] recognize that in LC experiments, $h$ is the dependent variable and $P$ is the independent variable. However, on applying these OP-type concepts to a VE material (the polymer, PA12), resulted in an unphysical fit, $h_f \sim 10^8$.

We note that these fits have achieved excellent $R^2$ values. However, fidelity to the phenomena, requires appropriate fitting residuals as well. Thus, our earlier work [77], [78], suggests non-applicability of the PL for polymeric systems, over the prescribed unloading range. Then, the corollary that the OP method could determine $h_c$ for polymers (including standards [61]–[65]), should be re-examined. Hence, more reliable methods are necessary.

AFM [67], [79]–[85], and SPM (scanning probe microscopy) [20], [59], [86]–[88] are methods, where the indenter tip itself scans the residual indent, and enables topological determination of the $A_c$. SEM [89] and optical photographs [90] of the residual imprints, have also been employed to estimate $A_c$. Johnson's spherical conical cavity model [92], modified to match the OP framework, yields $A_c$ values, correlating the indentation properties to the parameters of vertical sections of the residue imprints, at different loads [67], [82]. For transparent elastomers, horizontal optical full load views from beneath provide $A_c$ estimates, and vertical views from alongside yield $h_s$ [57], [91].

The pile-up indicated above as reported in the literature, has been discerned via topographical images of the residual nanoindentation imprint. Based on $h_{\mathrm{m}}^{\mathrm{PU}}$, the imaged maximum post-unload pile-up (PU) height (see [19], [20]) $A_c(h_c) \approx 24.5 \times \left(h_c + h_{\mathrm{m}}^{\mathrm{PU}}\right)^2$, or by disregarding sink-in, if pile-up is observed via SPM [20]; i.e., $A_c \approx 24.5\times\left(h_{\mathrm{m}} + h_{\mathrm{m}}^{\mathrm{PU}}\right)^2$. This method assumes that the entire post-unload pile-up, exists at full-load, and is in load-bearing contact with the tip. A more detailed method [16], [17], [21]–[25], considers the area of a sector of the circle of maximum height, $h_{\mathrm{m}}^{\mathrm{PU}}$, as the contact pile-up area, $A_c^{\mathrm{PU}}$, to be added to $A_c^{\Delta}$. Randall *et al.* [86] have considered as $A_c$, the entire SPM residual imprint, including the imprint of the additional circle-sector pile-up contact. Thus, the OP and Loubet methods for analysing polymer nanoindentation, yield significantly different estimates of the nanoindentation properties [93].

One reported claim of correspondence between the OP method and the imaged $A_c$ for polymers [94], needs to be scrutinized, because in that work, (i) its reported depths from the SPM images are inconsistent with the $P-h$ data, (ii) its reported $E_r$ and $H$ values are far removed from the corresponding OP-based values, widely reported [28], [95], [96] for the same polymers, and (iii) the $H$ estimates, based on unspecified image analysis of the indents, do not correspond with those when $A_c$ is estimated from SPM views [2], [13].

Thus, for glassy polymers, there is no evidence via any of direct full load optical views, or *in situ* observation or via rational and rigorous analyses of the imprints [2], that validates (i) any version of the OP method, to estimate $h_c$ from the unloading $P-h$ data, or (ii) the Loubet method, that the complete depth, in addition to the pile-up, is in load-bearing contact.

Physical phenomena during polymer nanoindentation have been considered by extending the OP method to estimate the decomposition of the VEP contributions, to polymer nanoindentation [97], [98]. In another perspective, since the analyses are during rapid unloading, after a significant hold time, only elastic and plastic components are considered; the VE components are inactive. Voyiadjis *et al.* [95], [96], [99], [100], have modelled such nanoindentation phenomena via shear transformations and shear transformation zones, and our group [78], has modeled via activation volume.

Our experimental observations on glassy polymer nanoindentation, provide a qualitative, model-free representation of the physical phenomena in the sample. These evidenced phenomena, could be part of predictive modeling or computational methods such as FEA, at a subsequent stage.

# 3. Materials and Sample Fabrication

## 3.1. Thermoplastics

Among thermoplastics, we investigate nanoindentation of commercial grade acrylic sheet (PMMA), and commercial PC. At room temperature, these polymers are glassy. Their high transparency indicates that they are amorphous. In the literature, they have been investigated well for bulk properties as well as via nanoindentation [90], [95], [96], [101]–[104].

## 3.2. Thermosets

We consider SU-8 epoxies, cross-linked to two extents, as representative glassy thermosets. The SU-8 2050 grade epoxy resin employed in our experiments and analyses, has been supplied by MicroChem Corporation™. The samples are fabricated on silicon (Si) substrate, by following the standard photo-lithography process [78] (Appendix A). The less crosslinked sample (called PB05) is 82% cross-linked, and the maximally crosslinked sample (called HB15), is 95.5% cross-linked. This difference in the extents of cross-linking, gives rise to a difference in responses to nanoindentation.

# 4. Experimental Protocol

## 4.1. Nanoindentation Experiments

Nanoindentation experiments have been performed with a diamond Berkovich tip via a nanoindenter (TI Premier, Bruker corporation™). All samples have been indented at a constant 0.1 /s $\dot{P}/P$ loading rate, up to two $P_{\mathrm{m}}$ levels, 1000 μN and 9000 μN. These are well beyond the Hertzian limits as well as the reported limits of indentation size effects [100], [105], and they are comparable with literature data [95], [96], [106], [107]. For each sample, we have performed four indents up to each $P_{\mathrm{m}}$. To overcome VE effects on unloading data, we provide sufficient holding time (70 s for cross-linked SU-8, 100 s for PC and 150 s for PMMA) for the deformation to reach a plateau, followed by a rapid rate unloading (2 s unloading for all samples, at both $P_{\mathrm{m}}$ values) [19], [108].

We pre-compute $h_b$ (i.e., calibrate the tip) at different experiment times (0 days to 600 days) (Appendix B), and account for thermal drift variation during indentation [109]. We also verify that (i) there are negligible thickness and roughness variations along the sample surfaces [110], [111], (ii) there are no pop-in or large load deflections in $P-h$ data during loading and unloading [112]

Since an epoxy is likely to adhere to the tip surface [113], [114], we have coated our fabricated PB05 and HB15 surfaces with 2 nm gold (which will contain ~ 14 layers of gold atoms) via a controlled and optimized sputter process (Orion Sputter PHASE, AJA international, Inc.). This gold coat acts as the barrier between the tip and sample surfaces, without affecting the sample's nanoindentation properties. The PC and PMMA samples remain uncoated in our experiments.

## 4.2. Sample Surface Analysis

We carry out this analysis in two ways. (1) *in situ* nanoindentation and (2) SPM

1. We have carried out *in situ* nanoindentation (Bruker Corporation™ indenter in Fei Quanta 3D FEG SEM) on PB05, with a diamond Berkovich tip. The PB05 surface has been coated with a 5 nm gold layer to counter adhesion effects and to permit direct observation during deformation.

2. After every nanoindentation, we have acquired residual indent images via SPM, employing the nanoindenter itself by scanning the imprint at an optimized frequency (for image clarity), i.e., at 0.8 Hz over areas, $7\mu m \times 7\mu m$ for PB05 and HB15 ($P_m$ = 1000 μN), $10\mu m \times 10\mu m$ for PC and PMMA ($P_m$ = 1000 μN), and $20\mu m \times 20\mu m$, for all samples indented up to $P_m$ = 9000 μN.

# 5. Experimental Observations and Results

We follow a step-wise analysis of the Berkovich nanoindentation response of glassy polymers.

- We first analyze and model the rationally estimated, elastic portion of the unloading $P - h$ data.
- Next, we examine explicit evidence of the visible nanoindentation features. We begin with a live observation via *in situ* indentation, of the response of an amorphous glassy polymer (PB05). Then we rigorously analyze the SPM images of the post-indentation imprints, and obtain directly, the contact and non-contact geometries at full load.

## 5.1. *h* – *P* Data: VE Effects and Elastic Recovery

Glassy polymers can be approximated as solid-like VE materials, to reflect their zero creep rate, beyond their implemented hold times. In their stress-strain plots for finite rate, loading and unloading, such materials exhibit high onset slopes (zero-time modulus, say $E_0$). As the process progresses, the instantaneous moduli (slopes) decrease, due to VE effects (e.g., see [115]). At high rates, the flow effects in their ideal loading or unloading curves, would be significantly reduced.

We approximate zero-time unloading with rapid unloading, which recovers part of the elastic component, of the full load VEP deformation. Additional recovery occurs over time, due to VE effects. Since unloading begins just after the hold period, the *set* rate is achieved by accelerating from constant load; the unloading ends with deceleration, as the load reaches zero. Hence, we first identify the approximate $E_0$ regime, from within the unloading step.

In our load function, we have *set* the nominal unloading time at 2 s (for both $P_{\mathrm{m}}$ levels). We find that the constant *set* unloading rate range is from to $\sim 0.95\,P_{\mathrm{m}}$ ($\sim 0.98\,P_{\mathrm{m}}$, for $P_{\mathrm{m}} = 9000$ μN) to $\sim 0.3\,P_{\mathrm{m}}$. This constant $\dot{P}/P_{\mathrm{m}}$ range, depends primarily on the nominal unloading time and very slightly on $P_{\mathrm{m}}$; it is independent of the sample material (see Supporting Information 1).

The non-zero time required for unloading adds VE effects to the $E_0$ modulus estimate. However, the magnitude of the relative unloading rate, $\dot{P}/P$, increases, as unloading progresses at the set rate. This increase in recovery rate, would be expected to resist the time-dependent phenomena. Hence, the VE effects during unload would be dominated by those during the deceleration in $\dot{P}/P_{\mathrm{m}}$ as the load decreases from $\sim 0.3\,P_{\mathrm{m}}$ to zero. We explore this idea progressively by (i) isolating the VE effects significantly and (ii) separating out the VE effects completely.

### 5.1.1. Modeling Unloading Data: Constant Rate Range

We first consider the hypothetical non-elastic residual depth, $h_{\mathrm{ne2}}$, corresponding to a 2 second unloading, *continuing* at the *set* rate, up to $P = 0$ μN. We then examine whether this first approximation, $\left(h_{\mathrm{m}}-h_{\mathrm{ne}2}\right)$, could correspond to the approximate $E_0$ elastic portion of the indentation displacement. From the conventional PL model for the LC unloading $h-P$ data:

$$h = B' P^{\frac{1}{m}} + h_{\mathrm{ne}2}^{\mathrm{PL}} \qquad \text{eqn. 1}$$

The estimates of $h_{\mathrm{ne}2}^{\mathrm{PL}}$ (Table 1) should be greater than the experimental $h_f$. However, they are negative, or significantly lower, and hence, they are unphysical. Thus, the PL model for the unload data, is inappropriate for glassy polymer systems.

Consequently, we consider a generalized PL (GPL) method which provides the $h_{\mathrm{ne}2}$ values listed in the last row of Table 1. The GPL model recognizes that in the relationship between $\ln P$ and $\ln\left(h-h_{\mathrm{ne}2}\right)$ the tangent slope (power exponent), $m$, varies along the curve. Thus, if the force metric, $F=\ln P$, and displacement metric, $D$, represents the models for $\ln\left(h-h_{\mathrm{ne}2}\right)$, $dD/dF = k_2F^2 + k_1F + k_0$ (discussion in Supporting Information 2). Thus, for all the four polymers,

$$h = gP^{\frac{D_F}{F}} + h_{\mathrm{ne}2} \qquad \text{eqn. 2}$$

$D_F = \int_0^F \left(dD/dF\right) dF = D\left(F\right) - D\left(0\right)$, and $g$ is a fitted parameter. We find that this GPL framework describes well the relationship between $h-h_{\mathrm{ne}2}$ and $P$. The relative residuals of this fitted equation are up to ~ 0.03%, with $R^2$ ~ 0.999. The $h_{\mathrm{ne}2}$ values obtained by fitting the entire constant $\dot{P}$ range data to eqn. 2, would be physically viable (See Supporting Information 2 and Appendix C for a graphical discussion). However, the variation in the exponent corresponds to an increasing recovery; i.e., unloading creep effects persist, even though the fraction of unload modeled is that at constant $\dot{P}$ and increasing $\dot{P}/P$. We discuss next, the complete separation of unload creep effects.

**Table 1:** Nanoindentation depths, and recovery estimates for $P_m$ = 1000 μN and $P_m$ = 9000 μN.

| $h$ (nm) | $P_m$ (μN) | PB05 | HB15 | PC | PMMA | Quartz |
|---|---|---|---|---|---|---|
| $h_m$ | 1000 | 405 ± 2 | 421 ± 1 | 565 ± 5 | 478 ± 8 | -- |
| | 9000 | 1370 ± 11 | 1375 ± 7 | 1860 ± 16 | 1537 ± 10 | 287 ± 2 |
| $h_f$ | 1000 | 191 ± 1 | 172 ± 1 | 300 ± 3 | 304 ± 7 | -- |
| | 9000 | 529 ± 15 | 473 ± 7 | 983 ± 15 | 904 ± 4 | 133 ± 0 |
| $h_{ne2}^{PL}$ | 1000 | 39 ± 5 | -198 ± 24 | 111 ± 9 | 322 ± 5 | -- |
| | 9000 | 347 ± 13 | 209 ± 6 | 810 ± 16 | 1033 ± 9 | 139 ± 1 |
| $h_{ne2}$ | 1000 | 211 ± 2 | 206 ± 5 | 335 ± 11 | 336 ± 13 | -- |
| | 9000 | 711 ± 9 | 697 ± 13 | 1119 ± 27 | 1062 ± 15 | -- |

### 5.1.2. Modeling Unloading Data: Unloading Onset Range

The unloading step, although rapid, still contains VE effects. In order to isolate these time-dependence effects, we consider a PL fitting (analog of eqn. 1) of only the unloading onset, represented by the *top* 10% of the unloading data at the set rate (from 0.95 $P_m$ to 0.885 $P_m$). For this analysis, in eqn. 1, we replace $h_{ne2}^{PL}$ with $h_{ne0}$. This is the expected residual depth after zero-time instant unload (corresponding to $E_0$); i.e., if the PL exponent, *m*, at $P = P_m$, continued up to $P = 0$. The fitted parameters are listed in Table 2. The fitted data span a very small range, and there can be some uncertainty in the nonlinearly fitted parameters; e.g., the nonlinear fitting resulted in a different set of parameters for PMMA at 3000 μN. Since with a broad consistency, the exponent, $m \sim 1.24 \pm 0.03$, fixing the PMMA (3000 μN) exponent $m$~1.25, still retains the fit quality ($R^2$~0.9995).

As expected, the fitted $h_{ne0}$ ($\sim(0.75 \pm 0.05)h_m$) $> h_{ne2}$ ($\sim(0.6 \pm 0.1)h_m$). We consider this hypothetical unload to be purely elastic, since its $P-h$ values exhibit similarities with the unloading $P-h$ data for the quartz standard nanoindentation (Appendix D); these are:

i. Similar PL exponents:

$m$(polymers) ~ 1.24 ± 0.03, $m$(quartz) ~ 1.28 ± 0.03, $m$(OP materials) ~ 1.35 ± 0.1) [2].

ii. For all polymers and for quartz, $S \propto \sqrt{P_m}$.

**Table 2**: PL Fitting of the top 10% of the constant rate unloading $P-h$ data

| | PB05 | | | HB15 | | |
|---|---|---|---|---|---|---|
| $P_m$ (μN) | 1000 | 3000 | 9000 | 1000 | 3000 | 9000 |
| $m$ | 1.23 ± 0.01 | 1.26 ± 0.00 | 1.23 ± 0.01 | 1.23 ± 0.01 | N.A. | 1.25 ± 0.01 |
| $B$ | 0.42 ± 0.03 | 0.34 ± 0.00 | 0.23 ± 0.02 | 0.46 ± 0.02 | N.A. | 0.24 ± 0.02 |
| $h_{ne0}$ | 289 ± 3 | 501 ± 15 | 996 ± 10 | 297 ± 1 | N.A. | 1023 ± 8 |
| $R^2$ | ~ 0.9996 | ~ 0.9995 | ~ 0.9999 | ~ 0.9995 | N.A. | ~ 0.9998 |
| | PC | | | PMMA | | |
| $P_m$ (μN) | 1000 | 3000 | 9000 | 1000 | 3000 | 9000 |
| $m$ | 1.21 ± 0.01 | 1.21 ± 0.00 | 1.23 ± 0.00 | 1.27 ± 0.02 | 1.25 (*fixed*) | 1.26 ± 0.00 |
| $B$ | 0.45 ± 0.02 | 0.32 ± 0.01 | 0.25 ± 0.01 | 0.45 ± 0.05 | 0.30 ± 0.00 | 0.25 ± 0.01 |
| $h_{ne0}$ | 432 ± 6 | 758 ± 14 | 1440 ± 17 | 372 ± 11 | 658 ± 10 | 1195 ± 9 |
| $R^2$ | ~ 0.9995 | ~ 0.9998 | ~ 0.9999 | ~ 0.9989 | ~ 0.9995 | ~ 0.9997 |

Thus, over the small fitting range of the rapid unloading data, we find corroborating phenomenological similarity between the polymers and quartz. Then, as for quartz, this PL unloading curve (constant $m$), qualitatively approximates $E_0$ elastic unloading in polymer nanoindentation.

Next, in order to estimate the contact and non-contact geometries, we directly observe and examine the surface response: (i) We observe an indentation process live, via *in situ* indentation. (ii) we analyze the topology of the post-unload indented surface via SPM.

## 5.2. Direct Surface Observation During and After Nanoindentation

Our live observation of polymer nanoindentation provides a first-hand glimpse of the events. Combining with SPM images, we provide preliminary geometry-based estimates of the load bearing contact area, $A_c$, to estimate the edge-imprint contact depth, $h_c$.

### 5.2.1. *In situ* Nanoindentation

We examine visually via *in situ* monitoring, the material surface behavior during nanoindentation. In Fig. 1, we see that as loading progresses followed by hold, the sinking indenter tip casts a gradually darkening shadow, over the entire surrounding region (see the video, Supporting Information 3). The extra darkening in the region bordering the tip surface, indicates that that part of the sample surface, enters a region of greater shadow, i.e., below the surface level region, further away from the tip, which stays visible. Therefore, we infer that throughout the loading and hold steps for PB05, sink-in occurs along the indenter's edges, as well as along its facets.

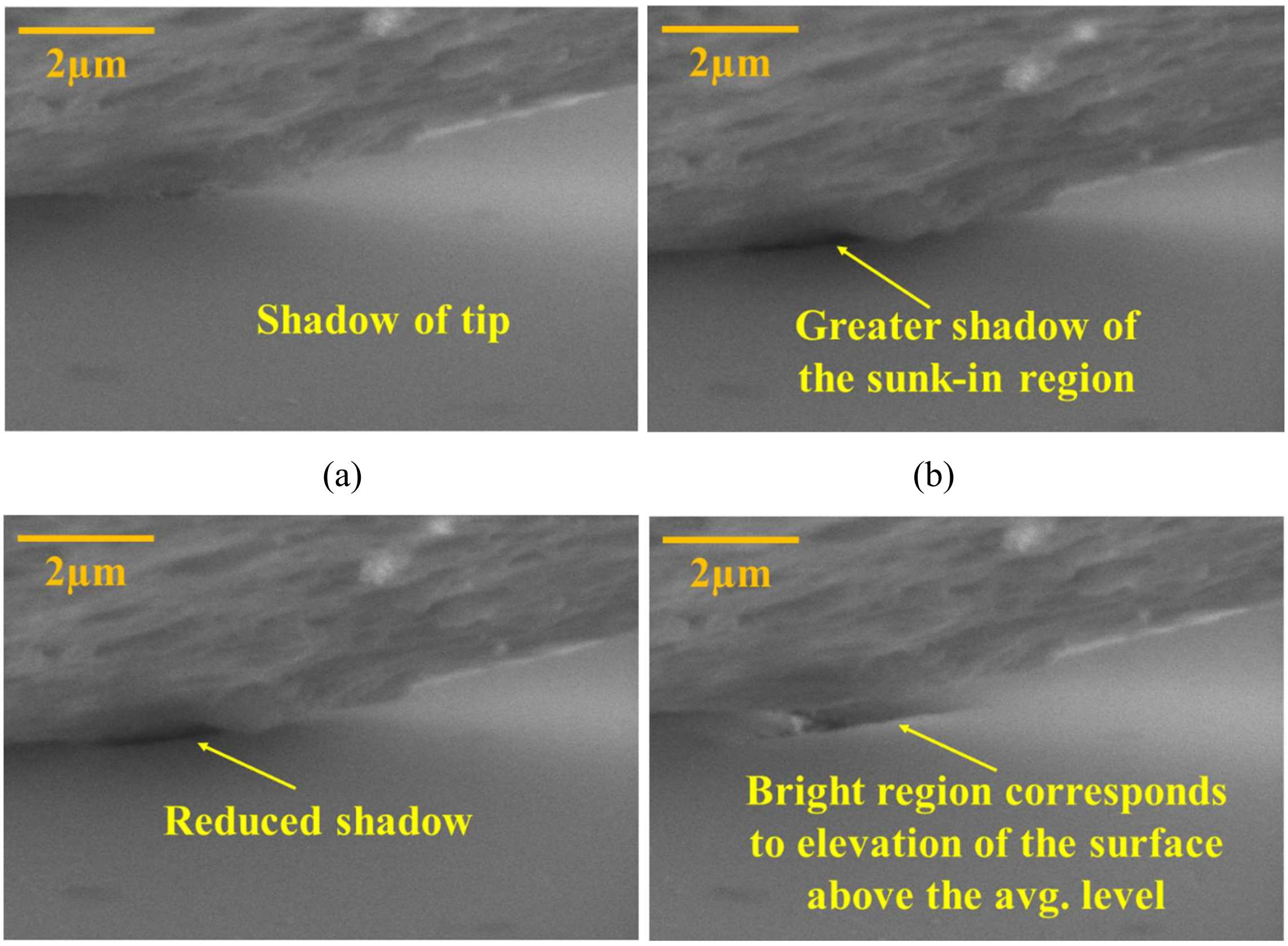

(a) (b) (c) (d)

**Figure 1:** SEM images of indentation events (a) beginning of loading, (b) applied load up to $P_m$, (c) during unloading, and (d) after complete unload for PB05.

During the rapid unloading step, the shadow cast by the indenter becomes less dark. This darkness reduction is greater in the regions very close to the surface-indenter contact, indicating that the sunk-in surface is rising. At the end of the unloading, there is a permanent imprint, corresponding to the non-elastic component of the deformation. In addition, the less dark boundary about the indented region becomes bright, corresponding to the post-unload pile-up above the unperturbed surface level.

Pile-up has been reported via *in situ* nanoindentation by other researchers [116]–[120]. In case of pliant or ductile systems, such as elastomers, pure materials, single crystals or materials at higher temperatures, examination of the *in situ* frames suggests that often, these materials exhibit only sink-in during loading. Visible pile-up, particularly in the recorded region, close to the tip, then occurs only during unloading. Such materials are effectively monolithic or homogeneous, and non-granular, with no preferential regions for fracture or dislocation. For such materials, explicit evidence is required to infer pile-up during loading. For composites, alloys, blends, fine-grained crystals, etc., there are interfaces within the bulk. These could be preferential regions for fracture or dislocation. For these materials, there is clear evidence of significant upflow close to the tip surface, even during loading. However, here too, a close examination is necessary, to determine the full-load pile-up in contact with the indenter tip, $h_c^{\mathrm{PU}}$, vis-à-vis the post-unload imaged maximum pile-up, $h_{\mathrm{m}}^{\mathrm{PU}}$.

We continue with our direct observation analysis by examining post-unload indentation imprints to estimate the contact and non-contact geometries.

### 5.2.2. Post-unload SPM of Nanoindentation Imprints

We analyze here the post-nanoindentation residual imprints acquired via SPM. In order to characterize these imprints, we note that as described in Appendix B, the blunted height, $h_b$, is linear with usage time (measured in days) [121], $h_b = pt + q$. In our SPM examination of the responses of our indented materials, we incorporate the $h_b$ values, corresponding to the experiment's usage day.

Indentation loading of the surface causes it to sink-in around the indenter tip. This sink-in decays with radial distance from the tip apex, and the surface ideally asymptotes to the unstressed zero-level. The sink-in depth, $h_s$, is thus the boundary of the contact between the material surface and the tip, which extends to $h_{\mathrm{m}}$. We identify via SPM, this contact region for Berkovich nanoindentation of glassy polymers. Figure 2(a)-(b) are the representative SPM imaged top view and vertical section, respectively, of the residual indentation imprints (see Supporting Information 4).

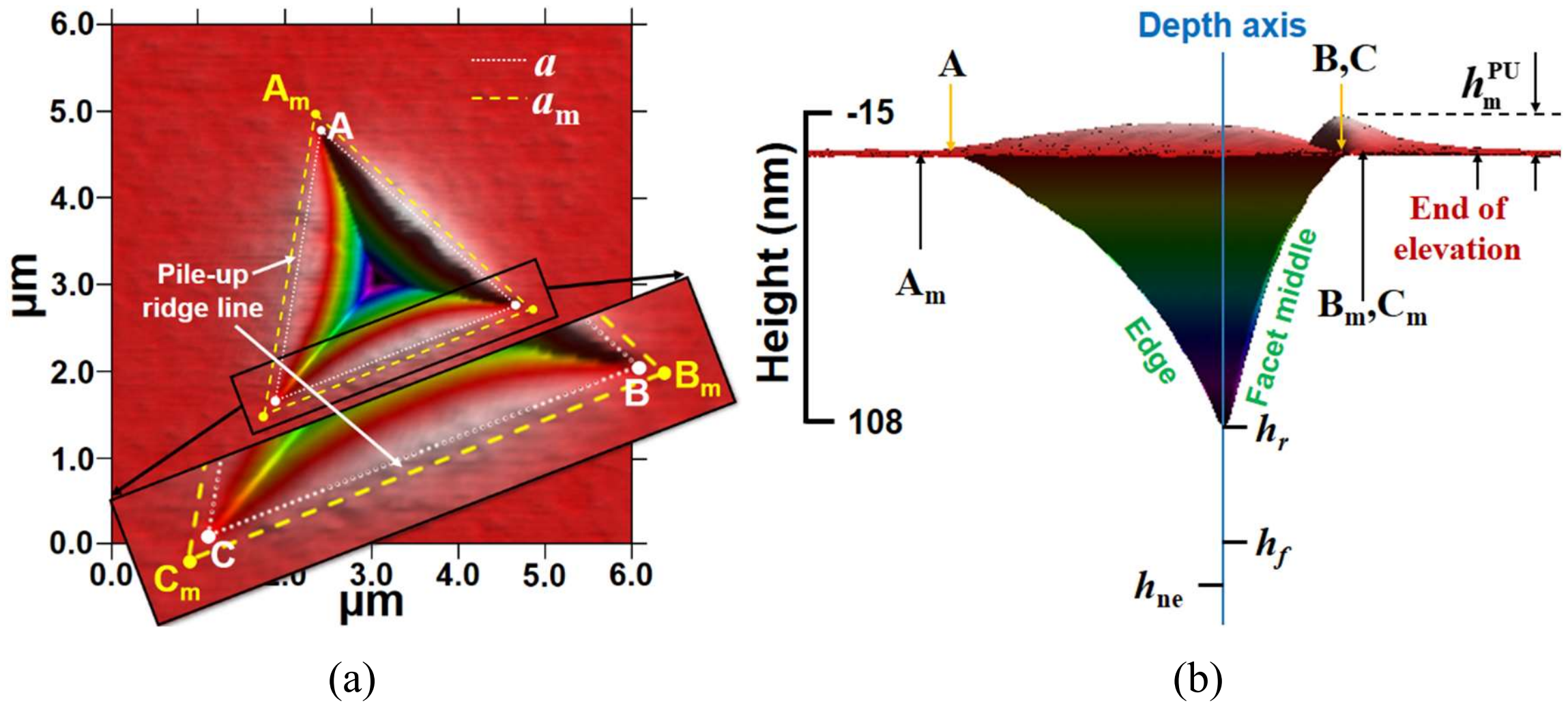

**Figure 2:** SPM image of PB05, $P_m = 1000\ \mu N$ (a) Top view of $\Delta ABC$ indicating the $a$-line = 2988 nm and the $a_m$-line = 3164 nm (b) vertical section.

High stress concentrations along the edges, cause instant yield and permanent imprints [2], [13], invariant for 12 hours after unload. In Fig. 2(a), the outermost tips of these imprints, $A$, $B$ and $C$, form the corners of the full load contact area. The constant height contours are concave outwards (bowed inwards), and have been considered as measures for $A_c$ previously (i.e., $A_c < \sqrt{3}a^2/4$) [20], [122]. However, such inferences are based on post-unload imprints, rather than full load estimates.

In order to unambiguously identify the actual full load contact profile, direct horizontal optical views from beneath the sample and vertical views from alongside the sample, though rare, would be most effective [57], [91]. For transparent amorphous materials such as soft silicone elastomers, the full load images indicate that the contact corners are the ends of the sharp edges of the indenter, with the in-between sides bowed inwards. The corresponding *in situ* Berkovich nanoindentation image, is consistent [117]. The visible deeper sink-in at the facet, begins further away from the tip surface, than the less deep sink-in at the edge does. For such soft elastic materials, the post-unload images retain only diffused edge imprints (no yield), with no clear post-unload pile-up.

In contrast, when the elastomer is coated with Au-Pd, the full load *in situ* images for the harder surface, are similar to those in Fig. 1. Stress concentration and yield impose permanent imprints at the edges, and at the apex. The sink-ins along the facet and at the edge are similar; there is a slight pile-up after complete unload, and the residue imprints are similar to those in Fig. 2.

The stress decreases along the facet [123], and the elastic contribution to the deformation increases. This increases the upflow along the facet middle, away from the edges, ranging from a greater recovery to an unloading pile-up (~ 0.8% – 8% of $h_{\mathrm{m}}$), about the triangle of $a$-lines ($\Delta\,\mathrm{ABC}$).

We first consider the geometric contact zone. We note that the inward bowed contours correspond to greater vertical recovery fractions, away from the edges [90]. Similarly, the vertical sections (Fig 2(b)) have convex-upwards profiles, along both, the edge as well as the facet middle; i.e., fractional upward recovery increases with radial distance from the apex. Thus the contours are measures of recovery, not of contact. The slight post-unload pile-up visible in our SPM images, also suggests that we could consider $\Delta\,\mathrm{ABC}$, to mark the geometric contact boundary for glassy polymers [2], [13], [59]; i.e., $h_c^{\Delta} = h_c + h_b = a/\left(2\sqrt{3}\tan\alpha\right)$. $A_c^{\Delta}\left(h_c\right) \approx 24.5\times\left(h_c^{\Delta}\right)^2$, is the horizontal imprint projection, [30], [66]–[72]. The projection based maximum depth, $h_{\mathrm{m}}^{\Delta} = h_{\mathrm{m}} + h_b$.

We find that for all our samples, $h_c^{\Delta} > h_c^{\mathrm{OP}}$, and the residual imprint area, $A_c^{\Delta} > A_c^{\mathrm{Q}}$, the OP-based quartz-calibrated area (Table 3). One simulation [45] has found similar low sink-in and linear depth contours, by fitting PMMA $P-h$ data, to a quadratic constitutive model, phenomenologically relating indenter load to apex displacement [124], [125]. We specifically observe that consistently, at the end of the hold period over the $P_{\mathrm{m}}$ range, $h_c^{\Delta} \propto h_{\mathrm{m}}^{\Delta}$. The ratio, $\kappa = h_c^{\Delta}/h_{\mathrm{m}}^{\Delta}$, does not change even as loading time (and thus $P_{\mathrm{m}}$) increases; i.e., it is independent of time-dependent VE effects. The $\kappa$ values for the various polymers are ~0.94 (PMMA), ~0.95 (PB05), ~0.96 (PC) (Fig. 3). For HB15, $\kappa$ varies slightly from ~0.86 to ~0.92, over the $P_{\mathrm{m}}$ range in this work. In contrast, $\kappa$(quartz)~0.7.

**Table 3:** $A_c^{\Delta}$ values from imprint vs $A_c^{Q}$ from OP tip area calibration on Quartz Standard.

| $P_{\mathrm{m}}$ (μN) | $A_c$ values (μm$^2$) | PB05 | HB15 | PC | PMMA |
|---|---|---|---|---|---|
| 1000 | $A_c^{Q}$ | 3.34 ± 0.03 | 3.44 ± 0.03 | 5.56 ± 0.10 | 4.18 ± 0.18 |
| | $A_c^{\Delta}$ | 3.85 ± 0.02 | 3.50 ± 0.01 | 7.11 ± 0.12 | 4.89 ± 0.08 |
| 9000 | $A_c^{Q}$ | 32.35 ± 0.59 | 32.22 ± 0.38 | 62.48 ± 1.25 | 42.96 ± 0.64 |
| | $A_c^{\Delta}$ | 43.68 ± 0.19 | 40.35 ± 0.23 | 75.96 ± 1.16 | 51.86 ± 0.29 |

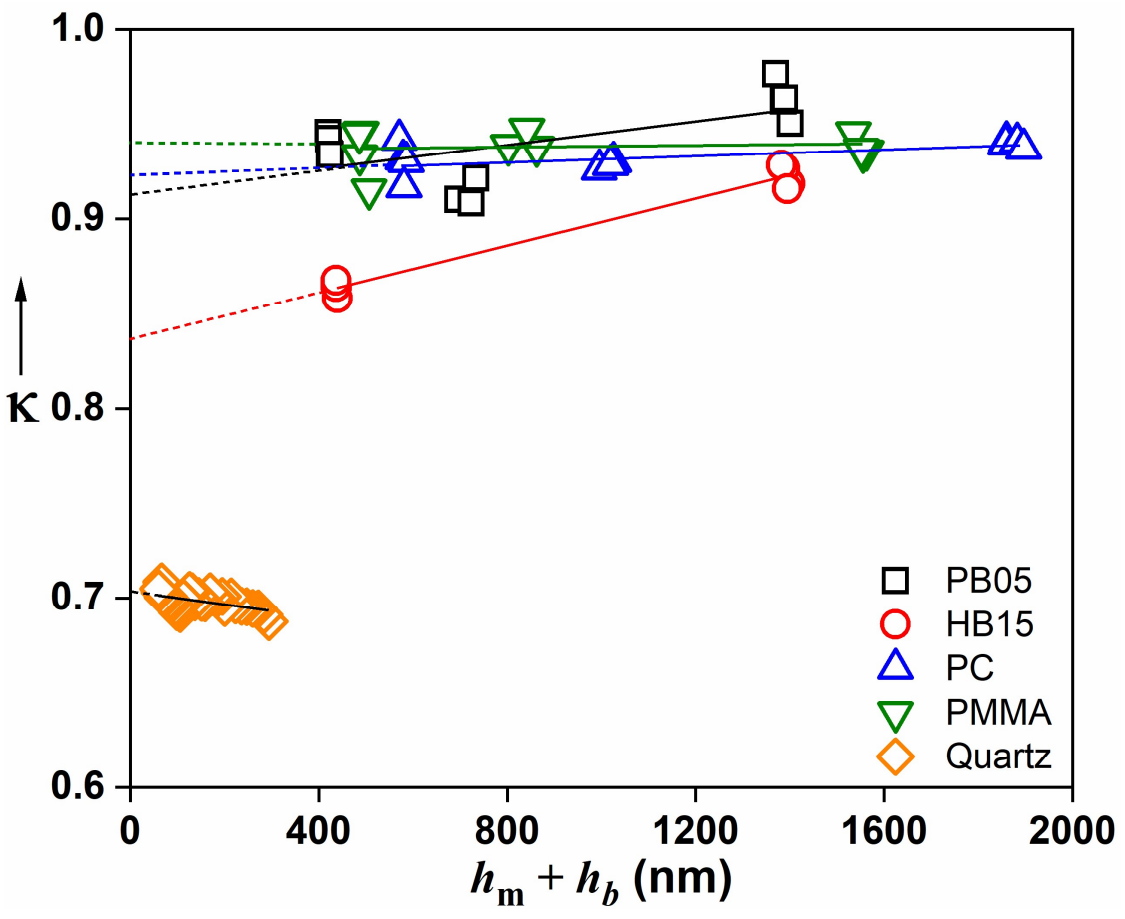

**Figure 3:** Contact depth fraction, $\kappa$ vs effective maximum depth, $(h_{\mathrm{m}}+h_b)$. $\kappa$ is almost constant over the entire depth range for quartz and the polymers. For HB15, $\kappa$ varies slightly.

For glassy polymers, $(h_f/h_{\mathrm{m}}) \sim 0.6$, which satisfies the requirement for the OP method validity ( $(h_f/h_{\mathrm{m}}) < 0.7$ ). Still, for these polymers, the OP method significantly overestimates $h_s$, indicating that the $(h_f/h_{\mathrm{m}})$ criterion is not sufficient. To progress, we first summarize the glassy polymer nanoindentation phenomena, observed via *in situ* nanoindentation and SPM:

a) the contact and sink-in along the edges, are clear and unambiguous

b) the contact fraction is significantly greater than in the case of OP materials,

c) sink-in occurs during loading, and pile-up usually occurs only during unloading

With these observations, we systematically infer phenomena from the bottom up:

deformation profiles $\rightarrow$ mechanisms $\rightarrow$ VEP material phenomena $\rightarrow$ interactions between the deformed region and its surroundings.

# 6. Bottom-Up Analysis of Phenomena

In this section, we analyze the experimental observations to infer the nanoindentation phenomena systematically, over multiple levels.

- First, combining the SPM image analyses with the ideal, elastic, Berkovich deformation profile estimates, we estimate for glassy polymers, a full load indent profile, including contact pile-up.
- Based on the deformation profile and the experimental observations, we propose a schematic of the deformation mechanism. This includes displacement patterns during indentation, within the material continuum, including the up-flow due to contact deformation.
- Combining next, the definition-faithful estimates of nanoindentation properties ($E_{r^*}$ and $H$) with the nanoindentation loading $P-h$ data, we develop a framework, to decompose the VEP deformation into its elastic, EP and VE components.
- This deconvolution of the various components, then provides insights into both, the constrained deformation and recovery and constraint-forced phenomena.

## 6.1. Deformation Profile during Berkovich Nanoindentation

We estimate first, the deformation profile for glassy polymers at full load, a preliminary descriptor of the material response during nanoindentation. We do so via two steps.

(i) We first propose a schematic for the response of an ideal elastic solid to ideal Berkovich nanoindentation, based on the previously proposed schematic for ideal conical indentation [9].

(ii) Considering the SPM-imaged surface topology, we extend the elastic Berkovich nanoindentation framework, to illustrate the response of amorphous VEP glassy polymers. We estimate the full load surface topology, i.e., the deformation profile, including the contact pile-up, $h_c^{\mathrm{PU}}$.

### 6.1.1 Ideal Elastic Nanoindentation by an Ideal Berkovich Tip

The Berkovich indenter has been considered to be an equivalent *cone* with half-angle 70.23° [1]. For a perfect conical indenter, the deformation pathways have been determined previously [9], [10]. A Berkovich tip possesses sharp edges and flat sides, rather than a curved surface. Previously [57], [91], in estimating the OP sink-in for elastomers, two cones were implicitly considered with half angles, $\theta$ = 77.03° on the edge side, and $\theta$ = 65.27° along the facet middle. Also, we have found that for glassy polymers, the baseline sink-in is uniform around the tip.

At this stage, we aim to employ easily useable equations for the topology (and not find new fundamental behavior-describing equations). Hence, we generalize the half-angle variation, all about an ideal Berkovich indenter, and consider it to be a system of continuously varying cones, with half-angle $\theta$ = 77.03° in the edge direction, decreasing to $\theta$ = 65.27° in the facet middle direction. There is a corresponding variation in the contact radius, $a_c$. Figure 4 illustrates this adaptation of the previously reported [9], [10] ideal elastic conical indentation, to our Berkovich tip approximation.

The difference between the equivalent cones and our Berkovich tip adaptation, is that the Berkovich surfaces of the varying cones, are not uniformly circular; they vary from being a sharp edge surface to a flat facet surface. In our model-free development, this cone variation approximation would not affect the final mapping to the empirically observed surface. This framework retains the required conditions for the indented surface topology, that the contact boundary, exhibits continuity with $r$ (radial distance from the tip apex), of the displacement and its derivatives. This makes our adaptation's resultant surface topology estimate, an objective and a functional approximation.

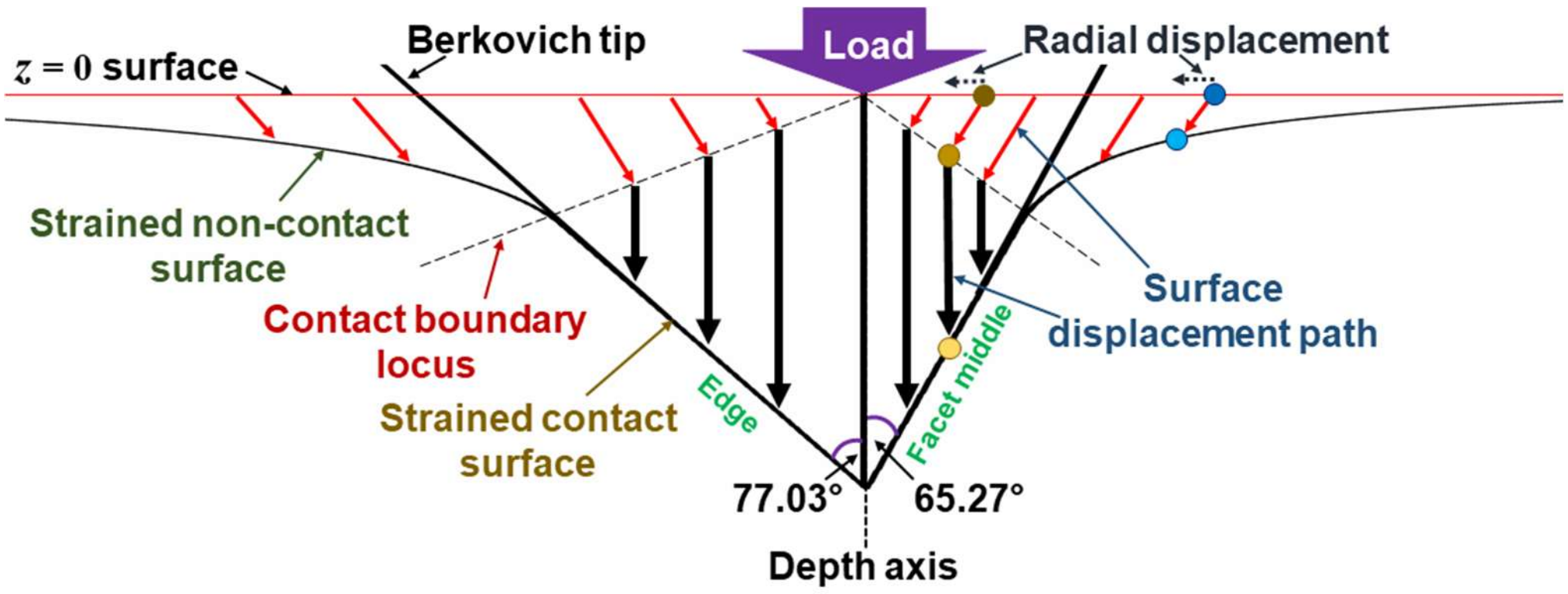

**Figure 4:** Berkovich indentation adaptation of the previously reported [9], [10] schematic for conical indentation of an ideal elastic solid; note the displacement paths of surface elements of the elastic medium. For $a_c > r$, sample radial displacement termini are shown in different shades of yellow; for $a_c < r$, they are depicted in different shades of blue.

The $a_c$ values range from $a_{c,ed} = a/\sqrt{3}$, in the edge direction, to $a_{c,fa} = a/2\sqrt{3}$, along the facet-middle direction. According to the LS analyses, the non-contact radial (inward) and axial (downward) displacements (i.e., sink-in profile, $h_s^p(r)$) are (see Supporting Information 5):

$$\text{Radial shift}\Big|_{z=0,r>a_c} = -\frac{1-2\nu}{4(1-\nu)}\frac{a_c^2 \cot\theta}{r} \qquad \text{eqn. 3}$$

$$h_s^p(r)\Big|_{r\geq a_c} = \cot\theta\left[a_c \sin^{-1}\left(\frac{a_c}{r}\right) + \sqrt{r^2 - a_c^2} - r\right] \qquad \text{eqn. 4}$$

As indentation progresses, $A_c$ increases, with a corresponding increase in $h_s$. As illustrated in Fig. 4, the radial and axial displacements occur until the surface touches the tip, following which, only axial displacement occurs. Eqn. 5 is the resultant expression for inward radial deformation in the contact region, with respect to the instantaneous $a_c$; the downward axial deformation follows the surface of the varying cones of the Berkovich tip.

$$\text{Radial shift}\Big|_{z=0,r<a_c} = \frac{1-2\nu}{4(1-\nu)} r\cot\theta\left[\ln\left(\frac{r/a_c}{1+\sqrt{1-(r/a_c)^2}}\right) - \frac{1-\sqrt{1-(r/a_c)^2}}{(r/a_c)^2}\right] \qquad \text{eqn. 5}$$

Eqns. 3 – 5 describe the flow paths for elastic deformation. The radial stretch gradient (along *r*) based displacement of the open surface occurs only up to contact. As indentation proceeds, additional surface is stretched. The radial stretch gradient along *z*, corresponds to a shear, as does the differential contact compression across *r*. This analysis for the ideal Berkovich nanoindentation of an ideal elastic material, is extended next, to the real Berkovich nanoindentation of compliant VEP glassy polymers.

### 6.1.2 Berkovich Indenter Deformation of Glassy Polymers

We estimate here the surface topology of amorphous VEP glassy polymers undergoing Berkovich indentation. We do so by superimposing suitably, the pile-up and SPM imaged geometric sink-in, onto the ideal elastic Berkovich deformation (eqns. 3 – 5, Fig. 4).

In Fig. 2(a)-(b), the $a_{\mathrm{m}}$-line corresponds to the *z*=0 surface level projection of tip cross-section for penetration $h_{\mathrm{m}}^{\Delta}$ at load $P_{\mathrm{m}}$. The post-unload SPM imaged surface in the $a_{\mathrm{m}}$-line region, is the deformed surface, which has recovered to the *z*=0 level. Every glassy polymer sample's SPM image exhibits a small pile-up, that begins inside the $a$-line, and continues past the $a_{\mathrm{m}}$-line.

Figure 1 and Supporting Information 3 clearly indicate that for PB05, only sink-in occurs during loading, and surface pile-up occurs at the end of unloading. Thus the pile-up is the likely result of the entire contact boundary recovering by $h_s$ as the non-contact region recovers by $h_s^p(r)$. We then identify the full-load contact portion of the measured pile-up, $h_c^{\mathrm{PU}} = h_c(\text{facet}) - h_c(\text{edge})$.

This pile-up effect is analyzed appropriately to add the additional contact area, $A_c^{\mathrm{PU}}$, the $h_c^{\mathrm{PU}}$ contribution to the *a*-line based $A_c^{\Delta}$. To estimate $h_c^{\mathrm{PU}}$, we begin with the measured pile-up at zero load, i.e., the maximum pile-up height, $h_{\mathrm{m}}^{\mathrm{PU}}$. As per Table 4,

a) For PB05, HB15 and PMMA, $h_{\mathrm{m}}^{\mathrm{PU}}/h_s^{\Delta} < 1.0$. For PC, $h_{\mathrm{m}}^{\mathrm{PU}}/h_s^{\Delta} \geq 1.0$ (consistent with PC exhibiting the highest $\kappa$); i.e., pile-up for PC might have begun during loading.

b) $h_{\mathrm{m}}^{\mathrm{PU}}/h_s^{\Delta}$ is greater for PB05 than for HB15, and for PC than for PMMA.

**Table 4:** $h_{\mathrm{m}}^{\mathrm{PU}}/h_s^{\Delta}$ for all samples for $P_{\mathrm{m}}$ = 1000 μN and $P_{\mathrm{m}}$ = 9000 μN.

| $P_{\mathrm{m}}$ (μN) | PB05 | HB15 | PC | PMMA |
|---|---|---|---|---|
| 1000 | 0.540 ± 0.075 | 0.095 ± 0.004 | 1.106 ± 0.083 | 0.515 ± 0.065 |
| 9000 | 0.854 ± 0.102 | 0.245 ± 0.019 | 1.096 ± 0.032 | 0.416 ± 0.031 |

To estimate $h_c^{\mathrm{PU}}$, we analyze the SPM-imaged topology data via the following steps.

I. *Outward shift of the z-axis*

The LS analyses provide eqn. 4 (the LS-*z* equation), to compute $h_s^p(r)$. We recall that for a given $h_s$, the purely elastic maximum penetration of an ideal conical indenter, $h_{\mathrm{e}} = h_s\left[\pi/(\pi-2)\right]$. Therefore, as depicted in the estimated profile in Fig. 5, we propose that the sink-in *z-r* relationships for our experiment, correspond to radially outward-shifted $z$-axes. Then the shifted Berkovich cones of depth $h_{\mathrm{e}}$, share the surface lines with the experiment cones. The non-contact recovery corresponding to the shifted Berkovich cones, is elastic and exhibits radial as well as axial recoveries. The contact deformation and recovery can be approximated to be mainly axial, since any radial displacement is mapped inward by eqns. 3 – 5, only up to the shifted *z*-axis. Since the radial shifts have been implemented up to $r \geq a_c$, and pile-up is up to $r > a_c$, this imprecision in the radial displacement, would not cause any errors in our analysis.

At this time, in the absence of a physics-based profile development, which accounts for the high $\kappa$ for glassy polymers, we employ this shift-based profile. This shift ensures that at the contact boundary, there is a continuity in the displacement and its derivatives (with respect to $r$), as built into the available eqns. 3 – 5. These profiles are also similar to those reported via FEA [57], [117] as well as via optically recorded video frames of PDMS indentation [9], [91], [130].

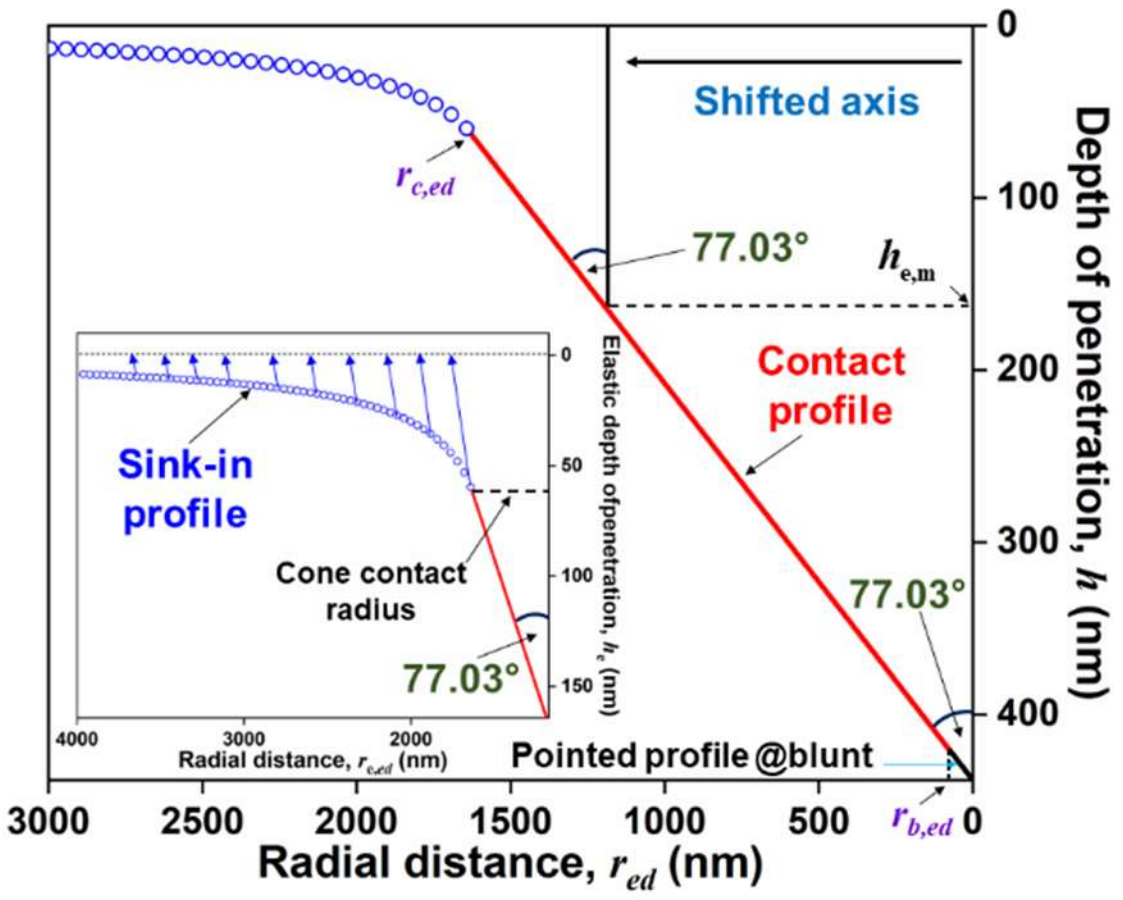

**Figure 5:** Edge side deformation profile for HB15 at full load. The $r$-axis is inverted, to facilitate visualization (e.g., with respect to Fig. 4 and subsequent figures).

II. *Estimate recovery in absence of VE effects*

The SPM image (Fig. 2(b)) indicates that the imprint apex is at depth, $h_r < h_f$ (recall $h_f < h_{\text{ne}2}$). We note that SPM imaging is completed in ~ 6 min after complete unload, during which, VE recovery occurs (e.g., see [81], [83], [84], [88]). This recovery continues even after 12 hrs, while $A_c^\Delta$ remains invariant. The residual level rises over time: (i) from the hypothetical $h_{\text{ne}0}$ or $h_{\text{ne}2}$, to the experimental $h_f$ (due to deceleration during the last 30% of the unload), and (ii) from $h_f$ to the imaged $h_r$, at zero load (see Supporting Information 6).

As illustrated in Fig. 4, elastic recovery contains axial and radial components. VE recovery is beyond the elastic recovery, and hence, is axial. Hence, analogous to the vertically upward elastic and VE recovery, the maximum pile-up also rises upward. $h_{\mathrm{m}}^{\mathrm{PU}}$ contains time-dependent VE effects, prior to which, the hypothetical elastic-recovery-based $h_{\mathrm{ne}2}^{\mathrm{PU}}$, is lower. Hence, to estimate $h_c^{\mathrm{PU}}$ from $h_{\mathrm{m}}^{\mathrm{PU}}$, we first estimate $h_{\mathrm{ne}2}^{\mathrm{PU}}$. We obtain 2 consecutive SPM profiles on HB15 and PC (~ 9 min apart), immediately following unload. For $r > a_c$, $\frac{h_{r1}^{p,\mathrm{PU}}(r)}{h_{r2}^{p,\mathrm{PU}}(r)} \approx \frac{(h_c - h_{r1})}{(h_c - h_{r2})}$; then $\frac{h_{\mathrm{ne}2}^{p,\mathrm{PU}}(r)}{h_r^{p,\mathrm{PU}}(r)} \approx \frac{(h_c - h_{\mathrm{ne}2})}{(h_c - h_r)}$ (Appendix E). This yields the hypothetical, geometry-based profile estimate, $h_{\mathrm{ne}2}^{p,\mathrm{PU}}(r)$, when apex recovery is $h_{\mathrm{ne}2}$. Despite $h_{\mathrm{ne}0}$ representing VE-free unloading, we consider $h_{\mathrm{ne}2}^{\mathrm{PU}}$ for estimating pile-up, because $(h_c - h_{\mathrm{ne}0}) < (h_c - h_{\mathrm{ne}2})$. Thus, the $h_{\mathrm{ne}2}$-based reduction estimate for $\left(h_{\mathrm{m}}^{\mathrm{PU}} - h_c^{\mathrm{PU}}\right)$, is conservative, and more appropriate, recognizing the uncertainties in this approach.

III. *Intersecting surface topology estimate with indenter tip profile:*

The recovery of the non-contact region is fully elastic. For glassy polymers, elastic recovery of the apex, is up to $h_{\mathrm{ne}2}$; for $r \le a_c$, there is residual displacement. For $r > a_c$, along the edge-line, $h_s^p(r) = 0$, and along the facet, pile-up occurs.

To exploit this sink-in recovery, we construct a full contact profile at displacement $h_c$, so that the full load sunk-in edge-line, becomes $h_s^p(r) = 0$ for $r > a_c$; this is also the $r > a_c$ profile at zero load. In the absence of pile-up for $r > a_c$, at both, zero load and for full contact displacement $h_c$, even the sunk-in facet middle line would rise to $h_s^p(r) = 0$. Then the pile-up profile along the facet middle line, $h_{\mathrm{ne}2}^{p,\mathrm{PU}}(r)$, corresponds to both, that at zero load, and to that for full contact displacement $h_c$. The extension of the contact line on the facet middle (slope=cot 65.27°) for full contact displacement $h_c$, rises from the $z$-axis at $h_c^{\Delta} = h_{\mathrm{m}}^{\Delta} - h_s^{\Delta}$ and meets the pile-up profile, $h_{\mathrm{ne}2}^{p,\mathrm{PU}}(r)$ for $r > a_c$ (Fig. 6(a)-(b)). $h_c^{\mathrm{PU}}$ corresponds to this intersection.

We consistently find that $h_{\mathrm{ne}2}^{\mathrm{PU}}$ ($h_{\mathrm{m}}^{\mathrm{PU}}$, corrected for VE effects) is near but away the geometric contact, making the contact pile-up, $h_c^{\mathrm{PU}} < h_{\mathrm{ne}2}^{\mathrm{PU}}$. The actual contact at $h_c^{\mathrm{PU}}$ would follow a smooth curve, to enable at the contact boundary, continuity in $h_{\mathrm{ne}2}^{p,\mathrm{PU}}(r)$ and its derivatives with respect to $r$. This development is consistent with previous researchers' *in situ* nanoindentation pile-up frames [116]–[120]. Also, one FEA approach to match $A_c$ during effective conical indentation ($\theta$ = 70.23°), considers tip bluntness as well as sink-in or pile-up [30]. Their schematic clearly indicates that $h_c^{\mathrm{PU}} < h_{\mathrm{m}}^{\mathrm{PU}}$, and is based on the intersection of the facet side surface topography with the indenter tip boundary. These validate our finding, about the radial location of $h_{\mathrm{m}}^{\mathrm{PU}}$.

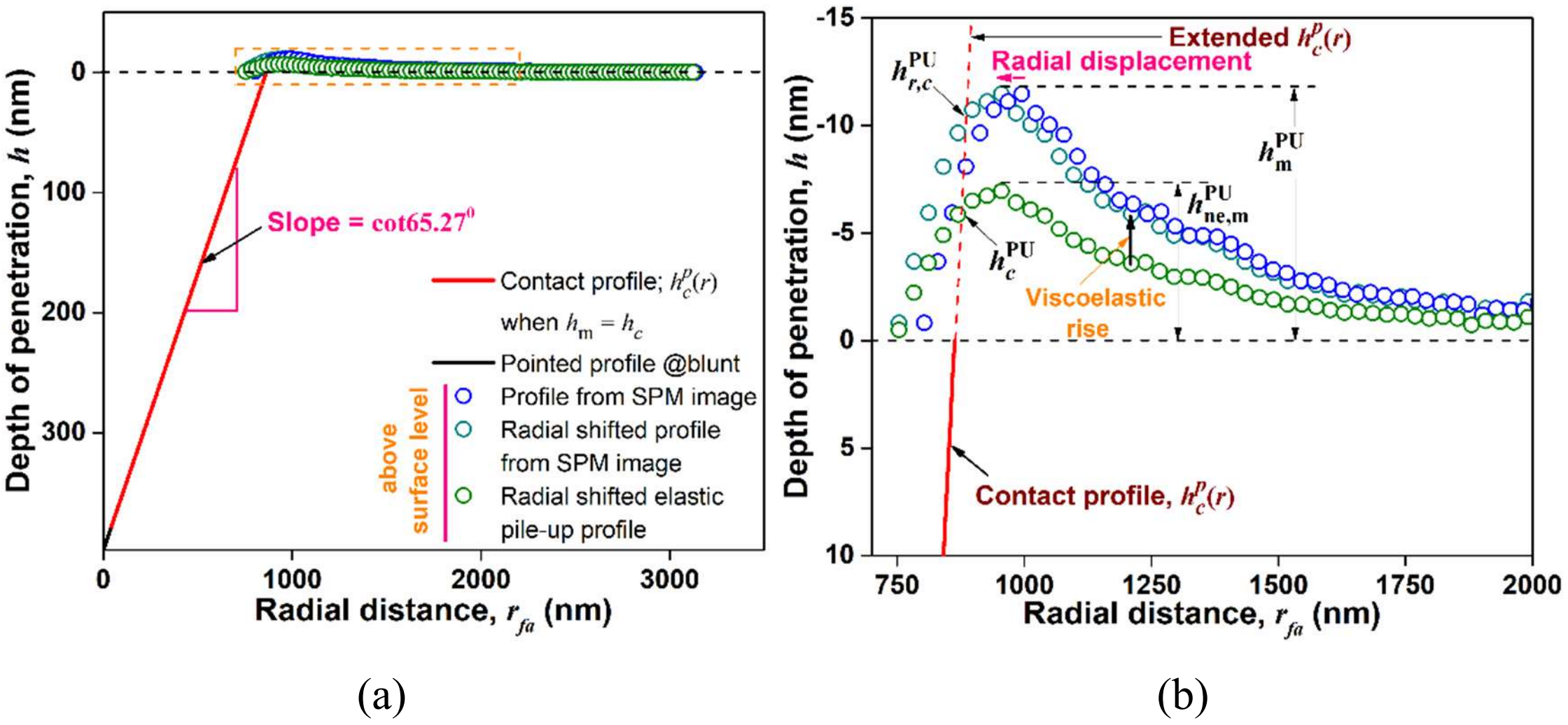

**Figure 6:** (a) Full contact profile at displacement $h_c$ at full load, (b) magnified indentation profile toward facet middle for one indentation of PB05 for $P_{\mathrm{m}}$ = 1000 μN. Note that $h_{\mathrm{ne}2}^{\mathrm{PU}}$ is beyond geometric contact ($h_c^{\mathrm{PU}} < h_{\mathrm{ne}2}^{\mathrm{PU}}$). The actual profile would be smooth to exhibit continuity at contact, in $h_{\mathrm{ne}2}^{p,\mathrm{PU}}(r)$ and its derivatives with respect to $r$.

IV. *Superimposing the* $h_{\mathrm{ne}2}^{p,\mathrm{PU}}(r)$ *on the shifted z-axis profile*

To estimate the full load profile along the facet middle, we consider the shifted eqn. 4, on which we superimpose the $h_{\mathrm{ne}2}^{p,\mathrm{PU}}(r)$ profile from Fig. 6(b). Figure 7 is a rational estimate of the likely surface topology at $P_{\mathrm{m}}$, which, if superimposed on the elastic and VE recovery, would yield our SPM imaged, zero-load profile. The case for HB15 (polymer with minimum pile-up) is depicted here. The estimated edge and facet-middle profiles for PC, which exhibits the maximum pile-up, are provided in Appendix F. In contrast to FEA reported earlier (e.g., [53], [56]), where the maximum pile-up is in contact with the indenter tip, our estimated profiles are similar to those suggested previously (e.g., [46], [87], [131], [132]). Thus, every step towards our surface profile estimate via minimal modifications of the currently available equations, is validated. The estimated profiles in Fig. 5 and Fig. 7, provide a basis for refinements in FEA and predictive modeling of polymer nanoindentation.

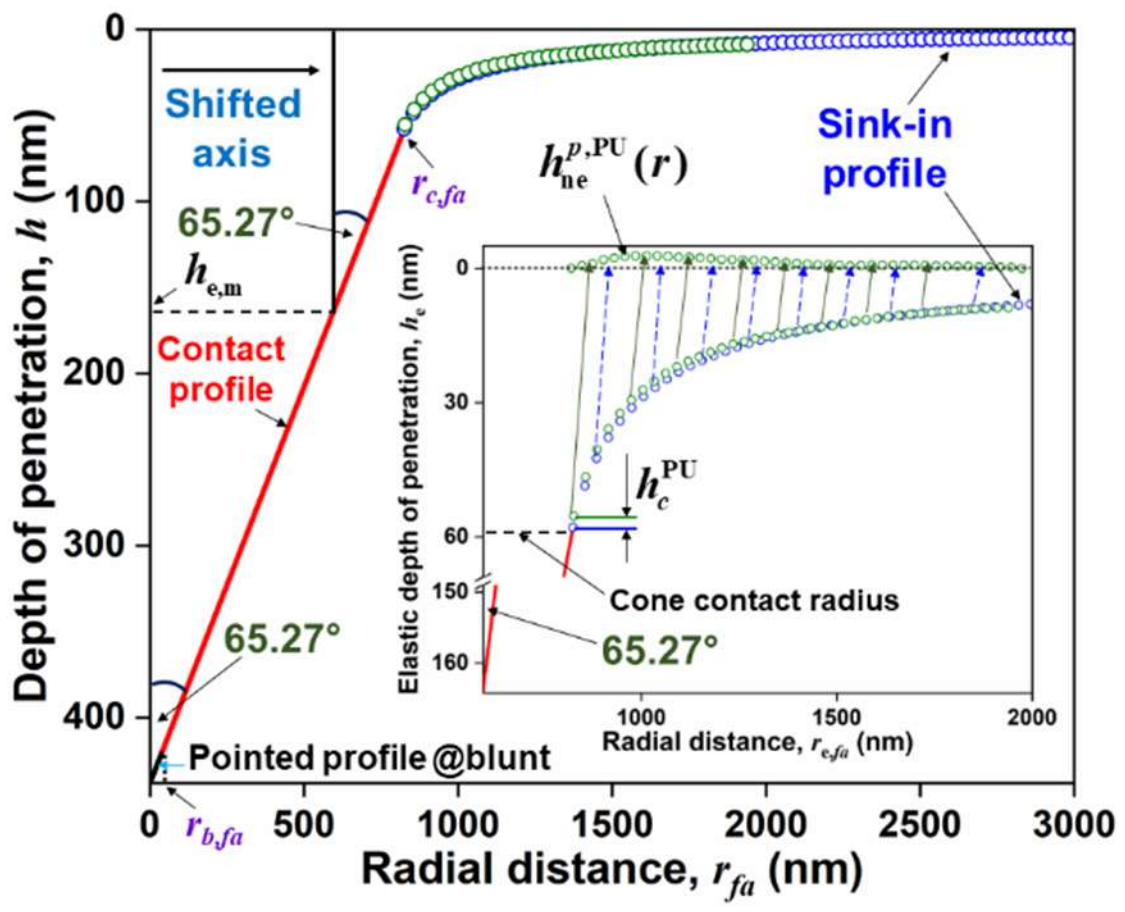

**Figure 7:** Deformation profile for HB15 at full load facet middle side to estimate $h_c^{\mathrm{PU}}$. The *z*-axis is shifted as in Fig. 5, and the non-contact displacements are radial and axial.

With pile-up $A_c = A_c^{\Delta} + A_c^{\mathrm{PU}}$ [14]–[25]. See Appendix G for $A_c^{\mathrm{PU}}$ calculation and Supporting Information 6 for corrections to errors due to tilt between the indenter tip and the indented surface [133]. $A_c^{\mathrm{PU}}$ increases $A_c$ for HB15 and PMMA by ~0.3% ($h_c^{\mathrm{PU}}$ basis) to ~2% ($h_{\mathrm{m}}^{\mathrm{PU}}$ basis).

For PB05 and PC, the respective $A_c^{\mathrm{PU}}$ fractions are ~2% and ~8%. Thus $\left(A_c\left(h_{\mathrm{m}}^{\mathrm{PU}}\right)-A_c\left(h_c^{\mathrm{PU}}\right)\right) \ll$ $\left(A_c\left(\mathrm{Loubet}\right)-A_c\left(\mathrm{OP}\right)\right)$; i.e., the choice between $h_c^{\mathrm{PU}}$ and $h_{\mathrm{m}}^{\mathrm{PU}}$, affects only slightly, the subsequent analyses for glassy polymers. However, the framework which considers $h_c^{\mathrm{PU}}$, provides an objective platform for future development. Also, based on these estimated topological profiles, we propose next, a schematic of the deformation phenomena.

## 6.2. Berkovich Nanoindentation Mechanism in Glassy Polymers

We describe here, the deformation mechanism and identify the deformation pathways within the material continuum. Based on the phenomena inferred above, through Figs. 2 – 7, we first propose for amorphous glassy polymers, a qualitative schematic of the full load nanoindentation profile (Fig. 8(a)-(b)). We reproduce from Fig. 2(a)-(b), the contact corners A, B, C, and the corners, $\mathrm{A_m}$, $\mathrm{B_m}$ and $\mathrm{C_m}$, for indentation depth $h_{\mathrm{m}}^{\Delta}$. At full load, the latter are on the tip edges, at the level of the unperturbed surface, above the deformed surface, but within the projection of the sink-in region. $\Delta\bar{\mathrm{A}}_{\mathrm{m}}\bar{\mathrm{B}}_{\mathrm{m}}\bar{\mathrm{C}}_{\mathrm{m}}$ is the $a_{\mathrm{m}}$-line triangle on the full load sample surface, beneath $\Delta\mathrm{A_m B_m C_m}$. At unload, the sample surface rises and the corners of $\Delta\bar{\mathrm{A}}_{\mathrm{m}}\bar{\mathrm{B}}_{\mathrm{m}}\bar{\mathrm{C}}_{\mathrm{m}}$, rise to reach $\mathrm{A_m}$, $\mathrm{B_m}$ and $\mathrm{C_m}$.

The elastic sink-in as per the OP or LS method, would have occurred up to $\mathrm{A_{OP}}$, $\mathrm{B_{OP}}$ and $\mathrm{C_{OP}}$ $\left(\text{i.e., } h_s^{\mathrm{LS}} \sim h_s^{\mathrm{OP}}\right)$. Reduction in the sink-in makes ΔABC the contact region boundary. Since for most polymers, pile-up occurs only during unloading, this full load sink-in is below the unperturbed surface level (yellow region). Table 4 and Appendix F indicate that pile-up for PC might have occurred during loading as well, as represented by the green zone. We discuss next, the causes of this reduced sink-in and pile-up, which would correspond to the upward-bulging yellow and green contact regions on the facets, above the $\Delta$ABC (i.e., $h_s^{\Delta}$) in Fig. 8(a)-(b). Their upper boundaries represent the outward-bowed contours beyond the *a*-lines, in Fig. 2(a); i.e., the locus of $h_c^{\mathrm{PU}}$, along the facet.

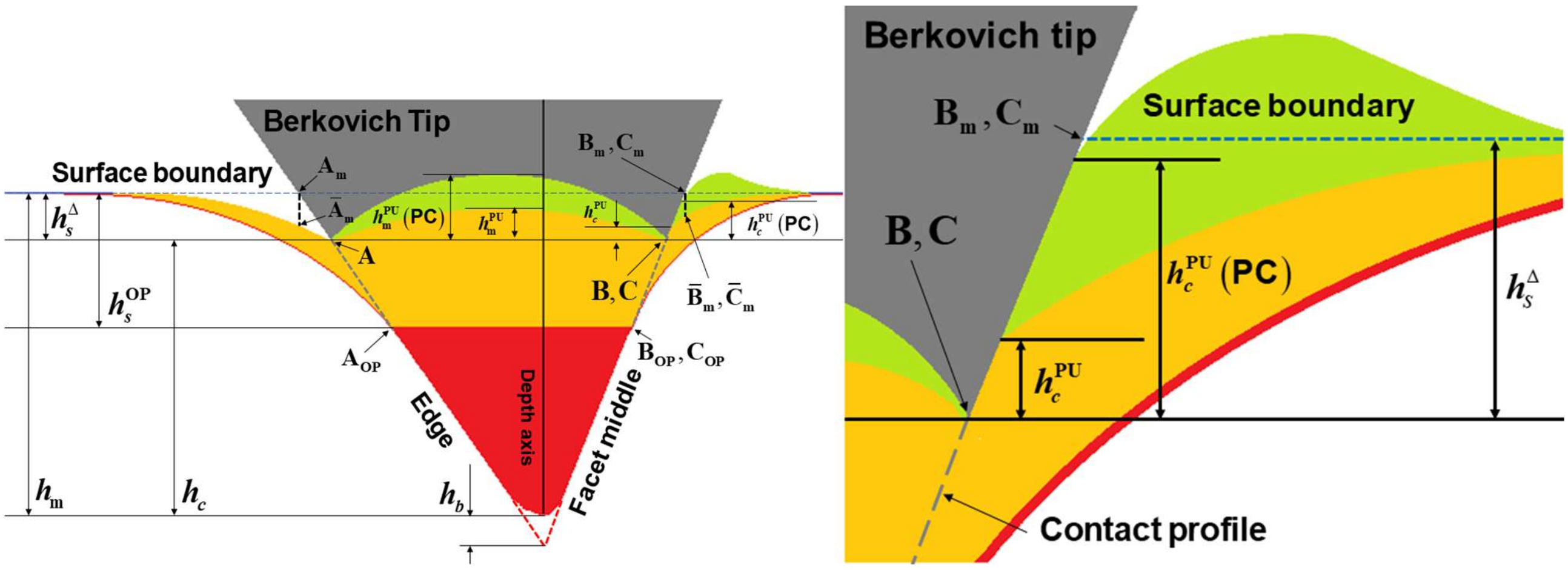

(a) (b)

**Figure 8:** (a) Schematic diagram of the deformation profile at $h_{\mathrm{m}}$ (and $P_{\mathrm{m}}$). (b) magnified contact pile-up at $h_{\mathrm{m}}$ (or $P_{\mathrm{m}}$) vis-à-vis the OP estimate (Red Curve), close to the zero level.

The polymers considered here are compliant materials; i.e., their respective elastic moduli, lie in the range 2 GPa – 3 GPa [77], [78], [115], [126]–[128], in contrast to the 20-200-fold greater moduli of the OP materials [2]. The greater compliance of polymers, enables greater indenter displacement. In glassy polymers, as the tip moves downward during loading, yield occurs immediately, particularly along the edges. The downward compressive stress at the interface, moves downward, the layer of the solid *wetting* the angled surface (of a conical or Berkovich tip) (Fig. 4).

The Poisson's ratios of the OP materials are in the range 0.08 (quartz) to 0.34 (aluminum), while The Poisson's ratios for glassy polymers are in the range 0.34 (SU-8) [129] to 0.39 (PMMA) [65]. The tendency for the sample's volume change due to the tip's downward motion, is resisted by its lateral Poisson expansion. In the inner regions below the tip umbrella, the lateral Poisson deformation is directed radially away from the tip, compressing laterally, the surrounding unstressed glassy polymer region. This laterally compressed outer region, then undergoes vertical Poisson expansion. The polymers' greater $\nu$, thus resists the volume reduction due to the greater penetration, and causes a greater upflow close to the indenter tip. This volume conserving upflow causes a greater $h_c^{\Delta}$ (and a high and invariant $\kappa$), i.e., a significantly reduced sink-in.

Unlike in the case of bulk uniaxial compression, the open vertical surface here (laterally directed surface normal), is radially far from the region compressed by the tip. Hence, this tendency toward graded lateral compression, drives the outer region's Poisson expansion vertically upward, towards the free upper surface (vertically upwards directed surface normal). This surface is nearer, and can accommodate the outer upward transfer of the inner lateral Poisson expansion. Thus, in competition to the downward tip-drag, there is upflow (Fig. 9).

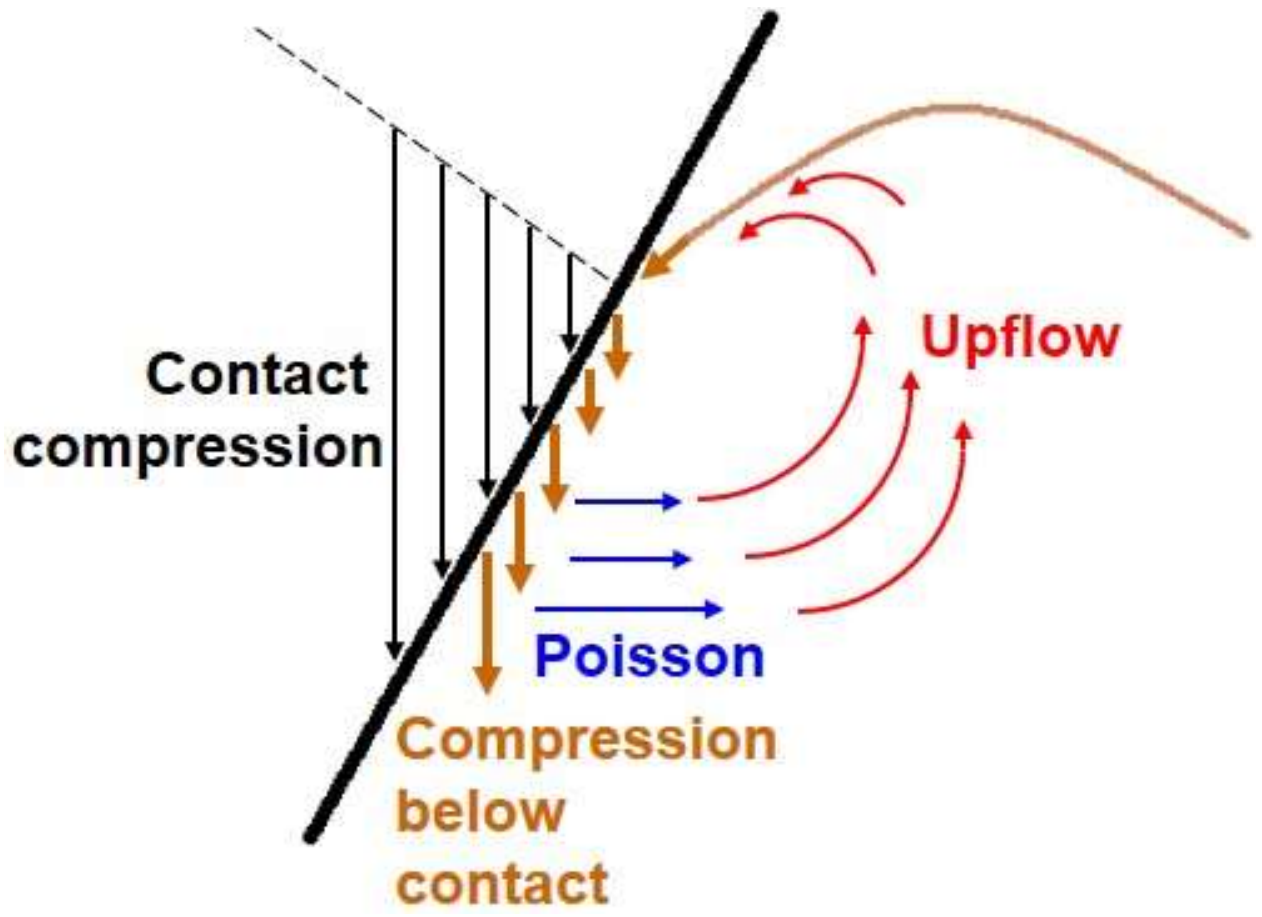

**Figure 9:** Schematic of transfer of lateral Poisson deformation to upflow. Description in text.

Along the edges, stress concentrations, cause the greatest downward compressive stresses. Hence, upward flow there, past the downward dragged contact layers, is the lowest. Thus, for our materials, this upward flow at the edges, reaches up to the lowest full load contact ($\Delta$ABC), and only causes reduction in $h_s$. Any additional upflow would be directed towards the facet middle, or radially away from the edge, so that any pile-up is with respect to the edge-contact.

Along the facet, the stresses are lower, since the indenting surface is flat. Hence, near the middle of the facet, in layers beyond the downward dragged contact layers, resistance to upward flow is the lowest, and results in a maximum elevation of the contact surface (Fig. 8(a)). The consequence of the reduced stress, is that the contact boundary, $h_c^{\mathrm{PU}} = h_c(\text{facet}) - h_c(\text{edge})$, is an arc, between the edges, across the facet (Appendix G); i.e., the upper boundaries of the yellow and green regions in Fig. 8(a)-(b). Analogous phenomena have also been proposed previously [46], [87], [131], [132].

This combination of downward movement along the tip and vertical upflow, causes $h_{\mathrm{m}}^{\mathrm{PU}}$ to be near but radially away from geometric contact, along the path of least resistance (Fig. 9). The indicated, downward and inward flows at the top, represent eqns. 3 – 5, caused by the downward pull of the tip. Then the contact pile-up, $h_c^{\mathrm{PU}} < h_{\mathrm{m}}^{\mathrm{PU}}$, as illustrated in Fig. 8. All the components of the mechanism described above and illustrated in Fig. 9, occur simultaneously, and throughout the nanoindentation. Then compressive loading strain is confounded by the upflow; the inward radial displacement combines with the lateral Poisson expansion. What is observed is the net response.

The frictionless surface usually considered for purely compressing flat surfaces as in a flat-punch tip or macroscopic compression plates, is to enable lateral Poisson expansion at contact. The downward compressive forces in conical and Berkovich indenters, being normal to the widening direction, prevent upward slip (upward flow at the tip contact). There would be no effect of contact friction for widening, since contact with the angled tip ceases, the instant the material moves laterally away.

Therefore, in the case of Berkovich indentation, the previous FEA-based suggestions of friction effects between the tip surface and the indented material (e.g., [47], [53], [56]), need to be re-examined. Consistent with our postulates, however, friction has been found to be inconsequential in some reports [41], [45], [48], or effective only if pile-up is significant [51], [54]. Hence, as opposed to friction, pile-up beyond contact is due to transfer of the lateral Poisson expansion of the inner region, to the vertical expansion of the outer region.

We combine our above analyses to explore the VEP phenomena during deformation.

## 6.3. Material Phenomena During Nanoindentation

In this section, we combine the estimate of the deformation geometry, with aspects of definition-faithful estimates of $H$, and $E_N$, to identify the deformation phenomena. We then estimate the hypothetical nanoindentation hardness, based on $P-h$ data with VE effects removed. Combining with the $h_{ne0}$ unloading $P-h$ data, we then deconvolute the VEP deformation, into VE, EP and elastic components.

### 6.3.1 Nanoindentation Properties: $E_r$ and $H$

In order to express nanoindentation in terms of VEP material phenomena, we first estimate for glassy polymers the nanoindentation properties, $E_r$ and $H$, as conventionally defined for EP materials. For this analysis, we employ the stiffness, $S$, a measure of the events at the immediate onset of unload, before the subsequent time-dependent phenomena occur. From the unloading $h-P$ data (Tables 1, 2), of each LC indent, $S = \left( dh/dP|_{P_m} \right)^{-1}$. Hence, from eqn. 2,

$$S = \left( \frac{gP^{\frac{D_F}{F}}}{P} \frac{dD}{dF} \bigg|_{P_m} \right)^{-1} \qquad \text{eqn. 6}$$

Since $S$ (Table 5) is the same, whether we obtain it via the PL ($h_{ne0}$ basis) or via the GPL ($h_{ne2}$ basis), the recovery onset, maps well with the criteria for the $E_0$ modulus. Combining eqn. 6 and $A_c = A_c^{\Delta} + A_c^{PU}$, yields our definition-faithful estimates of $E_r$ and $H$, (Table 6). (discussion vis-à-vis other methods [2], [19], [20], [24], [59], [73], [75], [76], in Supporting Information 7).

**Table 5:** $S$ (μN/nm) estimates for $P_m$ = 1000 μN and $P_m$ = 9000 μN

| $P_m$ (μN) | PB05 | HB15 | PC | PMMA |
|---|---|---|---|---|
| 1000 | 10.59 ± 0.09 | 11.02 ± 0.09 | 9.87 ± 0.04 | 12.22 ± 0.09 |
| 9000 | 30.63 ± 0.14 | 34.31 ± 0.06 | 27.97 ± 0.02 | 33.12 ± 0.07 |

One intuitively expects that $S$, $E_r$ and $H$, should increase with crosslinking. While this trend has also been observed through simulation (e.g., [134]) and bulk measurements of modulus (e.g., [135]–[139]), there are contrary reports as well [140]–[142]. We find that the definition-faithful estimates of $S$, $E_r$ and $H$, increase, on increasing the degree of crosslinking from the less cross-linked PB05 and almost fully cross-linked HB15. Following the trends for the uniaxial properties ([28], [65], [77], [115], [126]–[128]), our estimates of $S$, $E_r$ and $H$, for PC, are lower than those for PMMA.

**Table 6:** Nanoindentation properties for $P_m$ = 1000 μN and $P_m$ = 9000 μN.

| Property | $P_m$ (μN) | PB05 | HB15 | PC | PMMA |
|---|---|---|---|---|---|
| $E_r$ (GPa) | 1000 | 4.58 ± 0.04 | 5.05 ± 0.04 | 3.05 ± 0.04 | 4.38 ± 0.01 |
| | 9000 | 3.96 ± 0.03 | 4.68 ± 0.03 | 2.69 ± 0.02 | 3.81 ± 0.03 |
| $H$ (MPa) | 1000 | 237.7 ± 1.2 | 267.1 ± 0.3 | 120.7 ± 2.9 | 163.0 ± 2.5 |
| | 9000 | 191.1 ± 2.8 | 213.5 ± 3.1 | 105.1 ± 1.1 | 150.3 ± 2.2 |
| $E_N/E$ | 1000 | N.A. | 1.9 | 1.2 | 1.4 |
| | 9000 | N.A. | 1.7 | 1.1 | 1.2 |
| $H/\sigma_y$ | 1000 | N.A. | 2.8 | 1.7 | 1.4 |
| | 9000 | N.A. | 2.3 | 1.5 | 1.3 |

Next, we consider uniaxial compression properties, at $\dot{\varepsilon} = c\,\dot{h}/h = 0.5c\,\dot{P}/P$, the equivalent uniaxial strain rate [143]. $c$~0.09 correlates the notional nanoindentation strain rate, $\dot{\varepsilon}_N \sim \dot{h}/h$, to $\dot{\varepsilon}$ [77], [96], [144]. At this $\dot{\varepsilon}$, $\sigma_y$ values estimated by interpolating the literature data are ~93 MPa for HB15 (from [77], [78]), ~66 MPa for PC (from [127], [128]), ~120 MPa for PMMA (from [115], [126]). Similarly, the respective $E$ values are ~2.2 GPa for HB15, ~2.4 GPa for PC and ~2.7 GPa for PMMA. Partially crosslinked SU-8 (PB05) is not easily amenable to form micropillars since they are formed by washing away soluble, uncrosslinked material [77], [78]. However, the uniaxial deformation HB15 micropillars are similar in materials and synthesis, to the nanoindented HB15.

The conventionally defined nanoindentation properties of glassy polymers, are significantly different from those obtained by uniaxial testing (last two rows of Table 6). In contrast to the trends, $H \sim 3\sigma_y$ and $E_{\mathrm{N}} \sim E$, for EP materials [145], we find that for all polymers, $\sigma_y < H \leq 3\sigma_y$ and $E_{\mathrm{N}} > E$. Thus, for glassy polymers, there is a dichotomy in the trends between the two metrics of nanoindentation deformation, with respect to the corresponding uniaxial deformation properties.

For $E_{\mathrm{N}}$, Tranchida *et al.* [27], have introduced a phenomenological correction factor, $\chi = E_N / E$, to account for the significant deviation of $E_{\mathrm{N}}$ from $E$. In this approach, we need to know *a priori*, the $E$ values of the bulk material. Mokhtari *et al.* [146] account for this deviation of $E_{\mathrm{N}}$ from $E$, by equating $S$ to the slope at 42% of $P_{\mathrm{m}}$. While these corrections decrease $E_{\mathrm{N}}$ towards $E$, there are no reported corrections to increase $H$ towards $3\sigma_y$.

In the next section, we examine the clues from the dichotomy of the trends in our definition-based estimates of $E_{\mathrm{N}}$ and $H$, obtained from $P-h$ data and SPM-based topological analyses. We then, qualitatively decompose the VEP glassy polymer deformation by benchmarking against the $P-h$ relationship of the EP quartz standard.

### 6.3.2. Nanoindentation Deformation Phenomena: Elastic, EP, VEP Effects

For compliant glassy polymers, while the load-bearing contact fraction is invariant with increasing $P_{\mathrm{m}}$, $H$ consistently decreases. Unlike for the six OP materials, the VE (time-dependent) deformation exhibited by polymers, could be the most likely cause of this "indentation size effect" on $H$. One study considers such creep effects during loading, and then employs a generalized VE model along with the OP framework for $A_c$ [76]. Additionally, we have indicated above, the FEA study [45], which employs the quadratic VEP model [124], [125], also predicts such phenomena. In contrast, we provide a model-free approach to qualitatively address VE effects.

Since $\kappa$ does not vary with $P_{\mathrm{m}}$, i.e., with loading time, we assume (as a first approximation) that even during the loading period (subscript $L$), $\kappa = (h_{cL} + h_b)/(h_{\mathrm{m}L} + h_b)$, is a constant (or varies linearly, in case of HB15). This means that $h_c$ would be a function only of $h_{\mathrm{m}}$; i.e., a consequence of compliance-driven Poisson upflow.

Employing $h_{\mathrm{m}L}$ from the $P-h$ data of the experiments in Table 2, we estimate at the end of the loading period, $H_L = P_{\mathrm{m}} / \left(24.5(h_{cL} + h_b)^2\right)$. Figure 10 is the plot of $H_L$ against $\ln P_{\mathrm{m}}$. For HB15 only, we linearly interpolate the slight variation in $\kappa$ vs $(h_{\mathrm{m}} + h_b)$, to estimate the corresponding $H_L$.

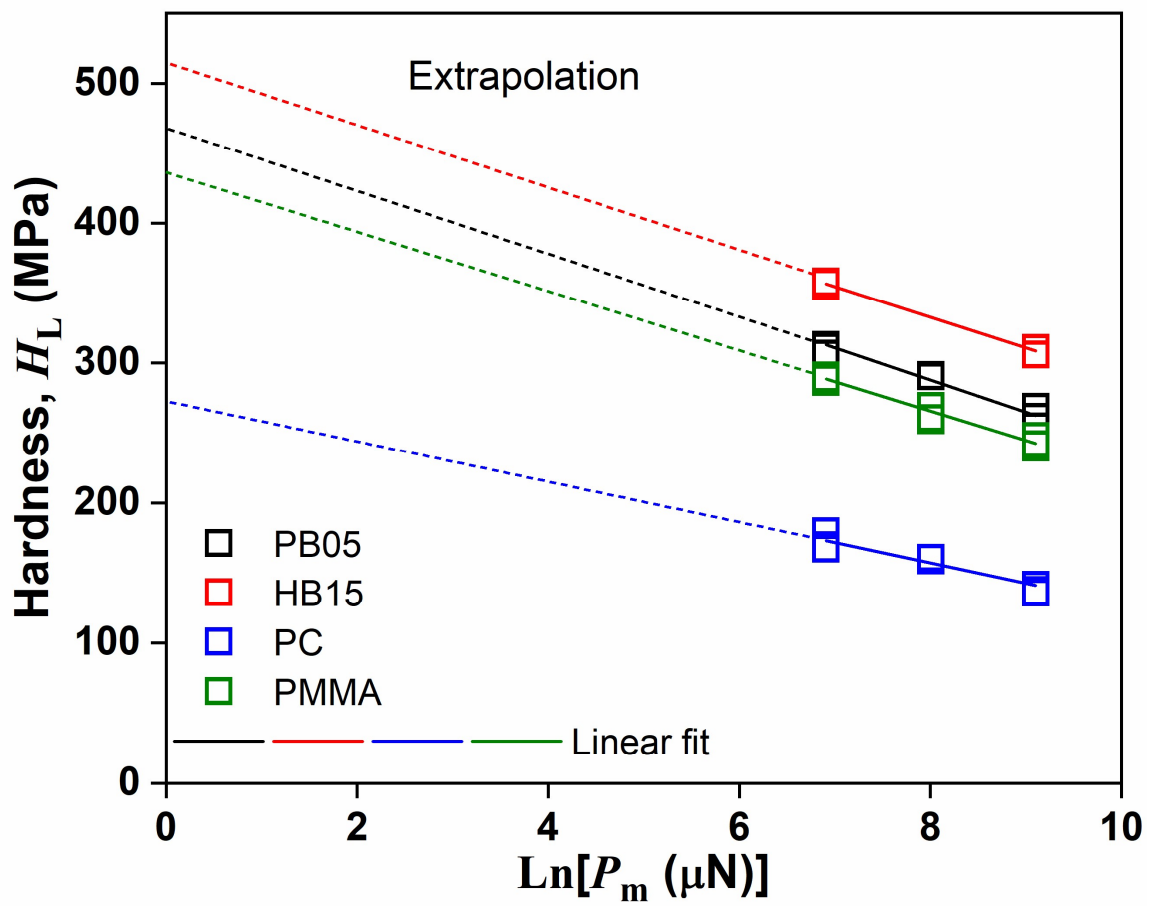

**Figure 10:** Estimate of EP (zero loading time) hardness, $H_{L0}$, by extrapolating the experimental loading hardness, $H_L$ to zero loading time $(\ln P_{\mathrm{m}} = 0)$.

The objective is to estimate the zero-time, limiting loading hardness, $H_{L0}$. We plot against $\ln P_{\mathrm{m}}$, because loading is at constant $\frac{d}{dt}(\ln P)$. With loading time for VE effects, proportional to $\ln P_{\mathrm{m}}$, zero loading time would occur for $\ln P_{\mathrm{m}}$=0. Since preload =1 μN, we express $P_{\mathrm{m}}$ in μN.

The $H_{L0}$ values are 507 MPa (HB15), 473 MPa (PB05), 433 MPa (PMMA) and 273 MPa (PC). Since $H_{L0}$ is the EP analog of polymer hardness, it is functionally similar to the hardness of the EP quartz standard (Appendix H). We thus exploit these phenomenological similarities between the EP hardness estimates and elastic unloading behavior of polymers and those of quartz.

The hypothetical EP loading plot for polymers is, $\left(h_{cEP}+h_b\right)=\left(\frac{P}{24.5\kappa^2 H_{L0}}\right)^{0.5}$. $\kappa=\frac{\left(h_{cEP}+h_b\right)}{\left(h_{EP}+h_b\right)}$, retains the volume conserving effect, for EP deformation as well, to estimate $h_{EP}(P)$. The hypothetical parabolic EP loading $P-h$ curve [15] for polymers, is from (0, 0) to $\left(\left(h_{EP,\mathrm{m}}+h_b\right),P_\mathrm{m}\right)$. The HB15 parabola is modified to accommodate the slight variation of $\kappa$.

At $P=P_\mathrm{m}$, the additional penetration, $\left(h_\mathrm{m}-h_{EP,\mathrm{m}}\right)$, would be due to the VE loading creep, which we subtract by translating the hypothetical elastic $h_{\mathrm{ne}0}$ unloading $P-h$ curve leftwards. Our consequent $P-h$ data estimates of the elastic and EP components, of polymer deformation, would map to those of conventional EP nanoindentation $P-h$ data; e.g., see [147]. Figure 11 illustrates our simplified deconvolution of the elastic, EP and VEP components during loading, followed by hypothetical elastic unloading and isolation of the VE loading and unloading creep.

We consider next, the significance of the EP hardness analog, $H_{L0}$, obtained by removing the softness effects of VE creep. $H\le 3\sigma_y$ (Table 6), while $H_{L0}>3\sigma_y$. $H_{L0}/\sigma_y$ values are ~5.4 for HB15, ~4.1 for PC and ~3.6 for PMMA. This deviation from $H_L\sim 3\sigma_y$ is significant, even after considering uncertainty in $H_{L0}$ (fitting $R^2$ ~0.97). Thus, the trend for hypothetical EP polymer hardness, becomes consistent with the trend, $E_\mathrm{N}>E$.

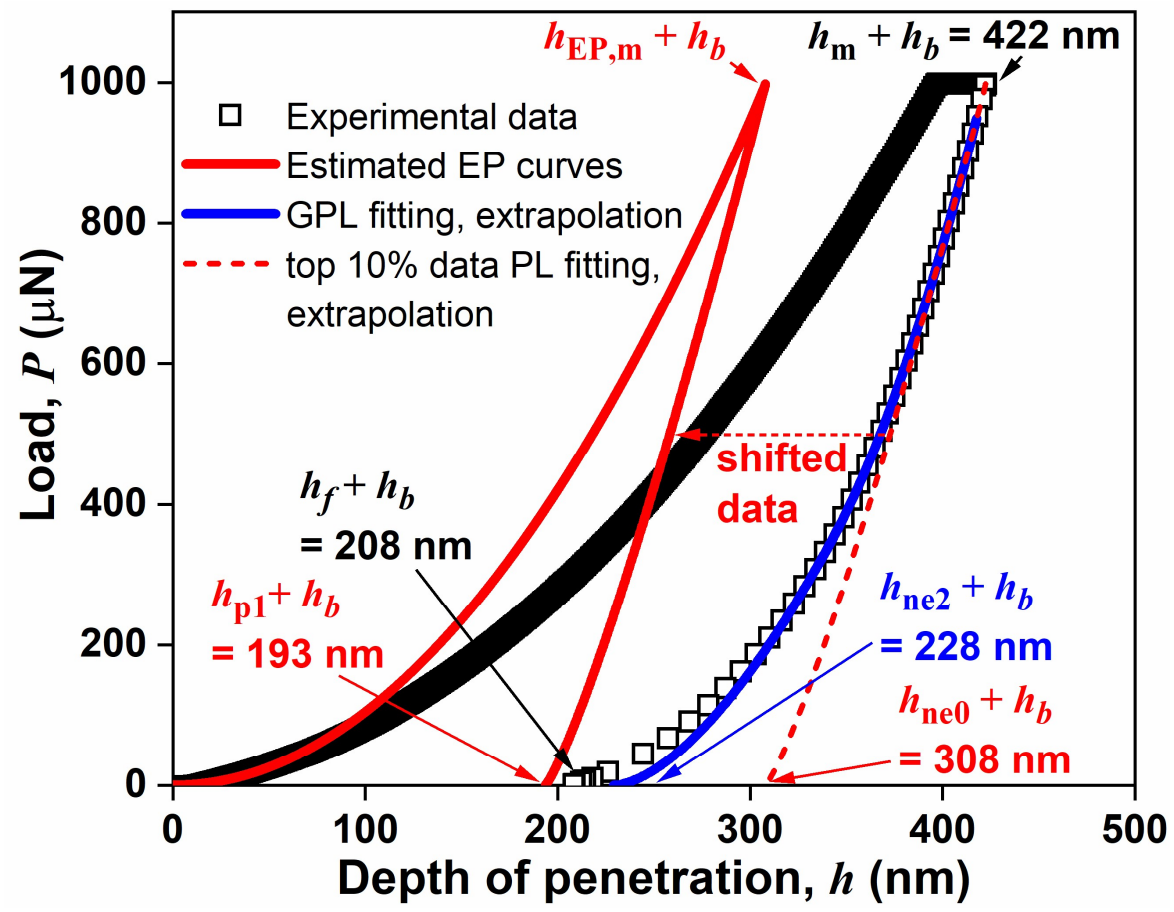

**Figure 11:** A simplified deconvolution of the nanoindentation deformation to facilitate visualization of its various contributions: elastic component ($h_{\mathrm{e}} = h_{\mathrm{m}} - h_{\mathrm{ne0}} = 116\,\mathrm{nm}$), time-independent plastic component ($h_{\mathrm{p}} + h_b$), EP component ($h_{\mathrm{ep,m}} + h_b = h_{\mathrm{p}} + h_b + h_{\mathrm{e}} = 308\,\mathrm{nm}$), remaining time-dependent VE component ($h_{\mathrm{m}} - h_{\mathrm{ep,m}} = h_{\mathrm{ne0}} - h_{\mathrm{p}} = 115\,\mathrm{nm}$).

A similar treatment on $E_{\mathrm{N}}$, would remove VE effects on it as well. However, since the dependence is $E_{\mathrm{N}} \sim \left(1/\sqrt{A_c}\right)$, as opposed to $H \sim \left(1/A_c\right)$, we find $E_{\mathrm{N}} > E$, even with the VE effects causing a a greater $A_c$. The consistent trends, $E_{\mathrm{N}} > E$ and $H_{L0} > 3\sigma_y$, for glassy polymers, suggest additional constraints from the unstressed surroundings, on the localized nanoindentation and recovery.

## 6.4. Constrained vs. Constraint Induced Deformation and Recovery

In this section we elucidate the effects of constraints during the localized glassy polymer nanoindentation deformation and recovery, vis-à-vis bulk uniaxial deformation.

The broad set of nanoindentation phenomena are a complex combination of localized VE, EP and elastic events. These localized events are constrained by the surrounding material. If we consider the hypothetical EP loading, and its corresponding hypothetical elastic unloading, then its post-unloading displacement $\left(h_{\mathrm{p}} + h_b\right)$, at $P$=0, is consistently lower than $\left(h_f + h_b\right)$. Any additional recovery in the apex residue level (after EP loading and elastic unloading) should not occur, since the apex residue consists only of plastic deformation. Any plastic deformation of an isolated body (such as a micropillar) would be permanent. In contrast, on reaching $h_p$, all local VE recovery of the apex zone is not complete; during post-unload SPM scans, the residue depth decreases to $h_r < h_p$.

Figure 12 provides a visualization of the likely events. There is a permanent imprint from the sharp edges, that radiates outward from the tip apex. The regions beneath the facets may not have yielded, but could be constrained from full zero-time recovery, since their boundaries comprise the yielded apex and edge regions. The yield regions beneath the edges are shallow, compared to a deep yielded region beneath the apex. On unload, the $E_0$ elastic recovery occurs everywhere in a localized sense, to the extent possible. The less strained outer regions with shallow yield edges, recover significantly (but not completely), while the apex region undergoes limited recovery.

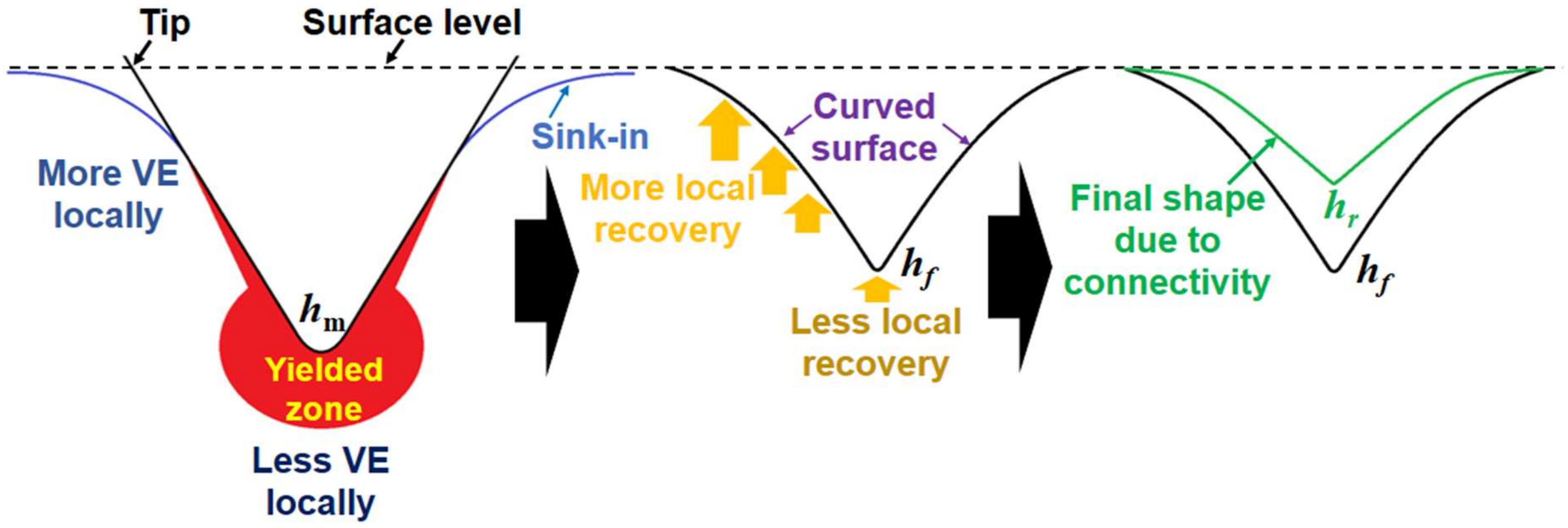

**Figure 12:** Visualization of the effect of the connection of the tip apex region of the residual imprint to its surroundings. The apex has completed VE recovery (with a deep permanently yielded zone underneath). The surrounding region with a shallow yield beneath the imprints of the edges and the even lesser yielded facets, continue to recover from its VE compression, and pulls the apex upwards, causing constraint-induced excess recovery.

However, due to VE effects, these less-yielded outer regions continue to recover further at zero load. We find that despite the progressively greater yielded content in the inner regions, the apex also continues to recover, from $h_p$ to $h_r$. Hence, the continued local recovery of the less-yielded regions surrounding the apex, “pulls up” the significantly yielded apex region; i.e., the greatly yielded apex region undergoes additional constraint-induced recovery.

Pile-up is another example of constraint-induced “excess” recovery. The edge sink-in elastically recovers to the $z$=0 level, pushing the less sunk-in facet profile to above the unperturbed surface level. The LS-described sink-in is also constraint-induced, since it occurs where there is no direct contact stress. Conversely, the regions away from the apex, asymptotically undergo zero strain, and constrain the deformation at the apex. Mirroring the post-unload constraint-induced recovery, the surroundings constrain the loading response. The lower stresses on the surrounding regions, tend to reduce deformations in contact regions. Thus, we infer that constrained deformation causes the trends, $E_{\mathrm{N}} > E$ and $H_{L0} > 3\sigma_y$.

We can understand this in terms of the radial distances, normalized by contact radius, $a_c$. Then the outer regions ($r/a_c > 1$) on glassy polymers, are relatively closer to the tip surface (compared to those on OP materials) because of the upflow. The well correlated compliance-induced increase in $A_c$ in glassy polymers, increases the required $P_{\mathrm{m}}$ for a given displacement. For a given deformation, the constraints, due to their normalized proximity, increase the required $P_{\mathrm{m}}$, even further. Thus, the greater upflow beneath the tip, could amplify the influence of the surroundings.

The physical origins of the amplified constraints could be due to the retained molecular connectivity across the contact radius. OP-type materials might not achieve such molecular connectivity, due to the greater normalized distances ($r/a_c$) and creation of slip-lines during nanoindentation [12].

Finally, based on our experimental observations, we provide briefly, two examples of suggested directions in developing predictive models and FEA frameworks. (i) A simulation [45] via a phenomenological quadratic constitutive model, relating indenter load to apex displacement [124], [125], has predicted low sink-in and linear depth contours, and fit the $P-h$ data. We suggest exploring a physical interpretation of this phenomenological model, in terms of the physical bulk parameters relating stress to strain and strain rate. (ii) We suggest also as an example, incorporation of deformation dependence of the Poisson's ratio [148]–[153]. This could account for the effects of the simultaneous coexistence of a wide range of polymer nanoindentation strains.

# 7. Conclusions

In this investigation, we have directly examined the topology, during and after nanoindentation of amorphous glassy polymers and analyzed the corresponding $P-h$ data.

_Experimental Observations_: Visual observation of nanoindentation deformation and post unload SPM analysis for all polymers, provide direct evidence, that for glassy polymers, the contact fractions are broadly independent of the load, and significantly greater than those predicted by the OP (Oliver-Pharr) method; $\kappa = h_c^{\Delta} / h_{\mathrm{m}}^{\Delta} \sim 0.86 - 0.95$, and correlates well with the individual polymer compliances. Sink-in occurs during loading, and observable pile-up is visible post-unloading.

Based on these observations, we have rationally identified material phenomena at various levels.

_Topology and Mechanism_: The experimental observations enable an estimate of the topological geometry of the Berkovich adaptation of conical indentation. The topological geometry suggests the deformation pathway from the indenter's downward displacement, via the Poisson widening (high $\nu$) of the inner regions, to the vertical Poisson expansion of the outer regions, leading to upflow. This upflow causes a simultaneously reduced edge-side sink-in with facet pile-up; i.e., high $\kappa$, with the contact pile-up height, $h_c^{\mathrm{PU}} < h_{\mathrm{m}}^{\mathrm{PU}}$, the SPM imaged maximum.

_Visco-elasto-plastic (VEP) Material Phenomena_: The unload segment of the $P-h$ data can only be described by a generalized power law (GPL), which indicates unloading creep; i.e., viscoelastic (VE) effects. The pure elastic (no VE effects) download estimate results from power law (PL) to only the approximate onset of the high rate download, is phenomenologically similar to quartz unload.

The definition-faithful nanoindentation modulus, $E_N$ and hardness, $H$, for glassy polymers (obtained from the SPM analysis and unload onset slope of the $P-h$ data) exhibit opposing trends, vis-à-vis the uniaxial bulk properties, $E$ and $\sigma_y$: $E_N > E$, but $H < 3\sigma_y$. Removing the creep effects from the $P-h$ data, yields the hypothetical zero-time elastoplastic (EP) loading hardness, $H_{L0} > 3\sigma_y$, a trend consistent with that of $E_N$. This combination of $H_{L0}$, and elastic unload, yields the equivalent EP $P-h$ contributions to the nanoindentation of VEP polymers, isolating the superimposed VE effects.

*Interactions between deformed region and surroundings*: For a broad range of EP materials, $E_N \sim E$, and $H \sim 3\sigma_y$. Our findings for $E_N$ and $H_{L0}$ for glassy polymers, then indicate that local deformation is constrained by the surroundings. This loading constraint inference mirrors the converse, that the constraints from the less yielded surroundings, induce extra VE recovery of the highly yielded apex region, by pulling it up, during unloading and at zero-load. Other constraints-induced phenomena include pile-up and sink-in.

Such phenomena that we have described, could be included in the development of physics-based predictive models and when building correct "user-described models" in FE simulation platforms.

# Acknowledgments

We would like to thank the Centre of Excellence in Nanoelectronics (CEN), IIT Bombay, for providing SU-8 epoxy and facilities for the SU-8 sample fabrication. The authors would like to acknowledge the Nanoindenter lab, Central facility, IIT Bombay for providing the standard quartz, and for the nanoindentation experiments.

# Funding

No funding was available for this project.

# Appendix A

## Structural characterization of cross-linked SU-8 via FT-IR

The standard photolithography process is illustrated in Fig. A.1. For PB05, we have carried out the post-exposure baking (PB) step for 5 min only. For HB15, we have performed 30 min of PB, after which, we have continued to the hard baking (HB) step for 15 min.

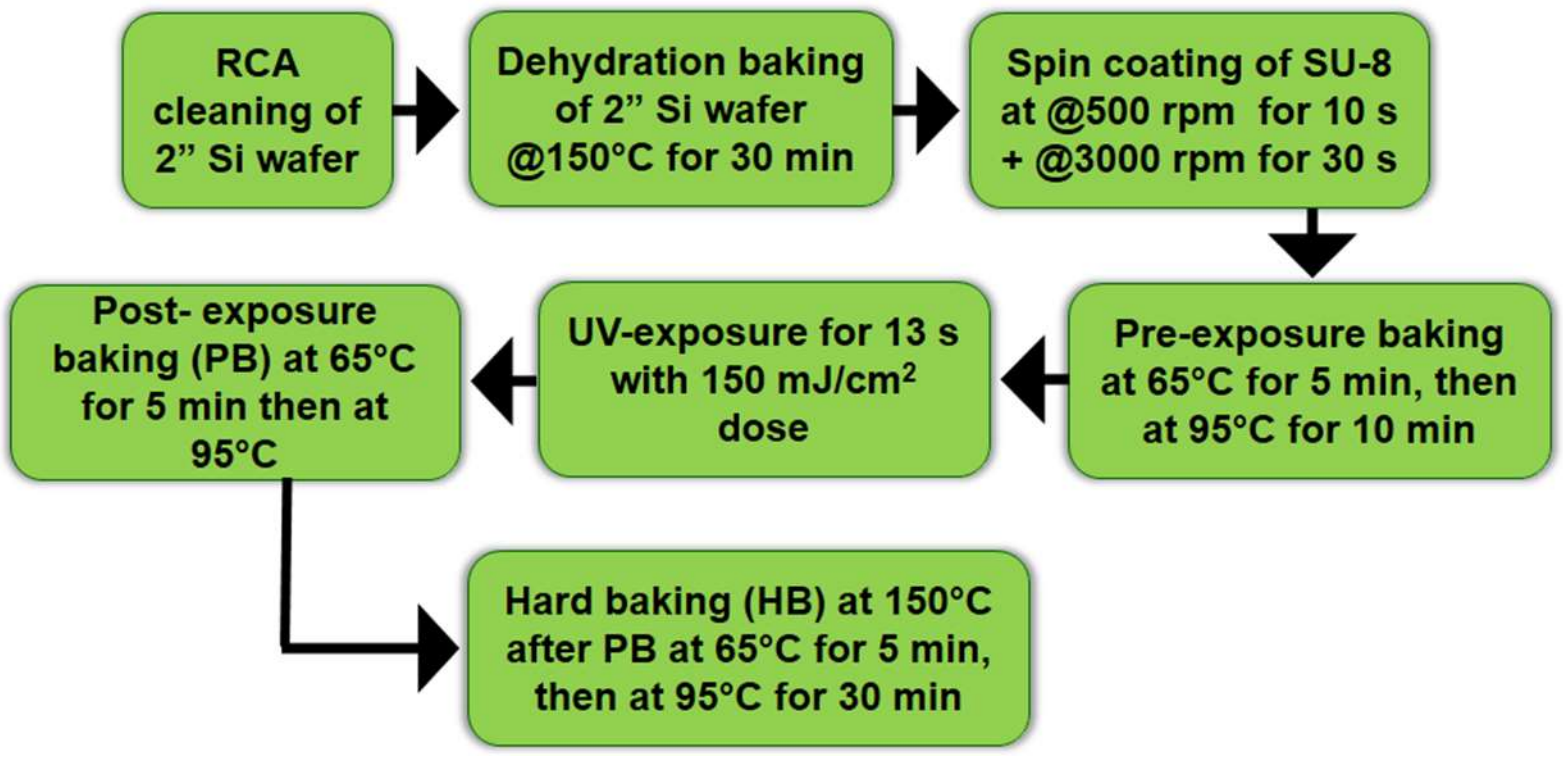

**Figure A.1:** Schematic of the steps involved in the standard photo-lithography process.

To evaluate the extent of cross-linking of SU-8, we have collected *in situ*, Fourier Transform Infrared Spectroscopy (FTIR) spectra at the baking temperature (PB and HB) [154], [155], by employing a 3000 Hyperion FTIR Microscope with Bruker Vertex 80 (Germany) spectrometer. We find that PB05 corresponds to, and HB15 corresponds to. We place the UV exposed spin coated SU-8 inside the furnace, ramp the temperature to 95°C, and hold for 5 min. This corresponds to the post-baked sample, PB05. To obtain the hard-baked sample (HB15), after 30 min of the PB step, we raise the temperature to 150°C, and hold for 15 min. The $I_{914}$ peak represents the epoxy group $(-C-O-C-)$ in the SU-8 monomer, and $I_{1608}$ peak is for the aromatic group, which does not take part in the cross-linking reaction [156]. With increase in extent of cross-linking, $I_{914}$ decreases, and would vanish when the SU-8 is fully cross-linked [157]. The extent of cross-linking (%) = $\left[(A_{BE} - A_{AB}) \times 100\% / A_{BE}\right]$ [158].

Here, $A_{\mathrm{AB}}$ is the value of the $I_{914}$ peak, normalized by the $I_{1608}$ peak, of the baked sample after UV-exposure. $A_{\mathrm{BE}}$ is the same ratio of the sample, before UV exposure.

# Appendix B

## Blunt height estimation of tip during each usage day

We have followed the conventional area calibration procedure [2] at each time, making 30 tip area calibration indents on the quartz standard. The maximum load, $P_{\mathrm{m}}$, varies from 100 μN for the first indent, progressively up to 10,000 μN, for the thirtieth indent. The *P-h* curves of the tip area calibration on quartz standard, provide the contact depth, $h_c$, as per the OP method. For each $P_{\mathrm{m}}$, the known $E_r = 69.6 \pm 3.5$ GPa, and $H = 9.3 \pm 0.9$ GPa, yield two $A_c$ values, $A_c^{Er} = \pi S^2 / 4E_r^2$ and $A_c^H = P_{\mathrm{m}}/H$, respectively. We consider the average of the two areas, recognizing that $A_c(h_c) \approx 24.5 \times (h_c + h_b)^2$ [30], [66], [68], [69], [159], to yield the fitted $h_b$. This fitting is implemented by self-consistently disregarding data in the low $h_c$ range $(h_c < R/3)$ [160], where $R = h_b\left[\sin\alpha/(1-\sin\alpha)\right]$, is the tip radius.

We have performed SEM analyses (model: dual beam FEI Carl Zeiss Auriga™) of the indenter tip on usage day 568, as shown in Fig. B.1(a) provides the direct estimate of $h_b$. Figure B.1(b) is a plot of the average of the two $h_b$ values as a function of usage days (when the quartz calibration was performed). We find that the linear fit of the average $h_b$ values, describes the trend very well, and the SEM $h_b$ value (~ 27 nm) lies very clearly on the linear fit.

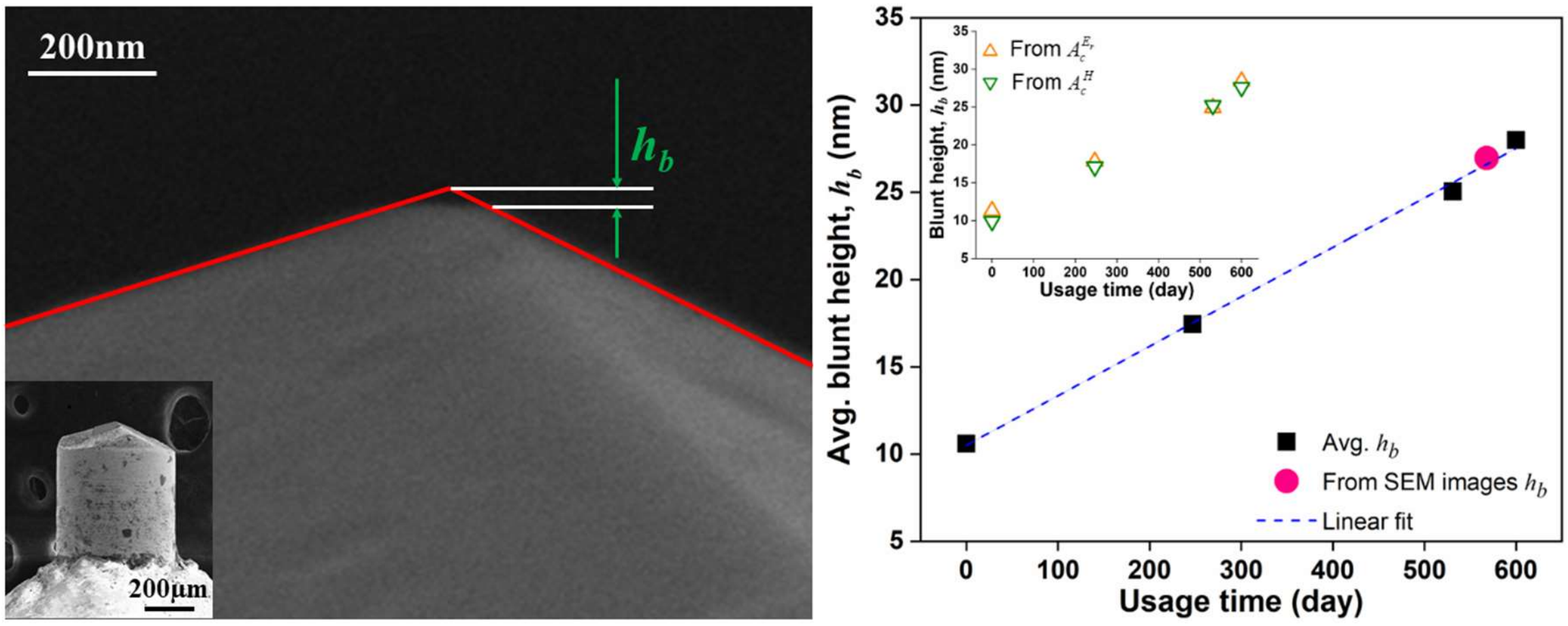

(a) (b)

**Figure B.1:** (a) SEM images of Berkovich tip and $h_b$ measurement and (b) $h_b$ values change by progressive utilization. The linear fit is $h_b = pt + q$.

# Appendix C

# PL fitting of *h* vs *P* unload data

The PL fit of $P$ vs $h$, appears adequate superficially ($R^2 \sim 0.998$). Although the relative residuals are significantly greater than those via our GPL method (Fig. C.1 and Supporting Information 2), they are still small, up to ~ 2%, even in the low $P$-$h$ region $\left(P \sim 0.20P_{\mathrm{m}}\right)$.

Hence, in order to understand the inadequacy of the PL equation fitting for glassy polymers (plotted in Fig. C.2(a)), we examine a magnified plot in the neighborhood of $h_{\mathrm{ne}2}$ (Fig. C.2(b)) and of the experimental $h_f$. Here, we find that the PL equation fitted curve indicates a non-zero load, consistent with the unphysical fitted $h_f$ value ($= h_{\mathrm{ne}2}^{\mathrm{PL}} = h|_{P=0}$).

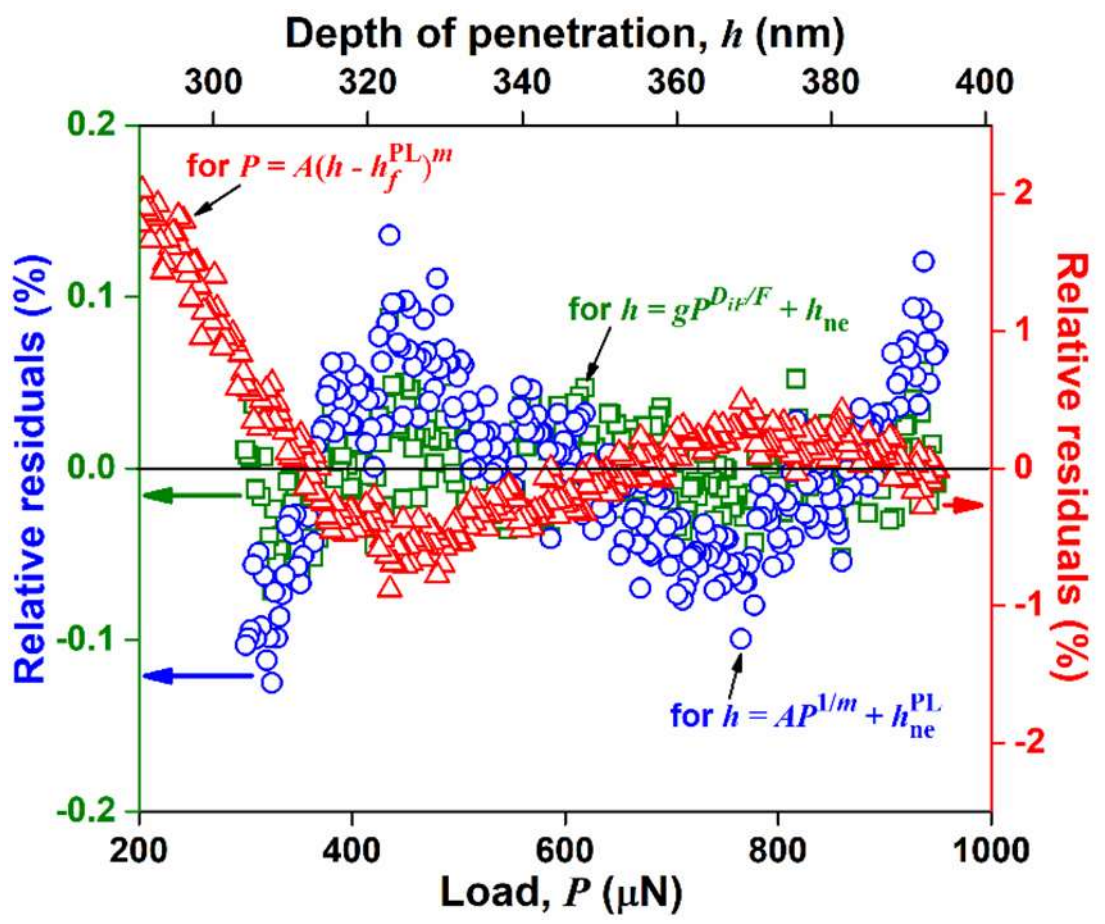

**Figure C.1:** Relative residuals from PL equation fitting for $P$ data (red), PL fitting for $h$ data (blue) and GPL fitting (green).

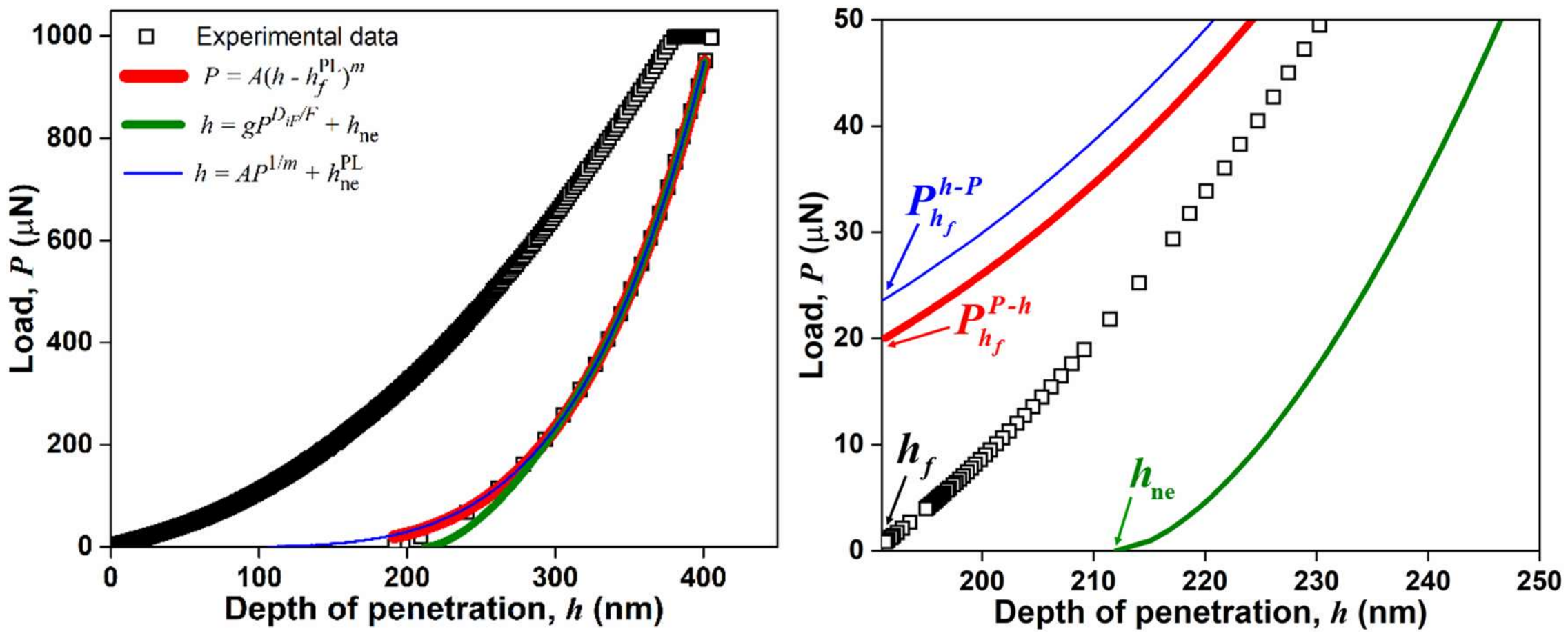

(a) (b)

**Figure C.2:** (a) $P$-$h$ curve with fitted equation for $P_m$ = 1000 μN and (b) magnified portion near the residual depth for PB05.

# Appendix D

# Elastic Nanoindentation Unloading of EP Quartz Standard

Here, we briefly describe the EP behavior of the quartz standard, so as to employ it as a benchmark, against which, the corresponding response of the polymers can be described. The data for quartz are available from the tip calibration experiments carried out on various days during the tip's lifetime. We have considered loading up to the maximum load, $P_m$ =10,000 μN, to represent all loads from 100 μN onwards. The unloading follows the PL, $P = B(h - h_f)^m$, (in the range from 0.2 $P_m$ to 0.95 $P_m$), where $B$, $m$, $h_f$ are the fitting constants; $h_f$ is the residual depth at $P$ = 0 μN. Hence, stiffness, $S = mP_m/(h_m - h_f)$. We fit $B = \sum_{i=0}^{3} b_i \times (h_m + h_b)^i$, and employ our empirical finding, $(h_f + h_b) = \sqrt{(P_m/0.48)}$; we find that $m$ varies linearly from ~ 1.25 (for $P_m$ ~2,000 μN) to ~ 1.31 (for $P_m$ ~10,000 μN). For the elastoplastic quartz standard, the stiffness, $S$, varies in a parabolic fashion vs $P_m$ (Fig. D.1(a)), as it does in case of the amorphous glassy polymers (Fig. D.1(b)).

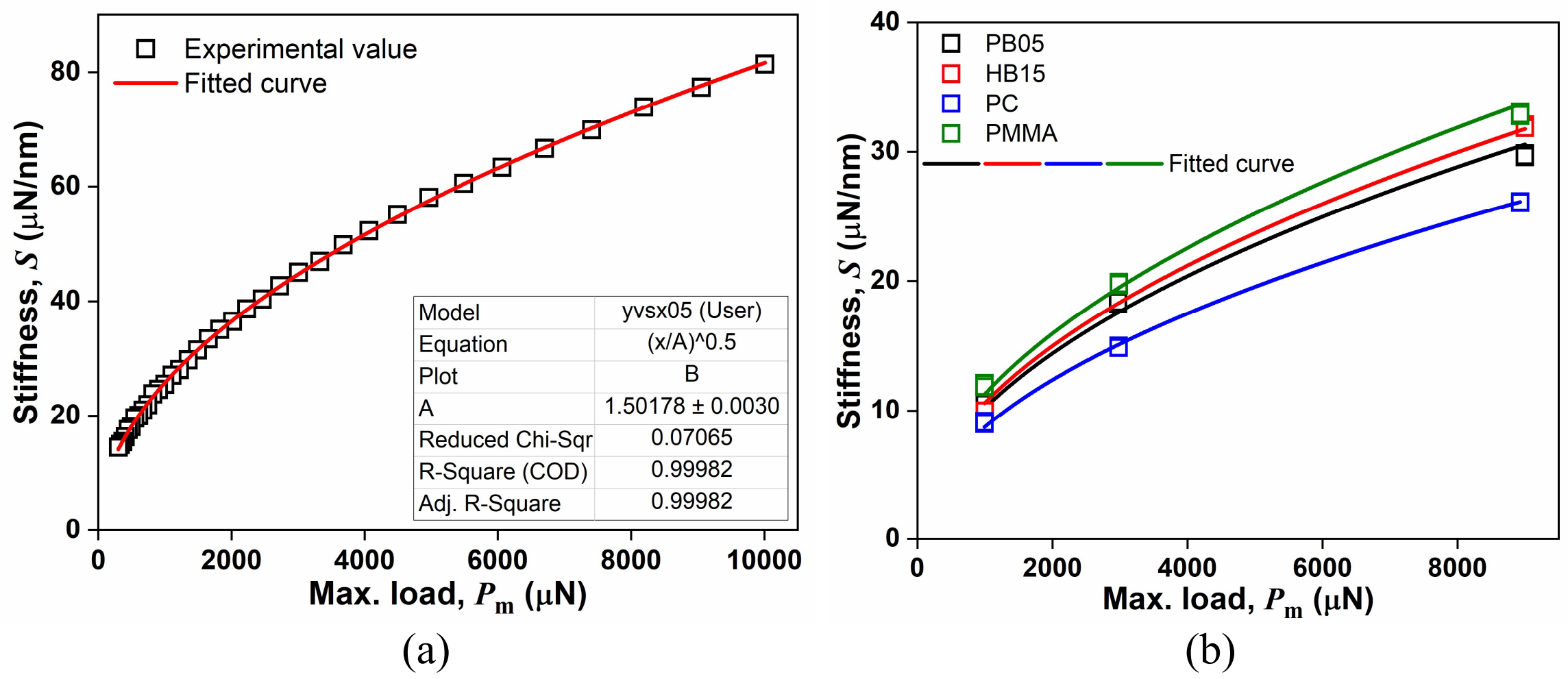

**Figure D.1:** Parabolic relationship of $S$ vs $P_m$ for (a) the quartz standard (b) glassy polymers.

# Appendix E

## Viscoelastic effects on imaged pile-up

Figure E.1(a) demonstrates via consecutive SPM profiles of HB15 and PC ($r > a_c$), at times $t_1$ and $t_2$ that $\frac{h_{r1}^{p,\mathrm{PU}}(r)}{h_{r2}^{p,\mathrm{PU}}(r)} \approx \frac{(h_c - h_{r1})}{(h_c - h_{r2})}$. The relative residuals (%) from proportional profiles for HB15 and PC are shown in Fig. E.1(b). It is seen that the residuals are random, scattered and within ± 15%.

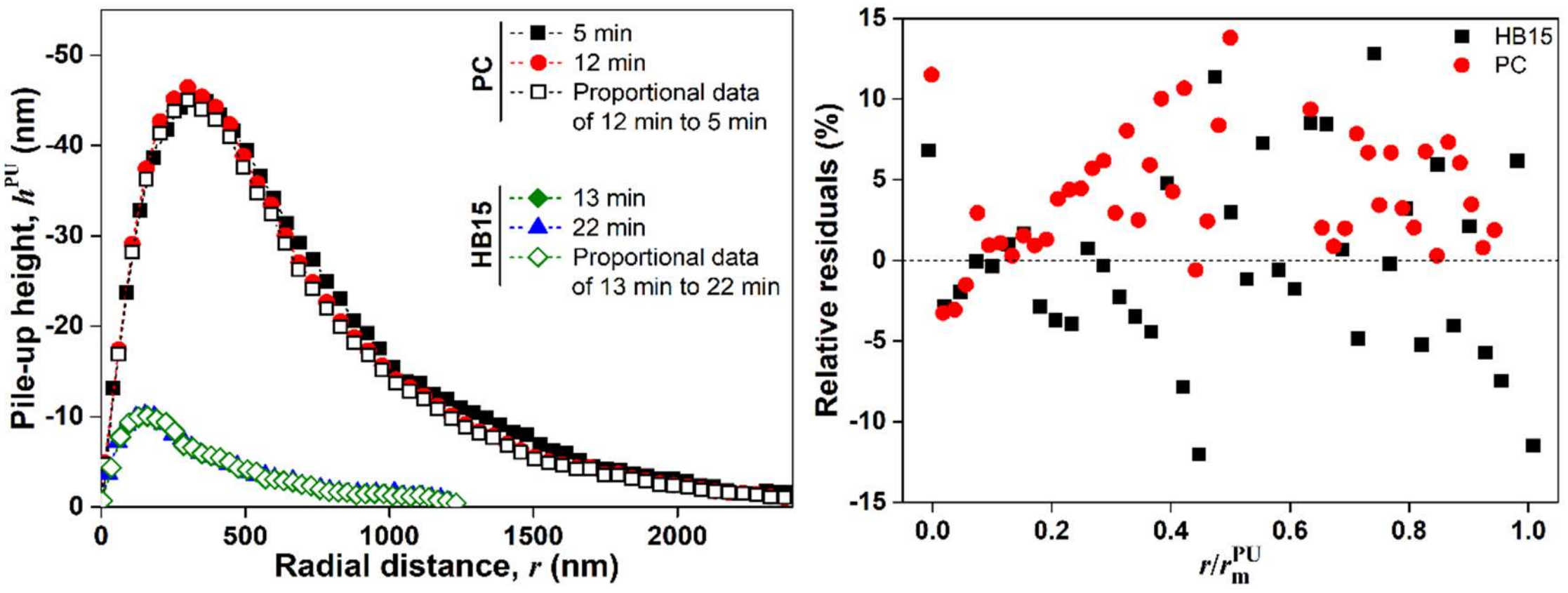

(a) (b)

**Figure E.1:** (a) Pile-up profiles for HB15 and PC. The mentioned times indicate the delay time when the images were acquired just after removal of load. (b) the relative residuals (%) from proportional profiles for HB15 and PC. Here, we have taken $r_{\mathrm{m,HB15}}^{\mathrm{PU}}$ = 1200 nm and $r_{\mathrm{m,PC}}^{\mathrm{PU}}$ = 2400 nm.

# Appendix F

## Polymer Nanoindentation: Surface Topology Estimate at $P_{\mathrm{m}}$

Figure F.1(a)-(b) is a rational quantification of the likely surface topology for PC, which would yield the profile obtained via the SPM imaging.

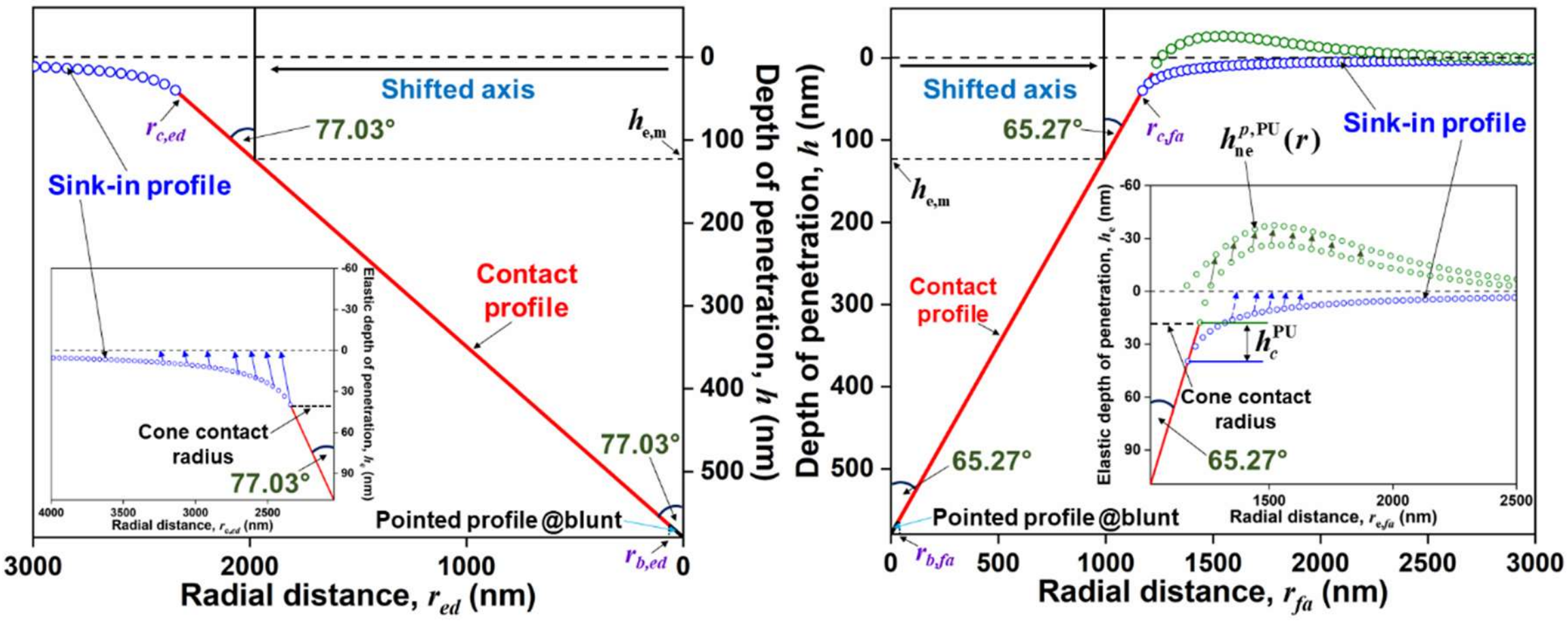

(a) (b)

**Figure F.1:** Deformation profile for PC at $P_{\mathrm{m}}$ (a) edge side, (b) facet side.

# Appendix G

## Effects of Pile-up on Contact Area

The increase in $A_c$, is the minor segment of a circle, bound by an arc (half-angle = $\eta$) and the chord (the $a$-line). The projected area of the elevation, i.e., the segment area is calculated as shown in Fig. G.1(a)-(b). The arc sector radius, $r^{\mathrm{PU}} = \left(h_c^{\mathrm{PU}}/2\right) + \left(a^2/8h_c^{\mathrm{PU}}\right)$. As $h_c^{\mathrm{PU}}$ increases, $r^{\mathrm{PU}}$ decreases and $\eta$ increases (Fig. G.1(c)). The $h_c^{\mathrm{PU}}$-based additional projected area (eqn. 8) is added to the imprint-based $A_c^{\Delta}$ [16], [17], [21]–[25]. The values for $r^{\mathrm{PU}}$, $\eta$ (radians), and $A_c^{\mathrm{PU}}$ for all the indentations, are listed in Supporting Information 6.

$$A_c^{\mathrm{PU}} = 3\left[\frac{\pi r^2 \eta}{360} - \left(\frac{a}{2}\frac{a/2}{\tan(\eta/2)}\right)\right]$$

eqn. G.1

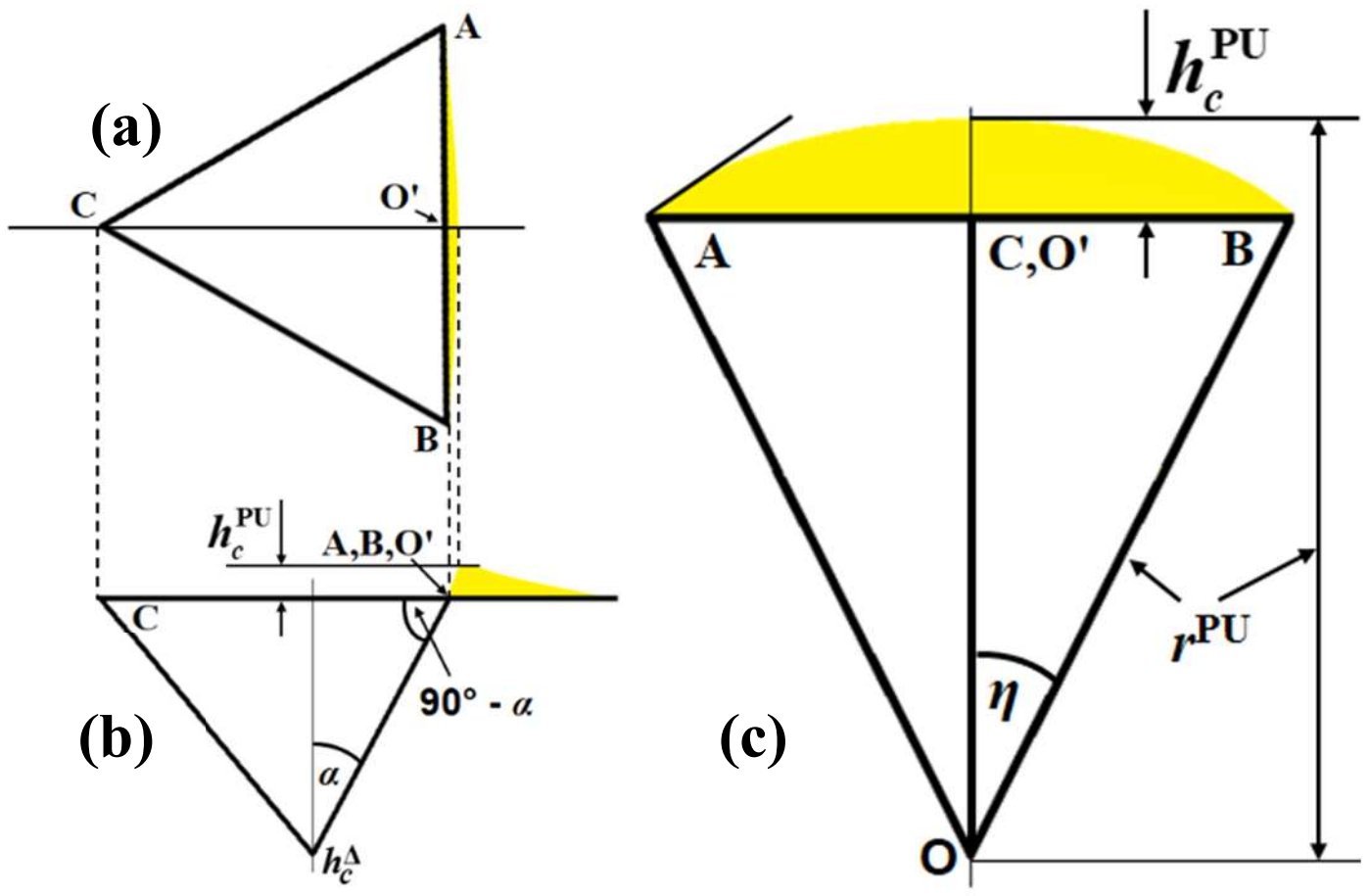

**Figure G.1:** Schematic depicting the deformation profile with pile-up arc: (a) top view (b) side view, and (c) the section of the hypothetical circle ($r^{\mathrm{PU}}$ and $\eta$) with the pile-up arc (arc tangent is indicated). Note that the depth scales in (a)-(b) and (c) are different; also, $r^{\mathrm{PU}} \gg h_c^{\Delta}$ and $\eta \gg \alpha$.

Our analysis estimates the $h_c^{\mathrm{PU}}$ on the facet via a proportional downscaling (Fig. 5(b)) of the SPM-based profile, followed by identifying the location of the contact. Directly considering the SPM-imaged $h_{\mathrm{m}}^{\mathrm{PU}}$ (instead of $h_c^{\mathrm{PU}}$), increases the $A_c$ estimates by ~ 1.5% for HB15 and by ~ 6.1% for PC.

# Appendix H

# Loading Data Analysis for Quartz Nanoindentation

We have considered loading up to the maximum load, $P_{\mathrm{m}}$ =10,000 μN, to represent all loads from 100 μN onwards. The loading curve for any tip usage day, is described by $P(\mu\mathrm{N}) \sim 0.11 \times \left(h(\mathrm{nm}) + h_b\right)^2$. This power law is expected for purely elastic materials [6]–[8] as well as for EP materials [15]. There is no additional displacement (no creep) during the hold period at $P_{\mathrm{m}}$, which is consistent with quartz's EP (very large viscosity) nature. Consequently, the average stress (instantaneous hardness, *H*) is constant during loading (i.e., $P(\mu\mathrm{N}) \sim 0.225 \times \left(h_c(\mathrm{nm}) + h_b\right)^2$). The contact depth, $h_c$, during loading at any given *P*, can be considered to be that at the onset of unloading from that *P*; i.e., we substitute $P_\mathrm{m}$=*P*, in the equation, $h_c = h_\mathrm{m} - \left(0.75 P_\mathrm{m} / S\right)$.

# Physical phenomena during nanoindentation deformation of amorphous glassy polymers

Prakash Sarkar[1], Prita Pant[1], Hemant Nanavati[2*]

[1]Department of Metallurgical Engineering and Materials Science, Indian Institute of Technology Bombay, Mumbai- 400076, Maharashtra, India

[2]Department of Chemical Engineering, Indian Institute of Technology Bombay, Mumbai- 400076, Maharashtra, India

*E-mail: hnanavati@iitb.ac.in

# Supporting Information 1

## Rate Variation during Unloading

Figure S.1 contains the plots of the *measured* normalized unloading rate, $\dot{P}/P_{\mathrm{m}}$, vs the normalized unloading $P$ values, $P/P_{\mathrm{m}}$, for all polymers as well as for the quartz standard. We observe that the constant *set* unloading rate range is from ~ 0.3 $P_{\mathrm{m}}$ to ~ 0.95 $P_{\mathrm{m}}$ (~0.98 $P_{\mathrm{m}}$, for $P_{\mathrm{m}}$ = 9000 μN). This constant $\dot{P}/P_{\mathrm{m}}$ range, depends primarily on the nominal unloading time and very slightly on $P_{\mathrm{m}}$, but is independent of the sample material.

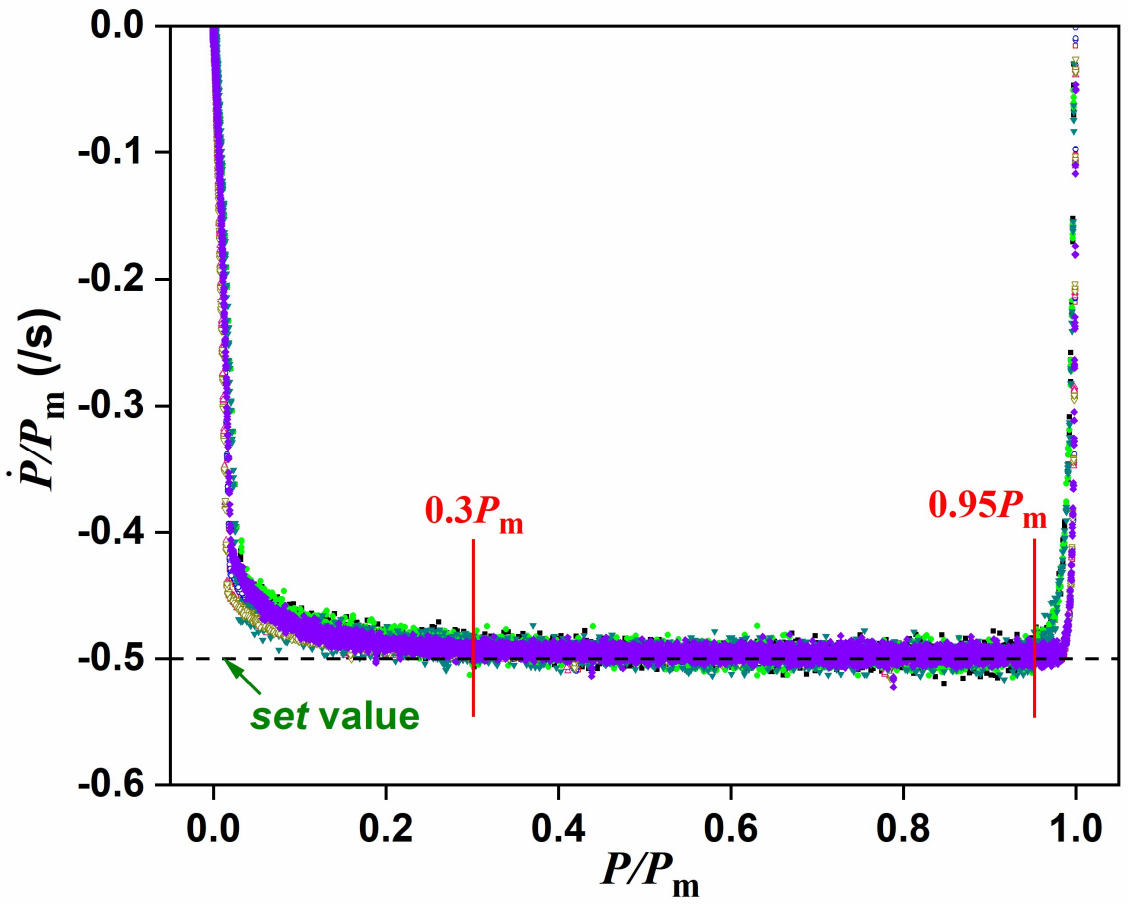

**Figure S.1:** Actual normalized unloading rate vs normalized unloading $P$ for all indents performed on PB05, HB15, PC, PMMA and quartz for $P_{\mathrm{m}}$ = 1000 μN and 9000 μN. The solid symbols are for $P_{\mathrm{m}}$ = 1000 μN and open symbols are for $P_{\mathrm{m}}$ = 9000 μN. For the unloading step, the loading rate is -ve; its absolute value increases downward.

# Supporting Information 2

## $h$ vs $P$ unload curve Analyses

### S.2.1. $h$ vs $P$ unload curve, GPL fitting parameters

The generalized power law (GPL) fitting has been described in eqn. 2. The corresponding fitted values are listed in Table S.2.1.

Table S.2.1: **GPL fitted constants and their ranges.**

| Sample | $P_m$ (μN) | $k_0$ | $k_1$ | $k_2$ | $1/m$ (Value or range) | |
|---|---|---|---|---|---|---|
| | | | | | $h = gP^{D_F/F} + h_{ne}$ | $P = B\left(h - h_f\right)^m$ |
| PB05 | 1000 | 8.15 ± 0.47 | -2.21 ± 0.01 | 0.16 ± 0.01 | 0.46 to 0.64 | 0.26 |
| | 9000 | 11.02 ± 0.30 | -2.35 ± 0.07 | 0.13 ± 0.00 | 0.45 to 0.57 | 0.27 |
| HB15 | 1000 | 7.94 ± 0.76 | -2.20 ± 0.23 | 0.16 ± 0.02 | 0.48 to 0.62 | 0.33 |
| | 9000 | 9.55 ± 0.65 | -1.96 ± 0.15 | 0.11 ± 0.01 | 0.40 to 0.56 | 0.34 |
| PC | 1000 | 5.12 ± 0.74 | -1.28 ± 0.22 | 0.09 ± 0.02 | 0.46 to 0.62 | 0.35 |
| | 9000 | 8.74 ± 0.67 | -1.80 ± 0.14 | 0.10 ± 0.01 | 0.45 to 0.59 | 0.36 |
| PMMA | 1000 | 1.51 ± 0.01 | -0.13 ± 0.01 | -- | 0.65 to 0.79 | 0.52 |
| | 9000 | 0.89 ± 0.01 | -0.03 ± 0.01 | -- | 0.57 to 0.60 | 0.54 |

$h_f$ values in all the literature methods are fitted values. They are either unphysical or significantly lower than the experimental values (Table S.2.2).

Table S.2.2: **Experimental and fitted $h_f$ (nm) values, via various methods.**

| References | $P_m$ (μN) | PB05 | HB15 | PC | PMMA |
|---|---|---|---|---|---|
| Oliver-Pharr [1] | 1000<br>9000 | 68 ± 3<br>320 ± 15 | 117 ± 1<br>504 ± 7 | 248 ± 5<br>933 ± 16 | 321 ± 5<br>1033 ± 9 |
| Kossman *et al.* [2] | 1000<br>9000 | 9 ± 6<br>93 ± 34 | -199 ± 18<br>-1177 ± 26 | 97 ± 4<br>443 ± 18 | 321 ± 6<br>1033 ± 9 |
| Liu *et al.* [3] | 1000<br>9000 | -1.8E6 ± 1.7E6<br>1.5E6 ± 1.8E6 | -3.2E6 ± 2.4E6<br>-7.0E6 ± 0.5E6 | -3.2E6 ± 0.5E6<br>-19.1E6 ± 7.0E6 | 1.1E6 ± 1.5E6<br>7.9E6 ± 2.5E6 |
| Experimental values | 1000<br>9000 | 191 ± 1<br>529 ± 15 | 172 ± 1<br>473 ± 7 | 300 ± 3<br>983 ± 15 | 304 ± 7<br>904 ± 4 |

## S.2.2. *h* vs *P* unload curves with GPL fitting

We have seen that the GPL framework describes well, the relationship between $h - h_{ne}$ and $P$, over the entire constant unload rate range, as shown in Fig. S.2.1, Fig. S.2.2(a)-(b), Fig. S.2.3(a)-(b) and Fig. S.2.4(a)-(b), for PB05, HB15, PC and PMMA, respectively. We find that the relative residuals of this fitted equation are small, scattered and random (up to ~ 0.03%), with $R^2$ ~ 0.999.

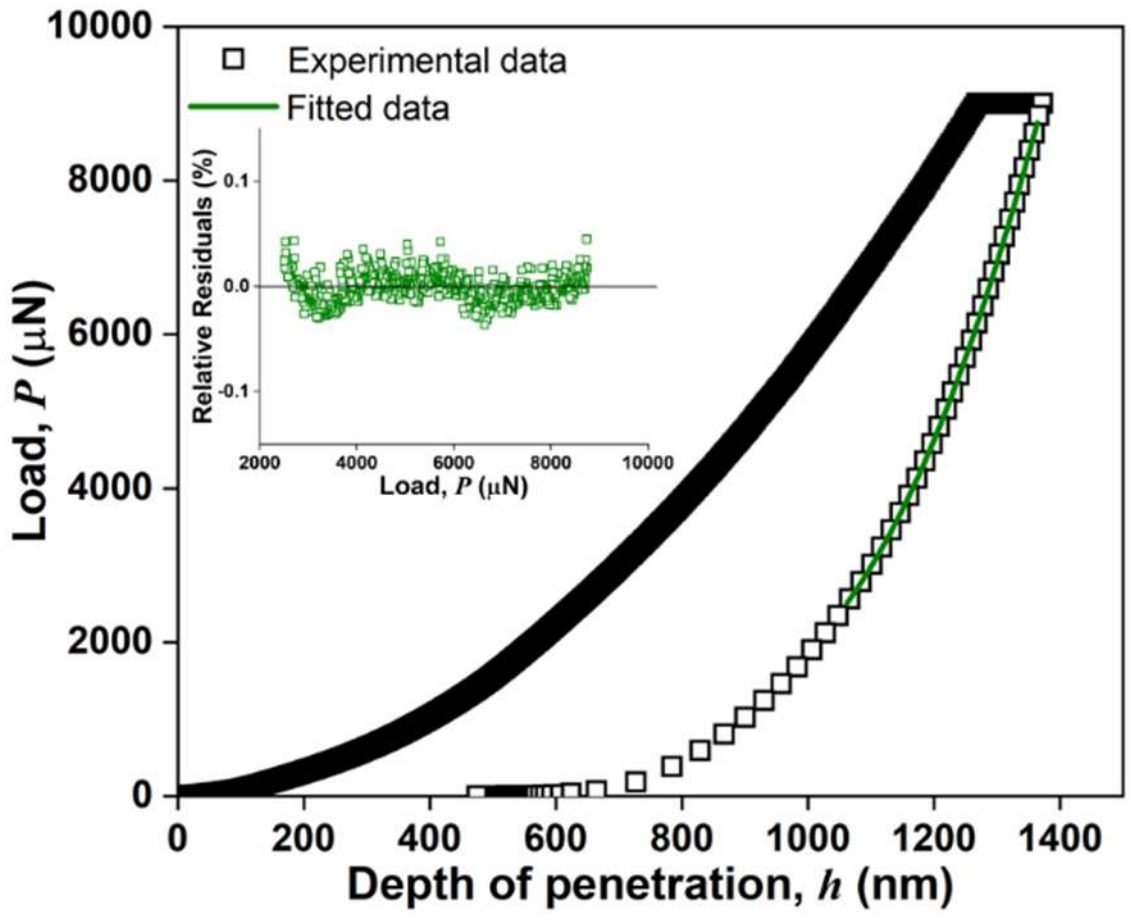

**Figure S.2.1:** $P$-$h$ curves for PB05 for $P_m$ = 9000 μN. Unloading curve fitted to GPL equation.

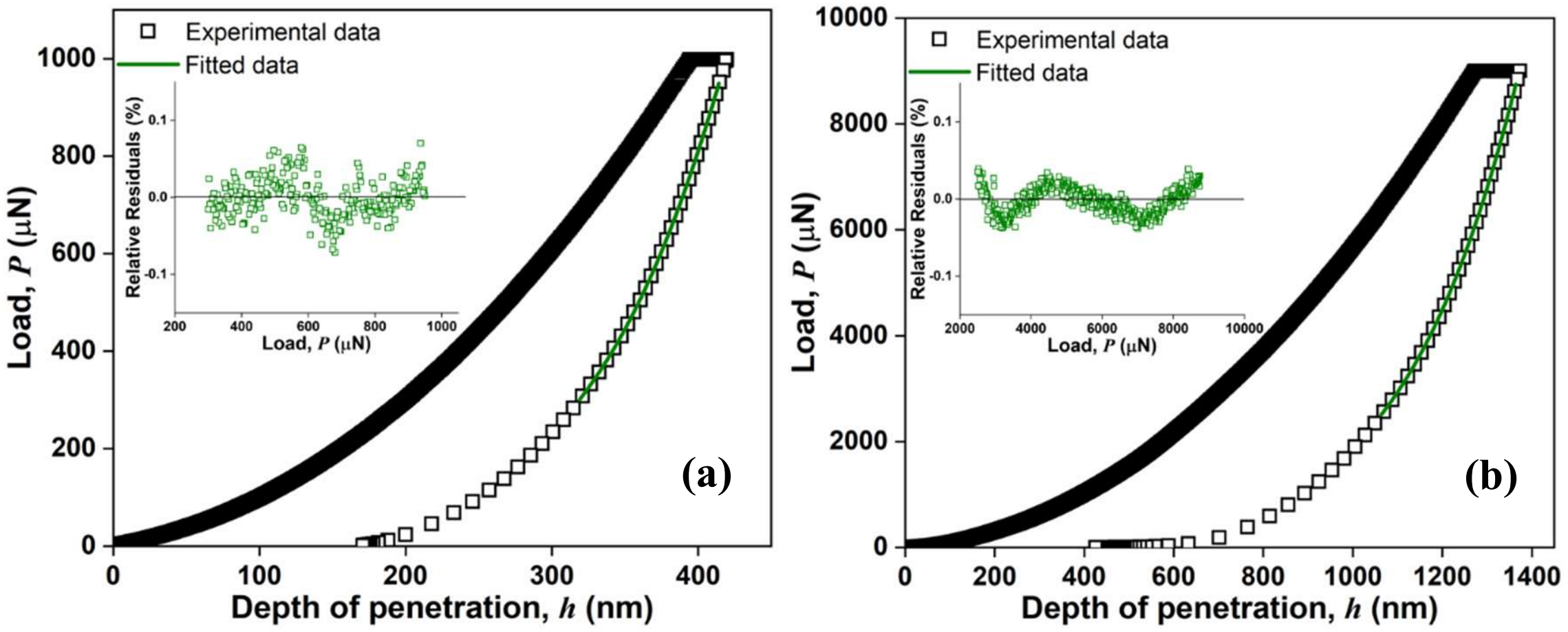

**Figure S.2.2:** $P$-$h$ curves for HB15 for $P_m$ = (a) 1000 μN, (b) 9000 μN. Unloading curve fitted to GPL equation.

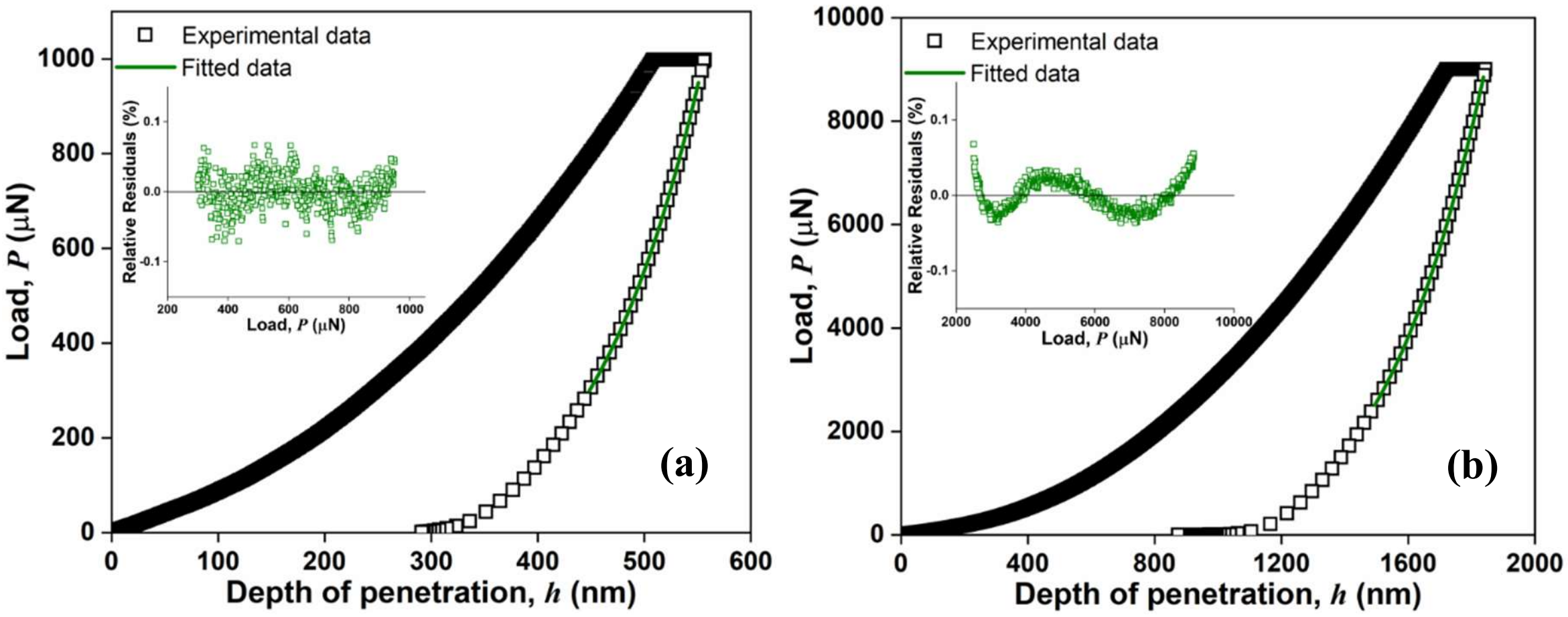

**Figure S.2.3:** $P$-$h$ curves for PC for $P_m$ = (a) 1000 μN, (b) 9000 μN. Unloading curve fitted to GPL equation.

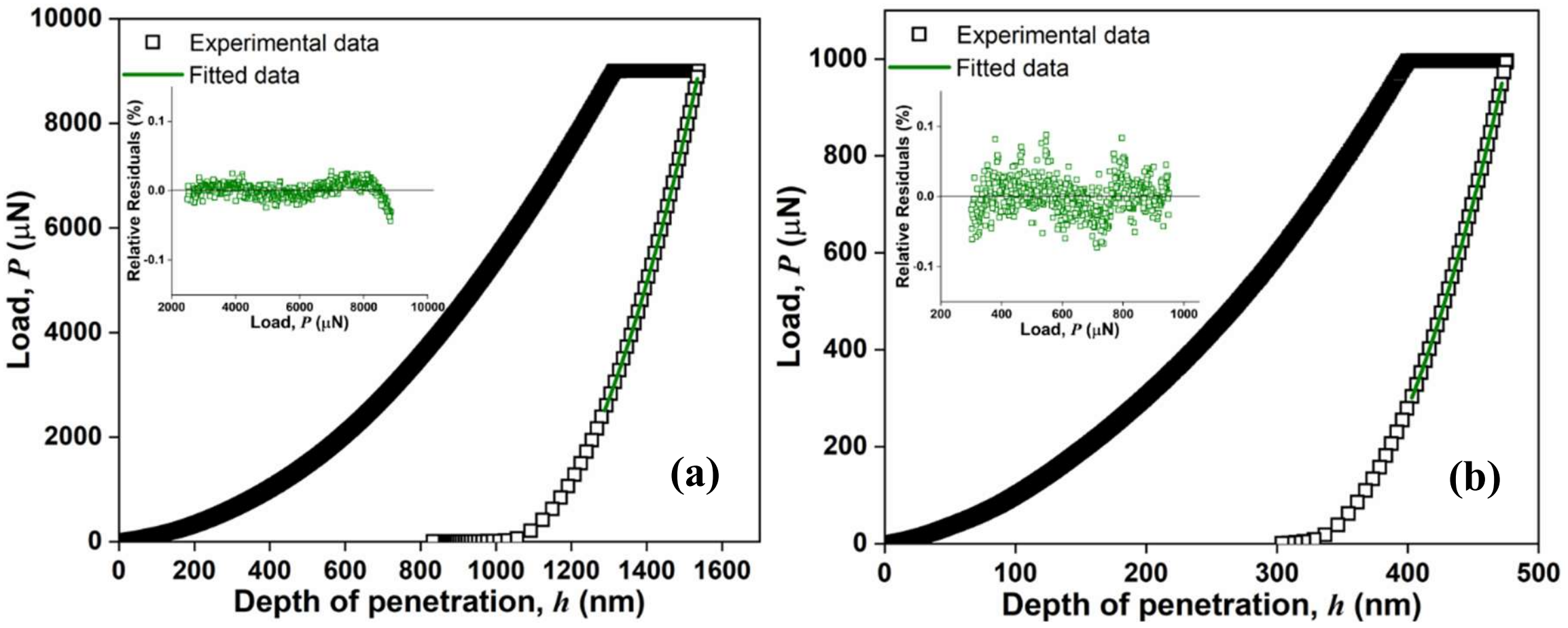

**Figure S.2.4:** $P$-$h$ curves for PMMA for $P_m$ = (a) 1000 μN, (b) 9000 μN. Unloading curve fitted to GPL equation.

# Supporting Information 4

## S.4.1. SPM images of residual imprints

The SPM top view image to compute $A_c$, are presented on PB05 for $P_m$ = 9000 μN (Fig. S.4.1), on HB15 for $P_m$ = 1000 μN (Fig. S.4.2(a)), for $P_m$ = 9000 μN (Fig. S.4.2(b)), on PC for $P_m$ = 1000 μN (Fig. S.4.3(a)), for $P_m$ = 9000 μN (Fig. S.4.3(b)), and on PMMA for $P_m$ = 1000 μN (Fig. S.4.4(a)), for $P_m$ = 9000 μN (Fig. S.4.4(b)).

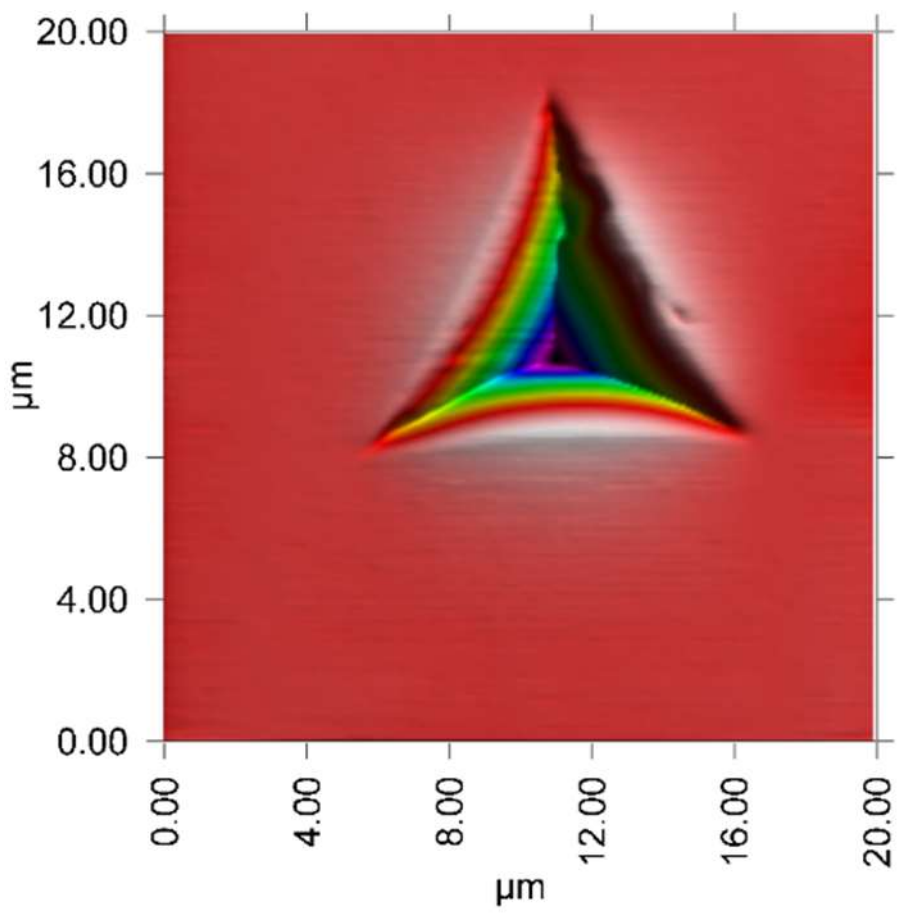

**Figure S.4.1:** Top view of residual imprint of PB05 for $P_m$ = 9000 μN.

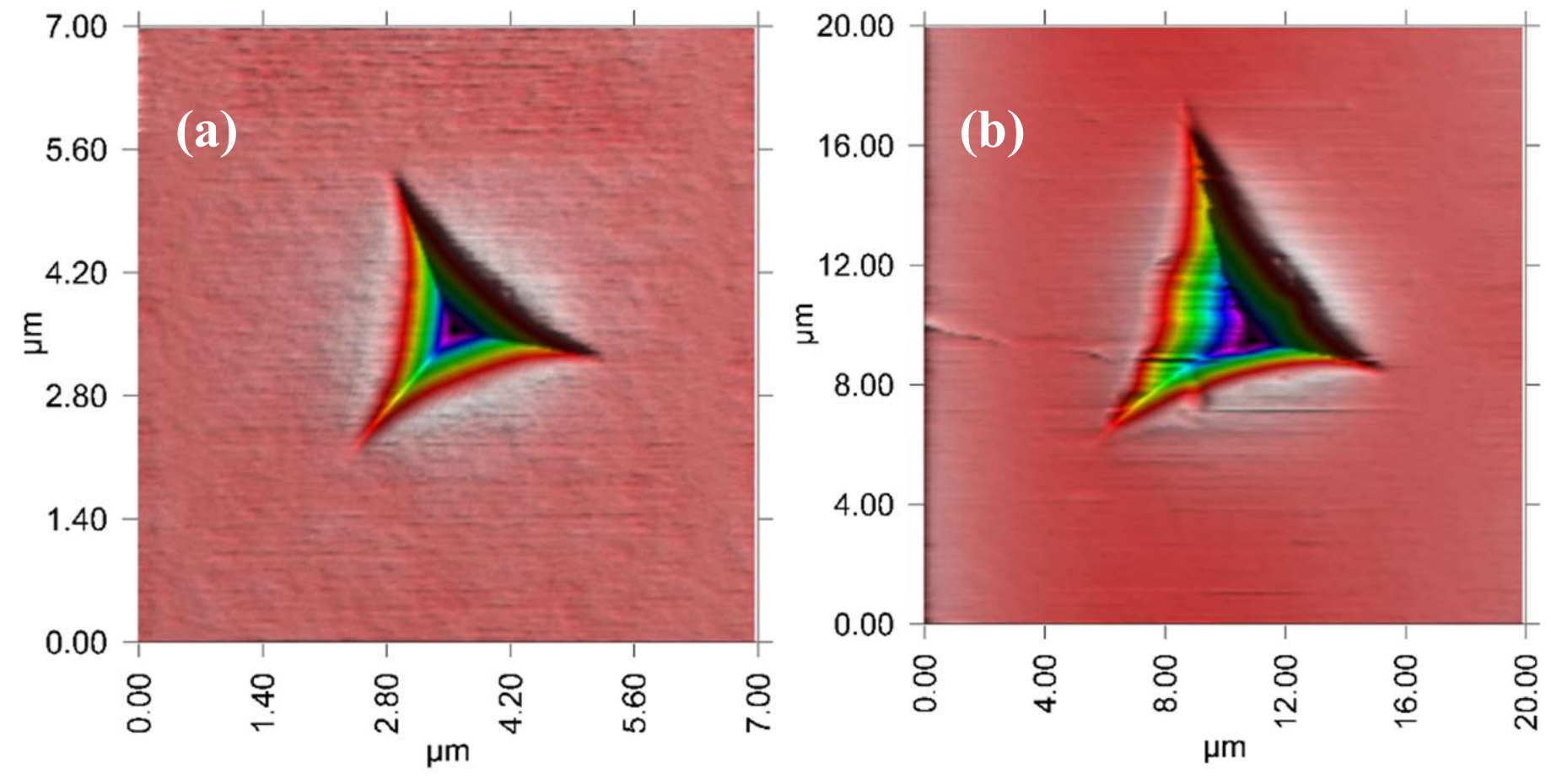

**Figure S.4.2:** Top view of residual imprint of HB15 for $P_m$ = (a) 1000 μN, (b) 9000 μN.

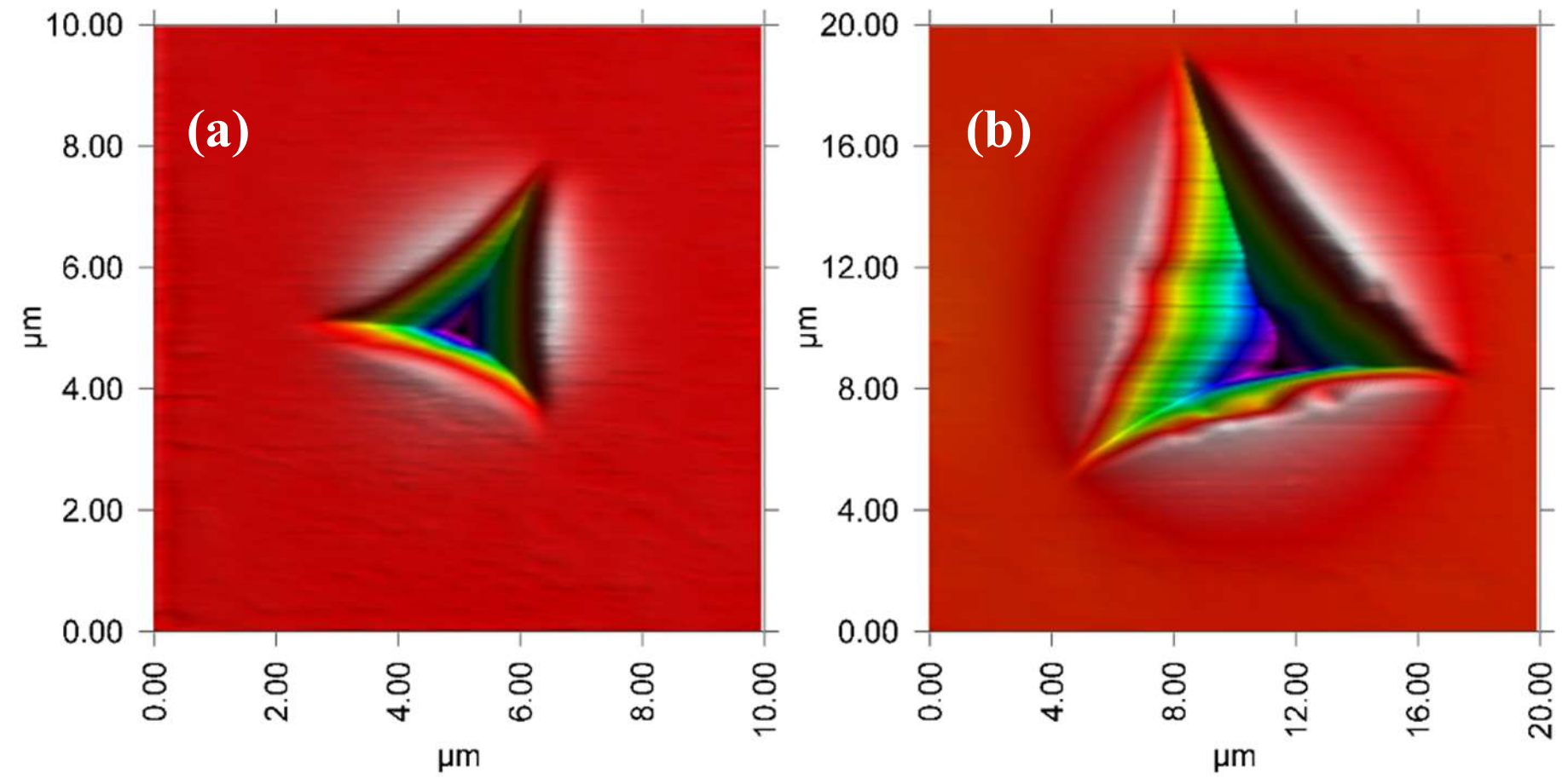

**Figure S.4.3:** Top view of residual imprint of PC for $P_m$ = (a) 1000 µN, (b) 9000 µN.

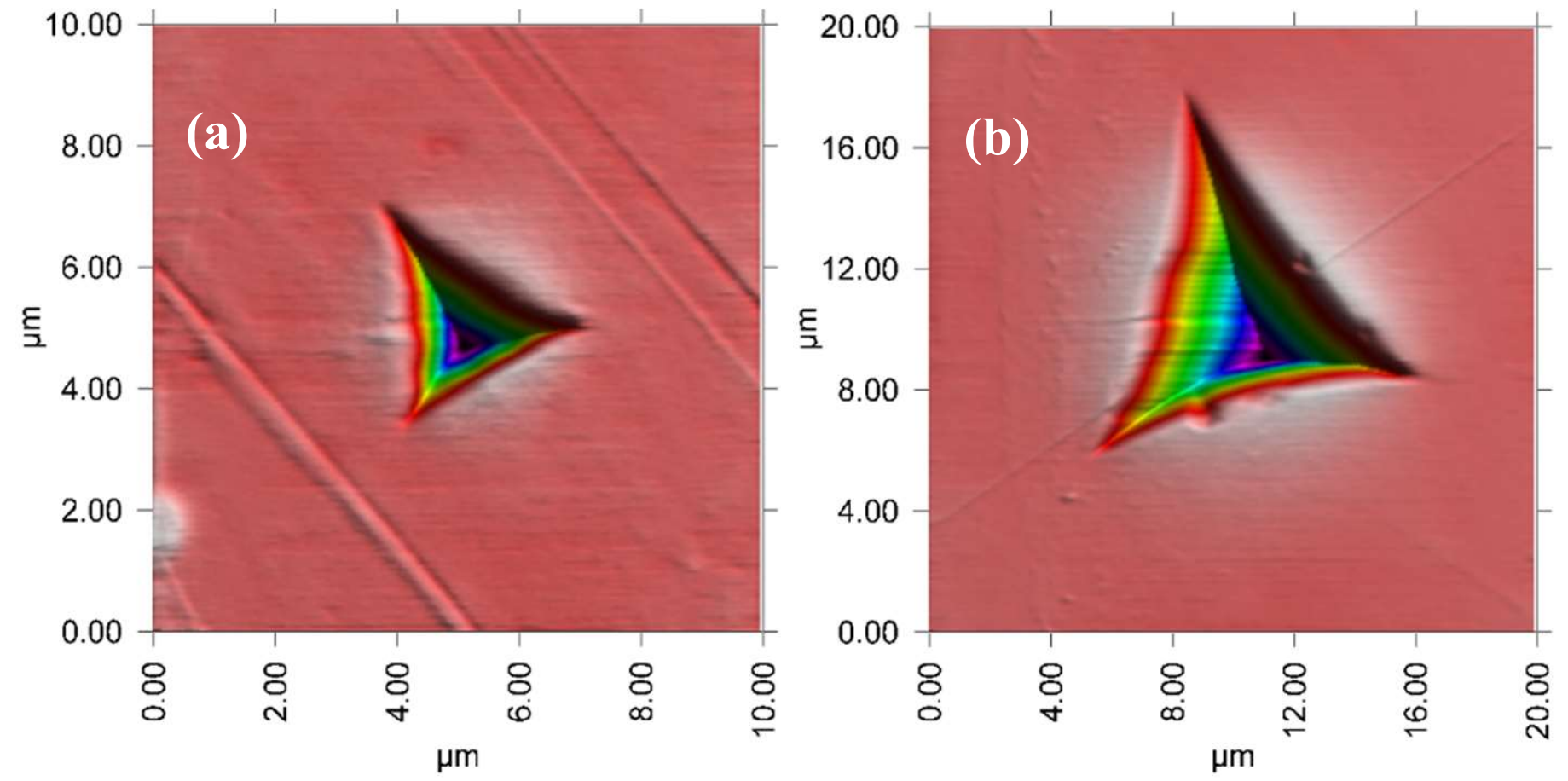

**Figure S.4.4:** Top view of residual imprint of PMMA for $P_m$ = (a) 1000 µN, (b) 9000 µN.

# Supporting Information 5

## S.5.1. Uniform $h_s$ during nanoindentation for an elastic medium

The mapping of the nanoindentation via a perfectly conical indenter to Berkovich nanoindentation has been described Figs. 4 – 7. The interpretation in Fig. S.5.1 retains the assumption of the effective circular cross-section, in providing the ideal $h_s$ estimate (eqn. 4). We consider a uniform sink-in during purely elastic nanoindentation, all about the triangle's perimeter.

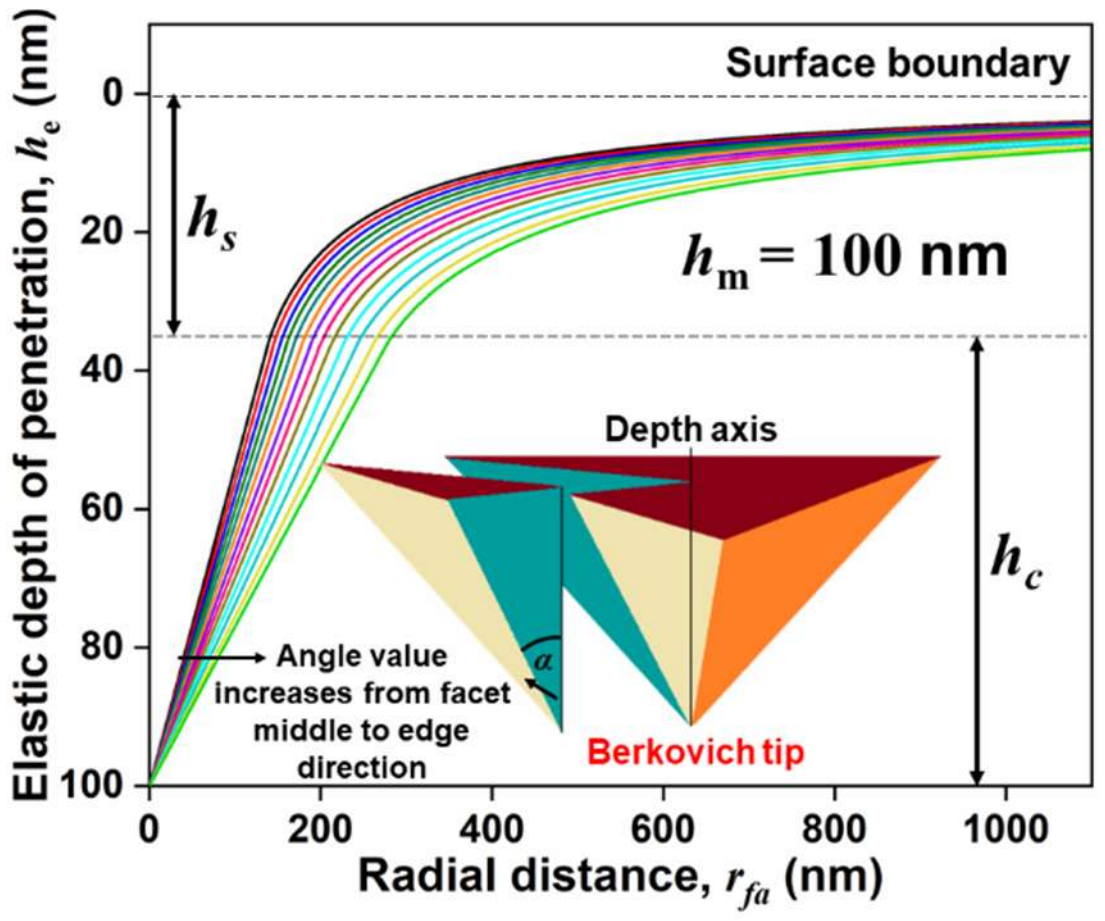

**Figure S.5.1:** Deformation profile at different angles from the facet middle direction to edge direction, for an elastic medium. $h_m$ =100 nm and the cone angle varies from 65.27° to 77.03°.

## S.5.2. Radial shift during deformation

Figure S.5.2 illustrates the inward (radial) surface displacement as described by eqns. 3 – 5 and Fig. 4.

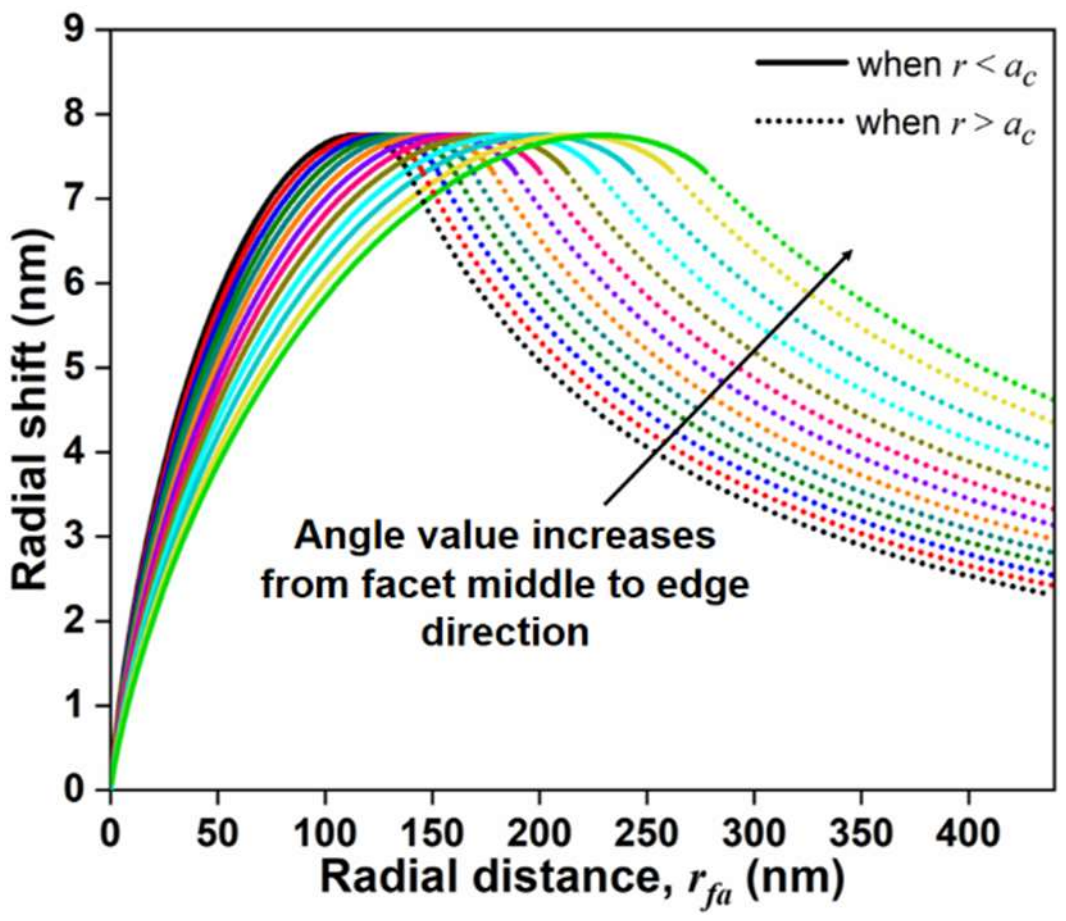

**Figure S.5.2:** Radial shift when $r < a_c$ (solid line) and $r > a_c$ (dotted line) from the facet middle direction to the edge direction, for an elastic medium, following eqn. 3 and eqn. 5. $h_{\mathrm{m}}$ =100 nm, $\nu$ = 0.35, and the cone angle varies from 65.27° to 77.03°.

## S.5.3. $h_s$ estimation from $S^{\mathrm{GPL}}$

We know that [4], for purely elastic deformation, $h_c / h_{\mathrm{m}} = 2/\pi$

$$\Rightarrow \frac{h_c - h_{\mathrm{ne}}}{h_{\mathrm{m}} - h_{\mathrm{ne}}} = \frac{2}{\pi}$$ (Corresponding to the elastic contribution to the total deformation)

$$\Rightarrow 1 - \frac{h_c - h_{\mathrm{ne}}}{h_{\mathrm{m}} - h_{\mathrm{ne}}} = 1 - \frac{2}{\pi}, \Rightarrow \frac{h_{\mathrm{m}} - h_c}{h_{\mathrm{m}} - h_{\mathrm{ne}}} = \frac{h_s}{h_{\mathrm{m}} - h_{\mathrm{ne}}} = \frac{\pi - 2}{\pi}$$

$$h_s = \frac{\pi - 2}{\pi}\left(h_{\mathrm{m}} - h_{\mathrm{ne}}\right) \qquad \text{eqn. S.5.1}$$

From eqn. 2 and eqn. 6, $S^{\mathrm{GPL}} = \left(\frac{h_{\mathrm{m}} - h_{\mathrm{ne}}}{P_{\mathrm{m}}} \frac{dD}{dF}\right)^{-1}, \Rightarrow h_{\mathrm{m}} - h_{\mathrm{ne}} = \left(\frac{S^{\mathrm{GPL}}}{P_{\mathrm{m}}} \frac{dD}{dF}\right)^{-1}$

Combining with eqn. S.5.1,

$$\Rightarrow h_s^{\mathrm{GPL}} = \frac{\pi - 2}{\pi}\left(\frac{S^{\mathrm{GPL}}}{P_{\mathrm{m}}}\frac{dD}{dF}\right)^{-1} \qquad \text{eqn. S.5.2}$$

# Supplementary Information 6

## S.6.1. Residual imprint analysis

As shown in Fig.S.6.1 for PB05 and HB15, SPM images indicate that the $a$-values (corresponding to instantaneous, permanent plastic deformation) remain unchanged even after 12 hrs. after unloading.

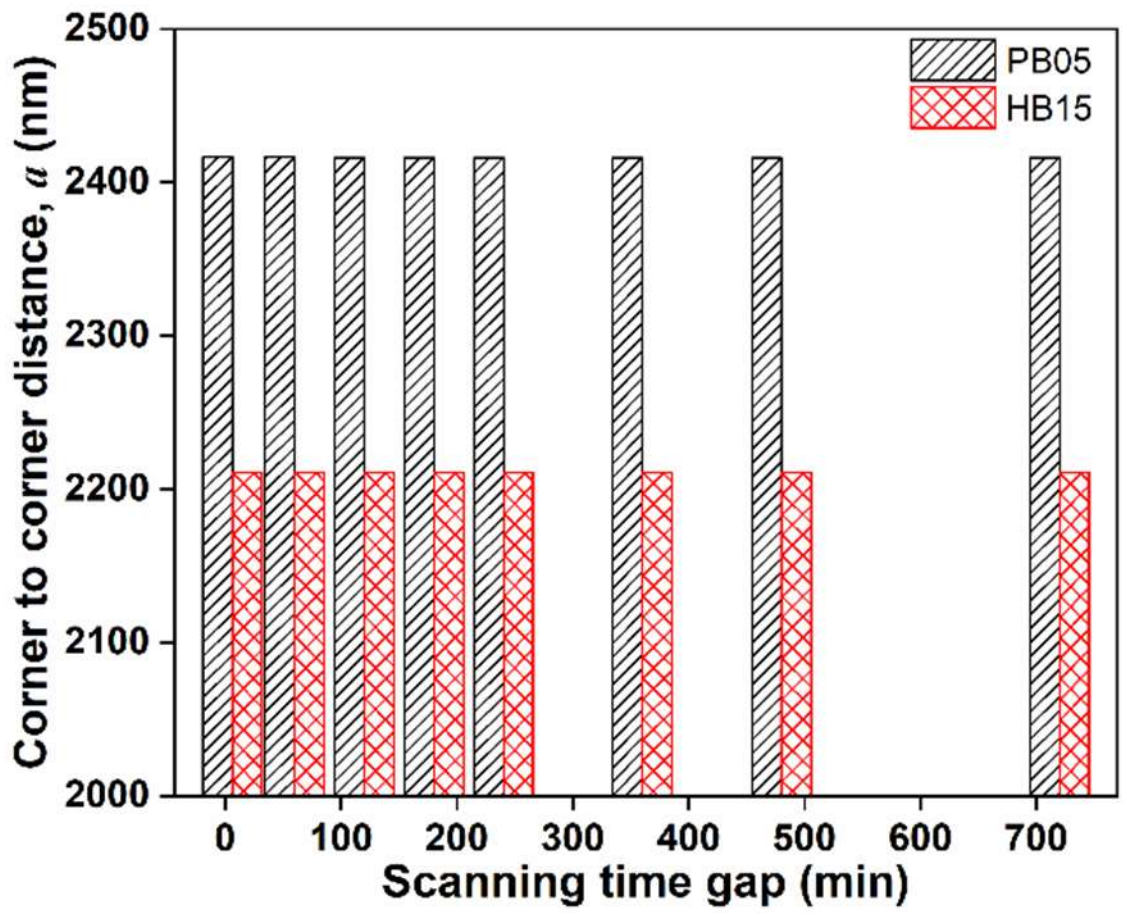

**Figure S.6.1:** $a$-line observation via SPM images over 12 hrs. duration for PB05 and HB15.

The viscoelastic variation in $h^{\mathrm{PU}}$ is illustrated in Fig. S.6.2(a)-(b) (PB05) and Fig. S.6.3(a)-(b) (HB15).

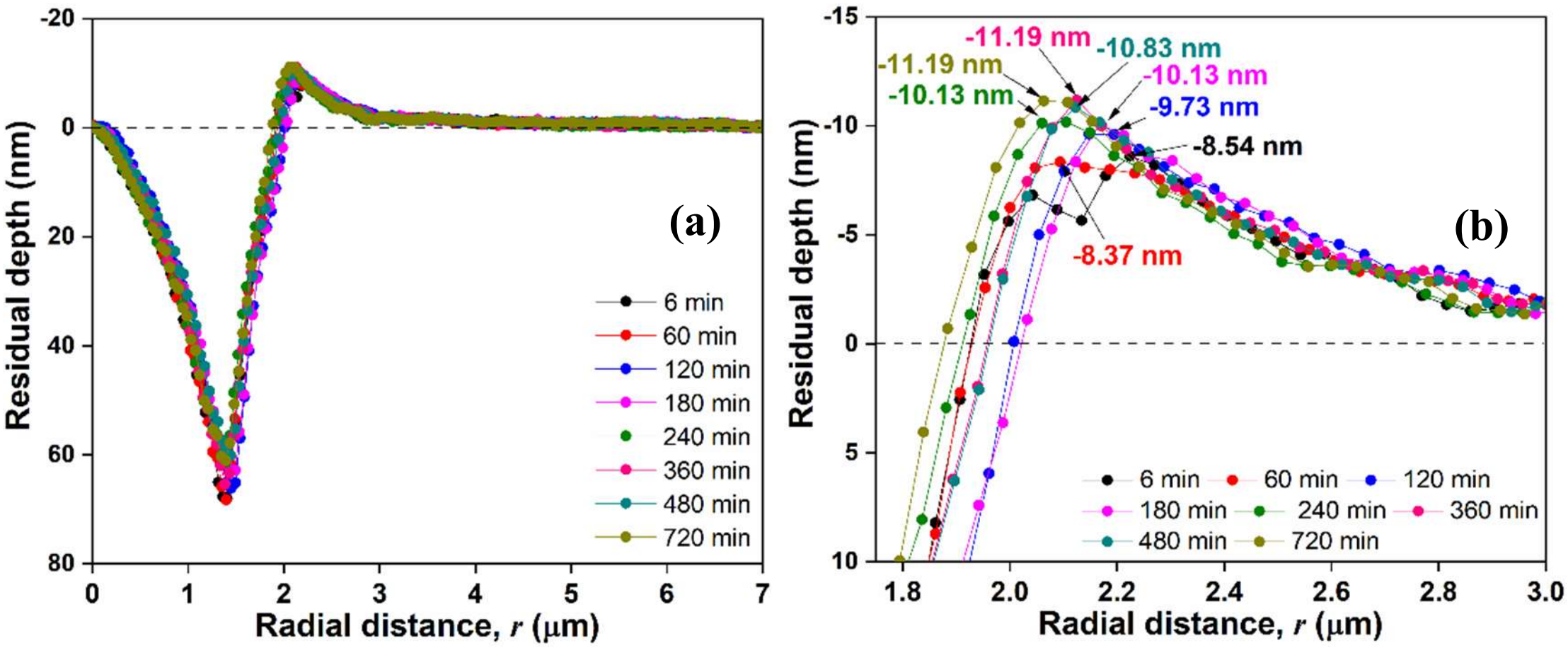

**Figure S.6.2:** Scan line of residual indent imprint for PB05 (a) full and (b) magnified pile-up region.

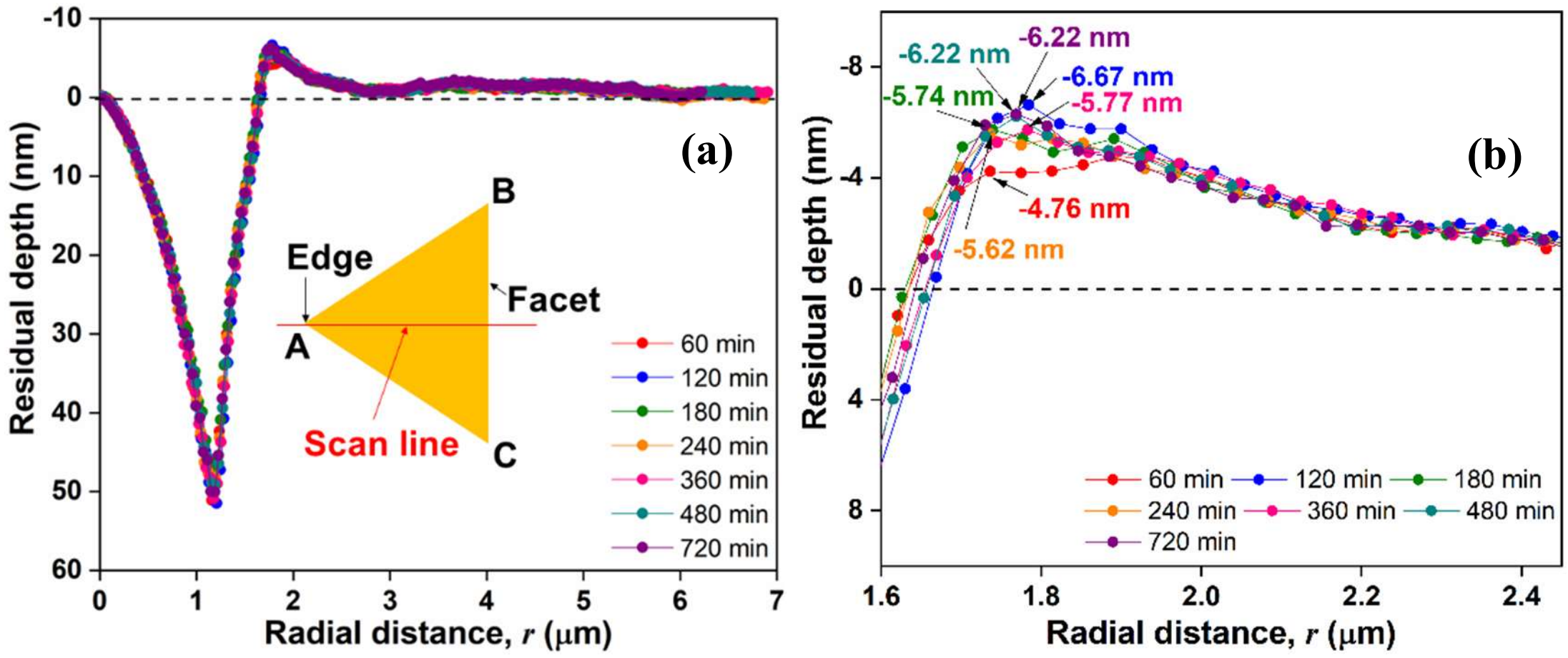

**Figure S.6.3:** Scan line of residual indent imprint for HB15 (a) full and (b) magnified pile-up region.

## S.6.2. Corrections to Estimates of $A_c$

V. *Pile-up parameters*

The values for $r^{\mathrm{PU}}$, $\eta$ (radians), and the projected area of the elevation, $A_c^{\mathrm{PU}}$, are listed in Table S.6.1.

**Table S.6.1:** The $r^{\mathrm{PU}}$, $\eta$, and $A_c^{\mathrm{PU}}$ values based on various pile-up measures.

| Parameters | Pile-up height (nm) | PB05 | | HB15 | | PC | | PMMA | |
|---|---|---|---|---|---|---|---|---|---|
| | | 1000 µN | 9000 µN | 1000 µN | 9000 µN | 1000 µN | 9000 µN | 1000 µN | 9000 µN |
| $r^{\mathrm{PU}}$ (mm) | $h_c^{\mathrm{PU}}$ | 0.22 ± 0.06 | 0.73 ± 0.15 | 1.04 ± 0.11 | 1.52 ± 0.21 | 0.14 ± 0.05 | 0.99 ± 0.12 | 0.33 ± 0.03 | 3.35 ± 1.75 |
| | $h_{\mathrm{ne,m}}^{\mathrm{PU}}$ | 0.15 ± 0.08 | 0.46 ± 0.05 | 0.34 ± 0.01 | 0.92 ± 0.03 | 0.06 ± 0.01 | 0.43 ± 0.04 | 0.28 ± 0.04 | 1.13 ± 0.12 |
| | $h_{r,c}^{\mathrm{PU}}$ | 0.13 ± 0.03 | 0.41 ± 0.08 | 0.54 ± 0.06 | 0.80 ± 0.12 | 0.10 ± 0.02 | 0.29 ± 0.08 | 0.13 ± 0.02 | 1.34 ± 0.76 |
| | $h_{\mathrm{m}}^{\mathrm{PU}}$ | 0.08 ± 0.01 | 0.26 ± 0.01 | 0.18 ± 0.00 | 0.41 ± 0.05 | 0.04 ± 0.00 | 0.18 ± 0.00 | 0.09 ± 0.01 | 0.38 ± 0.02 |
| $\eta$ (radian) | $h_c^{\mathrm{PU}}$ | 0.41 ± 0.09 | 0.41 ± 0.07 | 0.08 ± 0.01 | 0.19 ± 0.02 | 0.90 ± 0.25 | 0.39 ± 0.05 | 0.30 ± 0.03 | 0.12 ± 0.05 |
| | $h_{\mathrm{ne,m}}^{\mathrm{PU}}$ | 0.57 ± 0.03 | 0.63 ± 0.06 | 0.24 ± 0.01 | 0.30 ± 0.01 | 2.03 ± 0.37 | 0.88 ± 0.08 | 0.36 ± 0.05 | 0.28 ± 0.03 |
| | $h_{r,c}^{\mathrm{PU}}$ | 0.69 ± 0.13 | 0.73 ± 0.14 | 0.15 ± 0.02 | 0.35 ± 0.05 | 1.25 ± 0.22 | 1.39 ± 0.30 | 0.75 ± 0.09 | 0.30 ± 0.13 |
| | $h_{\mathrm{m}}^{\mathrm{PU}}$ | 1.06 ± 0.09 | 1.12 ± 0.03 | 0.46 ± 0.01 | 0.69 ± 0.08 | 2.70 ± 0.07 | 2.14 ± 0.06 | 1.10 ± 0.10 | 0.83 ± 0.05 |
| $A_c^{\mathrm{PU}}$ (µm$^2$) | $h_c^{\mathrm{PU}}$ | 0.07 ± 0.01 | 0.78 ± 0.13 | 0.01 ± 0.00 | 0.33 ± 0.04 | 0.28 ± 0.08 | 1.29 ± 0.14 | 0.07 ± 0.01 | 0.27 ± 0.12 |
| | $h_{\mathrm{ne,m}}^{\mathrm{PU}}$ | 0.10 ± 0.01 | 1.20 ± 0.13 | 0.04 ± 0.00 | 0.53 ± 0.02 | 0.63 ± 0.12 | 2.93 ± 0.23 | 0.08 ± 0.01 | 0.64 ± 0.07 |
| | $h_{r,c}^{\mathrm{PU}}$ | 0.12 ± 0.02 | 1.39 ± 0.26 | 0.02 ± 0.00 | 0.62 ± 0.08 | 0.39 ± 0.07 | 4.61 ± 0.96 | 0.17 ± 0.02 | 0.69 ± 0.31 |
| | $h_{\mathrm{m}}^{\mathrm{PU}}$ | 0.18 ± 0.02 | 2.13 ± 0.07 | 0.07 ± 0.00 | 1.22 ± 0.14 | 0.84 ± 0.01 | 7.13 ± 0.28 | 0.24 ± 0.02 | 1.89 ± 0.13 |

VI. *Tip tilt effects on contact area estimations*

Since the indenter tip is attached to the holder manually, some non-zero tilt between the surface and the tip, is inevitable. Figure S.6.4 illustrates the representative effects of tip- surface tilt that we have encountered in our experiments. Instead of the imprint being an equilateral triangle, we find two limiting cases of isosceles triangles.

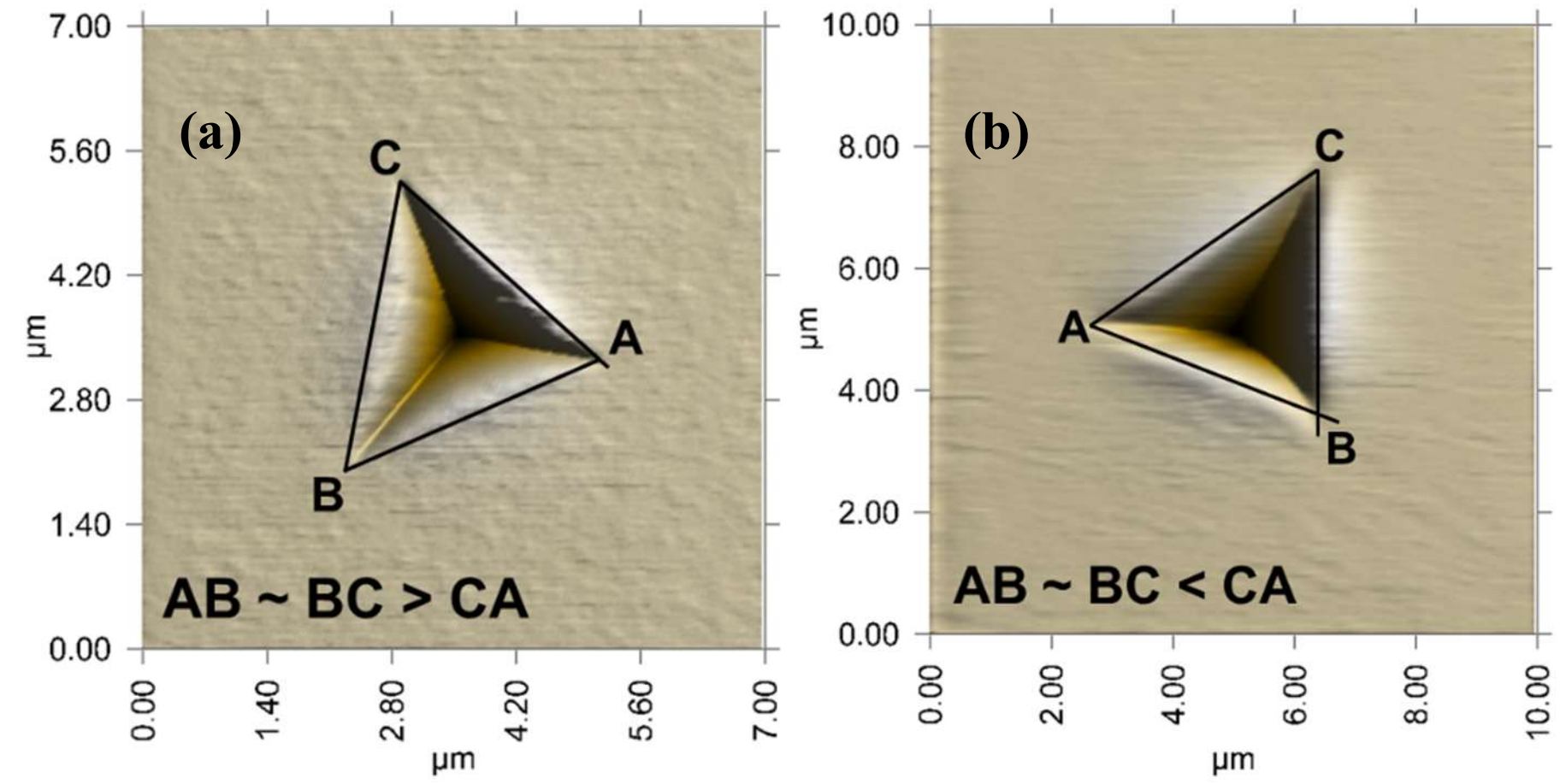

**Figure S.6.4:** Side edge length relation from top view SPM images for $P_{\mathrm{m}}$ = 1000 μN for (a) PB05, (b) PC.

The quantified corrections to the $A_c$ due to tilt, expressed in terms of the relative lengths of the imprint triangle, have been provided by Jakes *et al.* [5]. The lengths of triangle sides in our samples, deviate from equilateral values, by ~ 3% to ~ 16%. These deviations result in $A_c$ estimate corrections of < 1% for PB05 and HB15, ~ 1.5% for PC and ~ 3% for PMMA. Still, these corrections have been included here for completeness, and have been listed in Table S.6.2.

Table S.6.2: **Long side and short side ratio, and tilt correction factor when $P_m$ = 1000 µN and $P_m$ = 9000 µN.**

| Parameters | $P_m$ (µN) | PB05 | HB15 | PC | PMMA |
|---|---|---|---|---|---|
| Side edge | 1000 | AB ~ BC > CA | AB ~ BC > CA | AB ~ BC < CA | AB ~ BC < CA |
| | 9000 | AB ~ BC > CA | AB ~ BC > CA | AB ~ BC > CA | AB ~ BC > CA |
| $\frac{\text{Long side}}{\text{Short side}}$ | 1000 | 1.074 ± 0.007 | 1.065 ± 0.005 | 1.120 ± 0.004 | 1.164 ± 0.009 |
| | 9000 | 1.061 ± 0.015 | 1.036 ± 0.010 | 1.092 ± 0.016 | 1.140 ± 0.008 |
| Tilt correction factor | 1000 | 1.007 | 1.006 | 1.015 | 1.027 |
| | 9000 | 1.005 | 1.002 | 1.012 | 1.030 |

# Supporting Information 7

## S.7.1. Literature formulae to evaluate $E_r$ and $H$

The stiffness, $S$, is given by eqn. 6. For any polymer, the $S$ values from the different methods, all lie numerically, within ~ 10% of each other (Table S.7.1). However, for $P_\mathrm{m}$ = 1000 µN, we find, $S$ (PB05) > $S$ (HB15), via all methods except the GPL method. For $P_\mathrm{m}$ = 9000 µN, $S$ (PB05) > $S$ (HB15), only for the PL equation fitting. Also recognizing that the polymer nanoindentation $h_f$ values, fitted via all the literature methods, are unphysical. Hence, in general, only the data-faithful PL ($h_{\mathrm{ne}0}$ basis) or the GPL ($h_{\mathrm{ne}2}$ basis) methods yield $S$ (PB05) < $S$ (HB15).

Table S.7.1: **Calculated $S$ (µN/nm) estimates via various methods reported in the literature.**

| References | $P_\mathrm{m}$ (µN) | PB05 | HB15 | PC | PMMA |
|---|---|---|---|---|---|
| Oliver-Pharr [1] | 1000 | 11.53 ± 0.02 | 9.93 ± 0.03 | 9.13 ± 0.01 | 12.16 ± 0.06 |
| | 9000 | 31.86 ± 0.10 | 30.83 ± 0.02 | 26.45 ± 0.02 | 32.95 ± 0.04 |
| Kossman *et al.* [2] | 1000 | 11.86 ± 0.03 | 11.06 ± 0.05 | 9.88 ± 0.02 | 12.17 ± 0.06 |
| | 9000 | 32.96 ± 0.07 | 35.27 ± 0.10 | 28.81 ± 0.03 | 33.03 ± 0.03 |
| Liu *et al.* [3] | 1000 | 10.80 ± 0.05 | 10.32 ± 0.09 | 9.51 ± 0.16 | 12.28 ± 0.16 |
| | 9000 | 30.07 ± 0.10 | 32.79 ± 0.36 | 26.71 ± 0.40 | 33.70 ± 0.26 |
| Gong *et al.* [6] | 1000 | 11.11 ± 0.02 | 10.14 ± 0.04 | 9.30 ± 0.01 | 12.20 ± 0.04 |
| | 9000 | 30.78 ± 0.11 | 31.75 ± 0.04 | 27.21 ± 0.02 | 33.26 ± 0.02 |
| GPL method | 1000 | 10.59 ± 0.09 | 11.02 ± 0.09 | 9.87 ± 0.04 | 12.22 ± 0.09 |
| | 9000 | 30.63 ± 0.14 | 34.31 ± 0.06 | 27.97 ± 0.02 | 33.12 ± 0.07 |

## S.7.2. Literature formulae to evaluate $E_r$ and $H$

We have estimated $E_r$ and $H$, following various methods proposed by previous researchers [1]–[3], [6]–[10]which are tabulated in Table S.7.2.

Table S.7.2: **Formulae to evaluate $E_r$ and $H$ used by previous researchers.**

| References | Unload curve fitting equations | Contact area measurement equations |
|---|---|---|
| Oliver-Pharr [1] | $P = B\left(h - h_f\right)^m$ | $h_c = h_{\mathrm{m}} - \left(\varepsilon P_{\mathrm{m}}/S\right);\ A_c = C_0 h_c^2 + \sum_{i=1}^{8} C_i h_c^{2^{-(i-1)}}$ |
| Randall *et al.*[10] | $P = B\left(h - h_f\right)^m$ | $A_c = \sqrt{3}a^2/4$ |
| Hochstetter *et al.* [7] | $P = B\left(h - h_f\right)^m$ | $A_c = C_0\left[\zeta\left(h_{\mathrm{m}} - \left(P_{\mathrm{m}}/S\right) + h_b\right)^2\right]$ |
| Zhu *et al.* [8] | -- | $A_c = 3\left(\pi r^2 \eta/360\right) - \left[3.765k\left(h_{\mathrm{m}} + h_{\mathrm{m}}^{\mathrm{PU}}\right)\right]\tan\alpha + \left(\sqrt{3}/4\right)\left[7.53\left(h_{\mathrm{m}} + h_{\mathrm{m}}^{\mathrm{PU}}\right)\right]$ |
| Hardiman *et al.* [9] | $P = B\left(h - h_f\right)^m$ | $h_c = h_{\mathrm{m}} - \left(\varepsilon P_{\mathrm{m}}/S\right);\ A_c = C_0 h_c^2 + \sum_{i=1}^{8} C_i h_c^{2^{-(i-1)}}$ |
| Kossman *et al.* [2] | $h/h_{\mathrm{m}} = h_f/h_{\mathrm{m}} + G\left(P/P_{\mathrm{m}}\right)^m$ | $h_c = h_{\mathrm{m}} - \left(\varepsilon P_{\mathrm{m}}/S\right);\ A_c = C_0 h_c^2 + \sum_{i=1}^{8} C_i h_c^{2^{-(i-1)}}$ |
| Liu *et al.* [3] | $h = h_f + a_1 P^{1/2} + ..... + a_6 P^{1/64}$ | $h_c = h_{\mathrm{m}} - \left(\varepsilon P_{\mathrm{m}}/S\right);\ A_c = C_0 h_c^2 + \sum_{i=1}^{8} C_i h_c^{2^{-(i-1)}}$ |
| Gong *et al.* [6] | $P = P_{\mathrm{m}}\left[a_0 + a_1\left(h/h_{\mathrm{m}}\right) + a_2\left(h/h_{\mathrm{m}}\right)^2\right]$ | $h_c = h_{\mathrm{m}} - \left(\varepsilon P_{\mathrm{m}}/S\right);\ A_c = C_0 h_c^2 + \sum_{i=1}^{8} C_i h_c^{2^{-(i-1)}}$ |

## S.7.3. Comparison with $E_r$ and $H$ estimates from literature methods

By following the formulae and methods in Table S.7.2, we have evaluated $E_r$ and $H$. These estimates are listed in Table S.7.3 and Table S.7.4, respectively.

Table S.7.3: $E_r$ **estimates via methods followed by earlier researchers.**

| Ref. | PB05 | | HB15 | | PC | | PMMA | |
|---|---|---|---|---|---|---|---|---|
| | 1000 μN | 9000 μN | 1000 μN | 9000 μN | 1000 μN | 9000 μN | 1000 μN | 9000 μN |
| [1] | 5.59 ± 0.03 | 4.97 ± 0.06 | 4.74 ± 0.01 | 4.81 ± 0.02 | 3.43 ± 0.03 | 2.97 ± 0.03 | 5.32 ± 0.03 | 4.46 ± 0.03 |
| [10] | 5.21 ± 0.02 | 4.27 ± 0.02 | 4.71 ± 0.02 | 4.30 ± 0.01 | 3.08 ± 0.05 | 2.69 ± 0.02 | 4.87 ± 0.03 | 4.05 ± 0.01 |
| [7] | 5.42 ± 0.03 | 4.56 ± 0.05 | 4.65 ± 0.01 | 4.43 ± 0.02 | 3.10 ± 0.04 | 2.72 ± 0.03 | 4.72 ± 0.03 | 4.06 ± 0.03 |
| [9] | 4.56 ± 0.02 | 4.05 ± 0.04 | 3.86 ± 0.01 | 3.96 ± 0.02 | 2.73 ± 0.05 | 2.40 ± 0.02 | 4.49 ± 0.02 | 3.77 ± 0.03 |
| [2] | 5.72 ± 0.03 | 5.11 ± 0.06 | 5.17 ± 0.01 | 5.38 ± 0.02 | 3.65 ± 0.04 | 3.19 ± 0.03 | 5.33 ± 0.03 | 4.46 ± 0.03 |
| [3] | 5.30 ± 0.04 | 4.74 ± 0.05 | 4.89 ± 0.05 | 5.06 ± 0.07 | 3.53 ± 0.08 | 2.99 ± 0.03 | 5.35 ± 0.03 | 4.54 ± 0.06 |
| [6] | 5.42 ± 0.03 | 4.83 ± 0.06 | 4.82 ± 0.01 | 4.93 ± 0.02 | 3.47 ± 0.04 | 3.04 ± 0.03 | 5.34 ± 0.03 | 4.49 ± 0.03 |

Table S.7.4: $H$ **estimates via methods followed by earlier researchers.**

| Ref. | PB05 | | HB15 | | PC | | PMMA | |
|---|---|---|---|---|---|---|---|---|
| | 1000 μN | 9000 μN | 1000 μN | 9000 μN | 1000 μN | 9000 μN | 1000 μN | 9000 μN |
| [1] | 298.5 ± 3.1 | 278.3 ± 5.1 | 289.8 ± 2.3 | 279.3 ± 3.3 | 178.7 ± 3.4 | 143.0 ± 2.8 | 244.1 ± 2.0 | 207.9 ± 3.1 |
| [10] | 259.4 ± 1.1 | 206.0 ± 0.9 | 285.3 ± 1.0 | 223.0 ± 1.2 | 139.9 ± 2.3 | 117.6 ± 1.8 | 203.6 ± 3.3 | 172.2 ± 1.0 |
| [7] | 281.1 ± 3.0 | 235.3 ± 4.5 | 278.3 ± 2.4 | 237.2 ± 2.9 | 141.6 ± 4.4 | 119.9 ± 2.5 | 198.8 ± 1.6 | 172.8 ± 2.7 |
| [8] | 222.5 ± 2.1 | 174.7 ± 2.3 | 220.2 ± 1.2 | 181.5 ± 2.7 | 98.8 ± 1.2 | 86.1 ± 1.8 | 163.4 ± 0.8 | 142.0 ± 2.3 |
| [9] | 198.1 ± 1.6 | 185.1 ± 2.7 | 191.7 ± 1.0 | 189.0 ± 2.4 | 110.2 ± 1.6 | 93.6 ± 1.9 | 173.4 ± 1.0 | 148.5 ± 2.3 |
| [2] | 295.3 ± 3.0 | 274.9 ± 5.9 | 277.6 ± 2.3 | 266.4 ± 3.1 | 172.7 ± 3.9 | 139.3 ± 2.7 | 244.1 ± 2.0 | 207.7 ± 3.1 |
| [3] | 306.1 ± 3.2 | 284.5 ± 5.4 | 285.2 ± 1.2 | 273.1 ± 2.5 | 174.8 ± 3.3 | 142.6 ± 3.3 | 243.8 ± 2.1 | 206.5 ± 2.8 |
| [6] | 302.7 ± 3.1 | 281.9 ± 5.2 | 287.3 ± 2.4 | 276.2 ± 3.2 | 176.0 ± 3.9 | 141.7 ± 2.8 | 243.8 ± 2.0 | 207.3 ± 3.1 |

Figure S.7.1 indicates that the nanoindentation properties of the polymers examined in this work, estimated employing previous methods, can deviate up to ~ 50% from our estimates.

For PMMA and PC, all methods yield the expected trend of lower $E_r$ and $H$ values for PC, compared to those for PMMA. In case of literature methods for SU-8, only methods where $S$ is obtained by fitting $h$ vs $P$ relationships [2], [3], and only to our nanoindentation data at $P_{\mathrm{m}}$ = 9000 μN, indicate an increase in $E_r$ with crosslinking. Methods that consider higher estimates of $A_c$, such as via residual imprint [10] or by adding the $h_{\mathrm{m}}^{\mathrm{PU}}$ (maximum pile-up) effect to $A_{\mathrm{m}}^{\Delta}$ (considering the full tip in contact, with $h_s$=0) [8], [11], yield the similar trends in $H$.

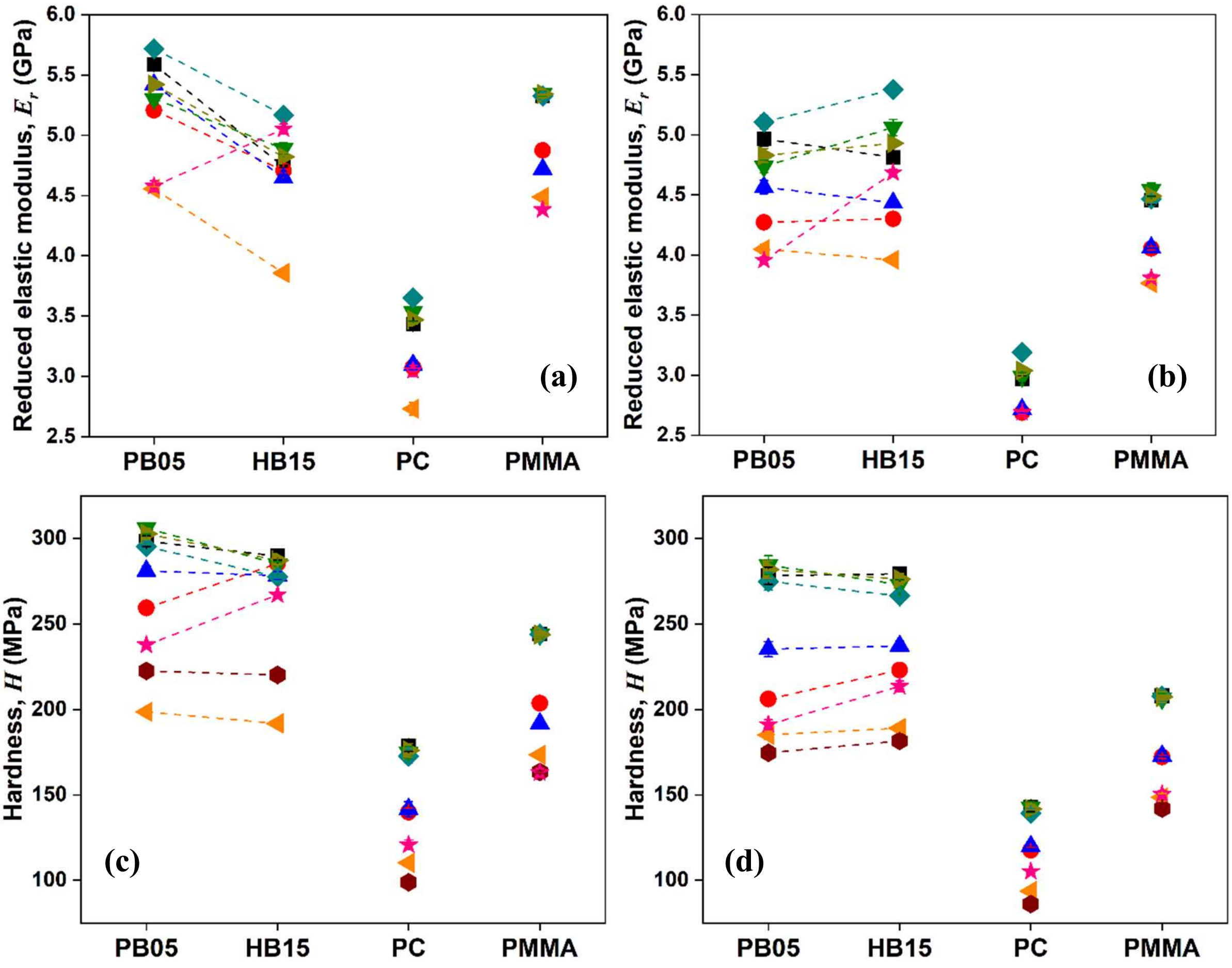

**Figure S.7.1:** $E_r$ values for (a) $P_{\mathrm{m}}$ = 1000 µN, and (b) $P_{\mathrm{m}}$ = 9000 µN and $H$ values for (c) $P_{\mathrm{m}}$ = 1000 µN, and (d) $P_{\mathrm{m}}$ = 9000 µN. Here, ■ [1], ● [10], ▲ [7], ▼ [3], ◆ [2], ◀ [9], ▶ [6], ⬢ [8], [11], and ★ Our data.

## S.7.4. Deviation of $E_r$ and $H$ from our estimated values

Figure S.7.2 and Fig. S.7.3 indicate that the $E_r$ and $H$ values of the polymers examined in this work, estimated employing earlier methods, deviate up to ~ 50% from our estimates.

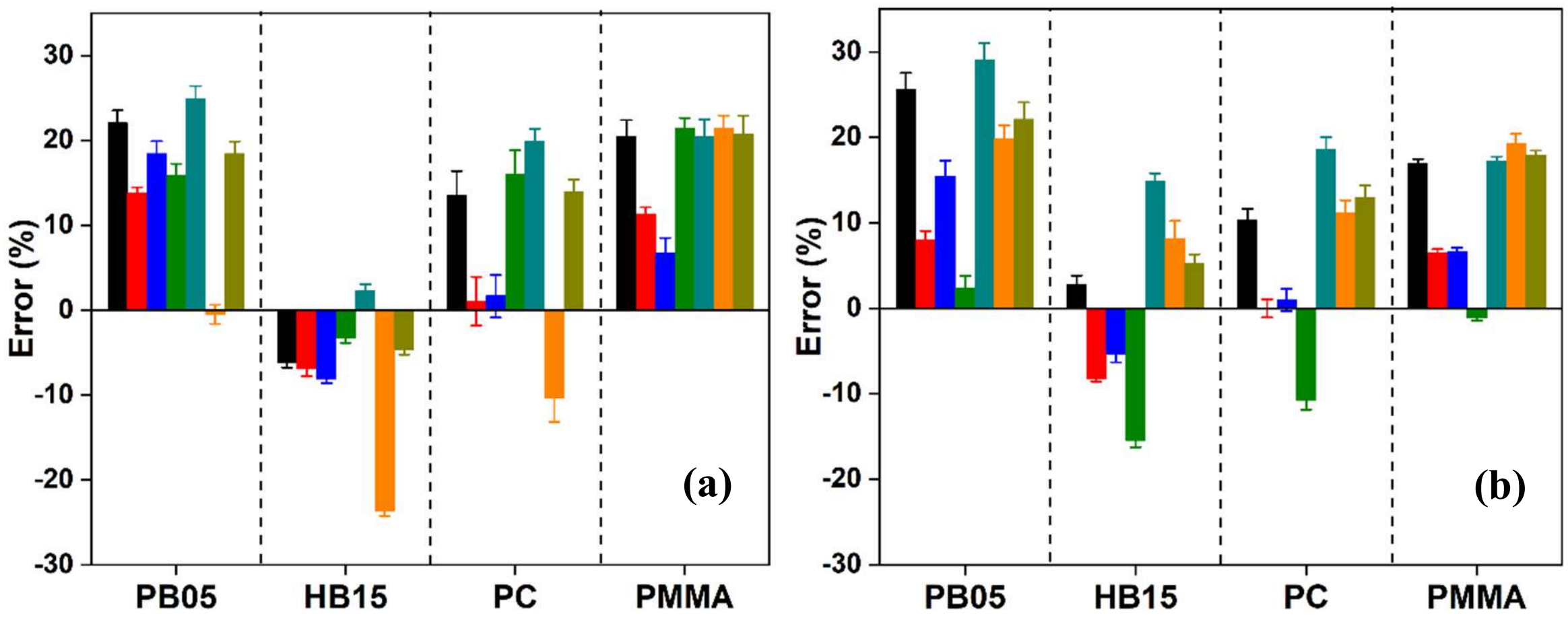

**Figure S.7.2:** Deviation of $E_r$ values for $P_m$ = (a) 1000 µN, (b) 9000 µN. Here, [1], [10], [7], [3], [2], [9], [6], and [8].

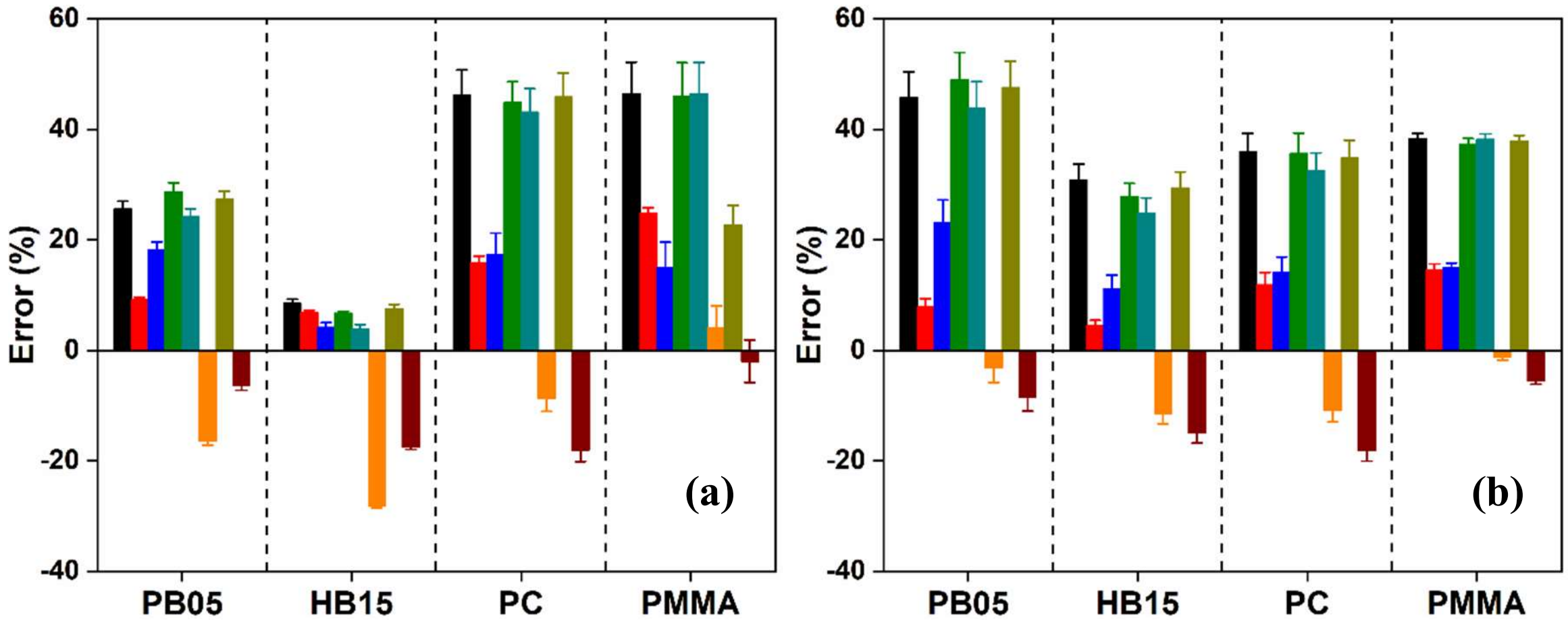

**Figure S.7.3:** Deviation of $H$ values for $P_m$ = (a) 1000 µN, (b) 9000 µN. Here, [1], [10], [7], [3], [2], [9], [6], and [8].